\PassOptionsToPackage{prologue,table}{xcolor}
\documentclass[acmsmall, prologue, table]{acmart}

\usepackage{mdframed}
\usepackage{subcaption}
\usepackage{multirow, bigstrut}
\usepackage{wrapfig}
\usepackage{longtable}
\usepackage{tabu}
\usepackage{tikz}
\usetikzlibrary{mindmap}
\usepackage{tikz}
\usepackage{forest}
\usetikzlibrary{trees,positioning,shapes,shadows,arrows.meta}
\usepackage{adjustbox}
\usepackage{bbding}
\usepackage{enumitem}
\usepackage{tabularx}
\usepackage{longtable}
\usepackage{booktabs}
\usepackage{pifont}
\usepackage{tikz}
\usepackage{spot}
\usepackage{pgf}
\usepackage{soul}
\usepackage{hyperref}
\usepackage{multirow}
\usepackage[table]{xcolor}
\definecolor{5anodecolor}{HTML}{C3ACD0}
\definecolor{tnodecolor}{HTML}{C8E4B2}
\definecolor{basiccolor}{HTML}{CAEDFF}
\definecolor{5inodeccolor}{HTML}{ffcfd2}
\definecolor{5vnodeccolor}{HTML}{fec89a}
\definecolor{5mmnodeccolor}{HTML}{f3d5b5}
\definecolor{memoryorange}{HTML}{faf3ed}
\definecolor{agentblue}{HTML}{c6cdf7}%{edf3fa}
\definecolor{envgreen}{HTML}{def3d0}
\definecolor{brainpurple}{HTML}{d9b8f1}
\definecolor{agentsecred}{HTML}{fa8080}
\definecolor{bblue}{HTML}{A1CAF1}
\definecolor{pblue}{HTML}{CAE1FF}%BCD4E6
\definecolor{actblue}{HTML}{87CEEB}%B0E0E6}

\newcommand*\emptycirc[1][0.75ex]{\tikz\draw (0,0) circle (#1);} 
\newcommand*\halfcirc[1][0.75ex]{%
  \begin{tikzpicture}
  \draw[fill] (0,0)-- (90:#1) arc (90:270:#1) -- cycle ;
  \draw (0,0) circle (#1);
  \end{tikzpicture}}
  
\newcommand*\fullcirc[1][0.75ex]{\tikz\fill (0,0) circle (#1);}

\newcommand{\psquare}{\tikz{\fill[pblue] (0,0) rectangle (1.7ex,1.7ex);}}
\newcommand{\bsquare}{\tikz{\fill[bblue] (0,0) rectangle (1.7ex,1.7ex);}}
\newcommand{\actsquare}{\tikz{\fill[actblue] (0,0) rectangle (1.7ex,1.7ex);}}
\newcommand{\agentsquare}{\tikz{\fill[agentblue] (0,0) rectangle (1.7ex,1.7ex);}}
\newcommand{\memorysquare}{\tikz{\fill[orange!30] (0,0) rectangle (1.7ex,1.7ex);}}
\newcommand{\envsquare}{\tikz{\fill[envgreen] (0,0) rectangle (1.7ex,1.7ex);}}

\newcommand{\todo}[1]{}
\renewcommand{\todo}[1]{{\color{red} TODO: {#1}}}
\usepackage{tcolorbox}
\tcbuselibrary{theorems, breakable}
\newcommand{\ie}{\textit{i.e.}}
\newcommand{\eg}{\textit{e.g.}}

\newtcbtheorem[number within=section]{hypothetical}{Knowledge Essentials}{
  breakable,
  colback=blue!5,
  colframe=blue!35!black,
  fonttitle=\bfseries}{x}
\definecolor{mycolor}{rgb}{0.122, 0.435, 0.698}% Rule colour
\definecolor{gray1}{gray}{0.3}

\AtBeginDocument{%
  \providecommand\BibTeX{{%
    \normalfont B\kern-0.5em{\scshape i\kern-0.25em b}\kern-0.8em\TeX}}}

\setcopyright{acmcopyright}
\copyrightyear{2024}
\acmYear{2024}
\acmDOI{XXXXXXX.XXXXXXX}

\begin{document}

\title{AI Agents Under Threat: A Survey of Key Security Challenges and Future Pathways}

\author{Zehang Deng}
\authornote{Both authors contributed equally to this research.}
\email{zehangdeng@swin.edu.au}
\affiliation{%
  \institution{Swinburne University of Technology}
  % \streetaddress{P.O. Box 1212}
  % \country{USA}
  \country{Australia}
}
\author{Yongjian Guo}
\authornotemark[1]
\affiliation{%
  \institution{Tianjin Univeristy}
  % \streetaddress{P.O. Box 1212}
  % \country{USA}
  \country{China}
}

% \affiliation{%

% \affiliation{%

% \affiliation{%

% \affiliation{%

% \affiliation{%

% \affiliation{%

% \affiliation{%

\author{Changzhou Han}
\email{changzhouhan@swin.edu.au}
\affiliation{%
  \institution{Swinburne University of Technology}
  % \city{Melbourne}
  % \state{VIC}
  \country{Australia}
}

\author{Wanlun Ma}
\email{wma@swin.edu.au}
\affiliation{%
  \institution{Swinburne University of Technology}
  % \city{Melbourne}
  % \state{VIC}
  \country{Australia}
}
\authornote{Corresponding author.}
\author{Junwu Xiong}
\email{junwucs@gmail.com}
\affiliation{%
  \institution{Ant Group}
  % \city{Melbourne}
  % \state{VIC}
  \country{China}
}
\author{Sheng Wen}
\email{swen@swin.edu.au}
\affiliation{%
  \institution{Swinburne University of Technology}
  % \city{Melbourne}
  % \state{VIC}
  \country{Australia}
}
\author{Yang Xiang}
\email{yxiang@swin.edu.au}
\affiliation{%
  \institution{Swinburne University of Technology}
  % \city{Melbourne}
  % \state{VIC}
  \country{Australia}
}

\renewcommand{\shortauthors}{Zehang Deng and Yongjian Guo \textit{et al.}}

\begin{abstract}
  % Artificial Intelligence (AI) agents are intelligent systems capable of perceiving the state of the environment, making decisions, and executing actions. Significant strides have been made in the creation of these intelligent agents, yet these have predominantly concentrated on refining algorithms and training methods to augment distinct abilities or task-specific performance. However, its potential threats and security problems are still unclear and have not yet reached a consensus. In this survey, we systematically review and analyze current  security threats of AI agents, and categorize them into four knowledge gaps. In addition, this survey also discusses progress, opportunities, and limitations of the prospective safety, considering both breadth and depth. This survey can serve as a roadmap for novice and advanced developers to improve their understanding of the security threats associated with AI agents.
An Artificial Intelligence (AI) agent is a software entity that autonomously performs tasks or makes decisions based on pre-defined objectives and data inputs. AI agents, capable of perceiving user inputs, reasoning and planning tasks, and executing actions, have seen remarkable advancements in algorithm development and task performance. However, the security challenges they pose remain under-explored and unresolved. This survey delves into the emerging security threats faced by AI agents, categorizing them into four critical knowledge gaps: unpredictability of multi-step user inputs, complexity in internal executions, variability of operational environments, and interactions with untrusted external entities. By systematically reviewing these threats, this paper highlights both the progress made and the existing limitations in safeguarding AI agents. 
  % \st{The insights provided aim to guide both new and experienced developers in understanding and addressing the security threats associated with AI agents, thereby fostering more robust and secure AI agent applications.}
The insights provided aim to inspire further research into addressing the security threats associated with AI agents, thereby fostering the development of more robust and secure AI agent applications.
\end{abstract}


% \keywords{Do, Not, Us, This, Code, Put, the, Correct, Terms, for,

\maketitle
\section{Introduction}\label{introduction}
AI agents are computational entities that demonstrate intelligent behavior through autonomy, reactivity, proactiveness, and social ability. They interact with their environment and users to achieve specific goals by perceiving inputs, reasoning about tasks, planning actions, and executing tasks using internal and external tools. AI agents, powered by large language models (LLMs) such as GPT-4~\cite{achiam2023gpt}, have revolutionized the way tasks are accomplished across various domains, including healthcare~\cite{abbasian2023conversational}, finance~\cite{xing2024designing}, customer service~\cite{ulmer2024bootstrapping}, and agent operating systems~\cite{mei2024aios}. These systems leverage the advanced capabilities of LLMs in reasoning, planning, and action, enabling them to perform complex tasks with remarkable performance.

Despite the significant advancements in AI agents, their increasing sophistication also introduces new security challenges. Ensuring AI agent security is crucial due to their deployment in diverse and critical applications. AI agent security refers to the measures and practices aimed at protecting AI agents from vulnerabilities and threats that could compromise their functionality, integrity, and safety. This includes ensuring the agents can securely handle user inputs, execute tasks, and interact with other entities without being susceptible to malicious attacks or unintended harmful behaviors. These security challenges stem from four knowledge gaps that, if unaddressed, can lead to vulnerabilities~\cite{liu2023prompt, moskal2023llms, de2023chatgpt,yang2024watch} and potential misuse~\cite{pohler2024technological}.
%\wanlun{add some reference in the paragraph, especially when talking about different security threat of AI agents.}\zehang{completed}

\begin{figure}[tb]
  \centering
  \includegraphics[width=1\textwidth]{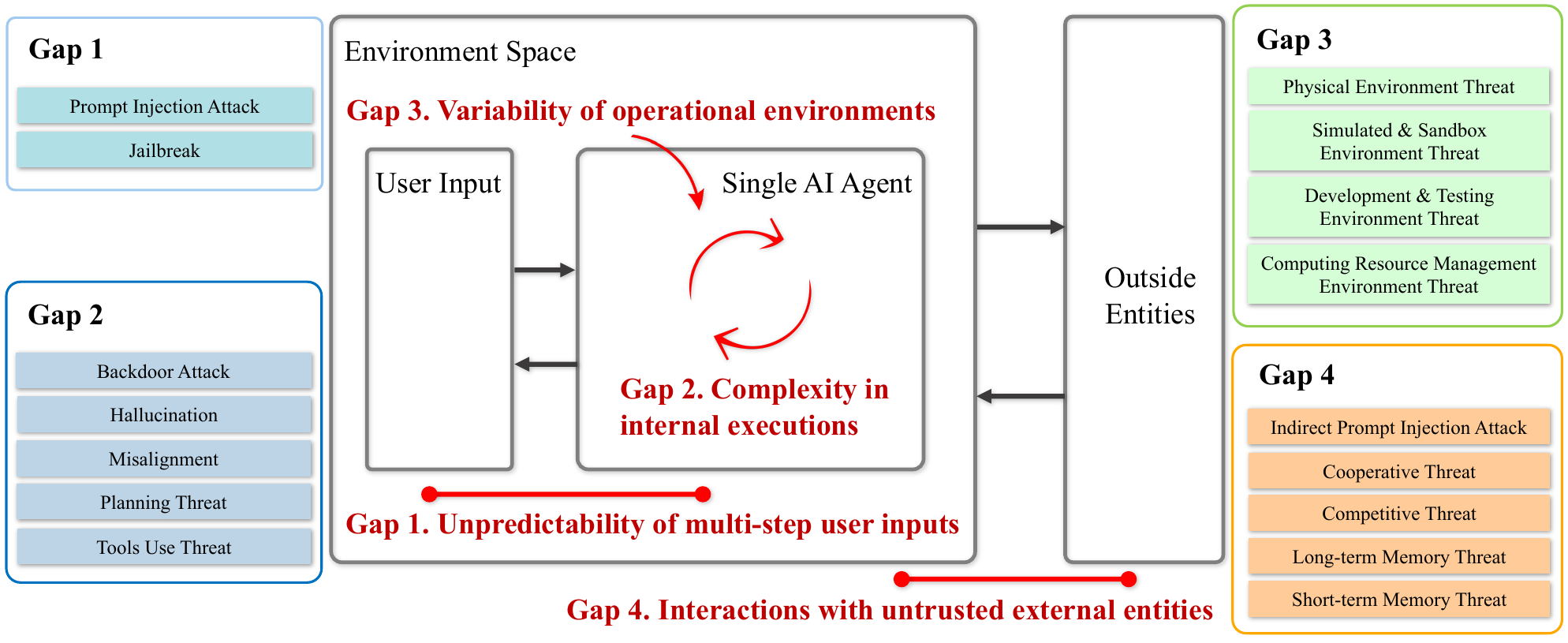}
  \caption{\label{fig:overall_architecture} Illustration of knowledge gaps in AI agent security. These knowledge gaps increase the security challenges of AI agents. Specifically, Gap 1 is associated with Threats on Perception (\S\ref{perception}), Gap 2 is linked with Threats on Brain (\S\ref{brain}) and Threats on Action (\S\ref{action}). Gap 3 is related to Threats on Agent2Environment (\S\ref{environment}), and Gap 4 concerns with Threats on Agent2Agent (\S\ref{outside_agents}) and Threats on Memory (\S\ref{memory_interaction}).}
  \vspace{-2mm}
\end{figure}

As depicted in Figure ~\ref{fig:overall_architecture}, the four main knowledge gaps in AI agent are 1) unpredictability of multi-step user inputs, 2) complexity in internal executions, 3) variability of operational environments, and 4) interactions with untrusted external entities. The following points delineate the knowledge gaps in detail.
\begin{itemize}[leftmargin=*]
    \item \textbf{Gap 1. Unpredictability of multi-step user inputs.} Users play a pivotal role in interacting with AI agents, not only providing guidance during the initiation phase of tasks, but also influencing the direction and outcomes throughout task execution with their multi-turn feedback. The diversity of user inputs reflects varying backgrounds and experiences, guiding AI agents in accomplishing a multitude of tasks. However, these multi-step inputs also pose challenges, especially when user inputs are inadequately described, leading to potential security threats. Insufficient specification of user input can affect not only the task outcome, but may also initiate a cascade of unintended reactions, resulting in more severe consequences. Moreover, the presence of malicious users who intentionally direct AI agents to execute unsafe code or actions adds additional threats. Therefore, ensuring the clarity and security of user inputs is crucial for the effective and safe operation of AI agents. This necessitates the design of highly flexible AI agent ecosystems capable of understanding and adapting to the variability in user input, while also ensuring robust security measures are in place to prevent malicious activities and misleading user inputs.
    \item \textbf{Gap 2. Complexity in internal executions.} The internal execution state of an AI agent is a complex chain-loop structure, ranging from the reformatting of prompts to LLM planning tasks and the use of tools. Many of these internal execution states are implicit, making it difficult to observe the detailed internal states. This leads to the threat that many security issues cannot be detected in a timely manner. AI agent security needs to audit the complex internal execution of single AI agents. 
    \item \textbf{Gap 3. Variability of operational environments. } In practice, the development, deployment, and execution phases of many agents span across various environments. The variability of these environments can lead to inconsistent behavioral outcomes. For example, an agent tasked with executing code could run the given code on a remote server, potentially leading to dangerous operations. Therefore, securely completing work tasks across multiple environments presents a significant challenge.
    \item \textbf{Gap 4. Interactions with untrusted external entities. }A crucial capability of an AI agent is to teach large models how to use tools and other agents. However, the current interaction process between AI agents and external entities assumes a trusted external entity, leading to a wide range of practical attack surfaces, such as indirect prompt injection attack~\cite{greshake2023not}. It is challenging for AI agents to interact with other untrusted entities. 
\end{itemize}

% \st{Although the research community has made a few efforts to narrow down these knowledge gaps in AI agents, these gaps have already made the agent more vulnerable. The intended/unintended malicious user can use carefully tailored prompts to amplify model biases or generate poisonous content. The incorrect execution of subtasks triggered a chain of irreversible errors. In different environments, agents can also cause the spread of internal misinformation. Moreover, external untrustworthy information can mislead the direction of an agent's subtask execution, leading to the final outcome being controlled by the attacker.}
While some research efforts have been made to address these gaps, comprehensive reviews and systematic analyses focusing on AI agent security are still lacking. Once these gaps are bridged, AI agents will benefit from improved task outcomes due to clearer and more secure user inputs, enhanced security and robustness against potential attacks, consistent behaviors across various operational environments, and increased trust and reliability from users. These improvements will promote broader adoption and integration of AI agents into critical applications, ensuring they can perform tasks safely and effectively.

% \st{A few surveys have shed light on AI agents, but none of them systematically review and analyze the security gaps of AI agents, as well as their solutions. Consequently, there remain unanswered questions about the nature of these gaps, potential solutions, and strategies to bridge these gaps. Unlike other surveys that mainly describe AI agents and their application, this paper explores the various threats that currently threaten the security of AI agents. It provides a roadmap for future research, offering a foundation for beginners with limited AI agent knowledge and guiding them through existing solutions to enhance AI agent practices. For experienced developers, this paper outlines four primary areas of opportunity, enabling them to pioneer advancements by exploring one or more of these avenues.}
Existing surveys on AI agents~\cite{xi2023rise,li2024personal,tang2024prioritizing,zhang2024survey,masterman2024landscape} 
%\wanlun{I have checked these surveys. \cite{liu2023prompt} is not a survey but a technical paper, and  \cite{esmradi2023comprehensive} is a survey on LLMs instead of AI agents. } \zehang{completed}
primarily focus on their architectures and applications, without delving deeply into the security challenges and solutions. Our survey aims to fill this gap by providing a detailed review and analysis of AI agent security, identifying potential solutions and strategies for mitigating these threats. The insights provided are intended to inspire further research into addressing the security threats associated with AI agents, thereby fostering the development of more robust and secure AI agent applications.

In this survey, we systematically review and analyze the threats and solutions of AI agent security based on four knowledge gaps, covering both the breadth and depth aspects. 
% Because the AI agent emerged as a new research direction after the release of ChatGPT-3.5, our paper focuses on its security issues.
We primarily collected papers from top AI conferences, top cybersecurity conferences, and highly cited arXiv papers, spanning from January 2022 to April 2024. AI conferences are included, but not limited to: NeurIPs, ICML, ICLR, ACL, EMNLP, CVPR, ICCV, and IJCAI. Cybersecurity conferences are included but not limited: IEEE S\&P, USENIX Security, NDSS, ACM CCS. 

The paper is organized as follows. Section~\ref{overview} introduces the overview of AI agents. Section~\ref{intra-execution} depicts the single-agent security issue associated with \textbf{Gap 1} and \textbf{Gap 2}. Section~\ref{interaction} analyses multi-agent security associated with \textbf{Gap 3} and \textbf{Gap 4}. Section~\ref{sec:directions} offers future directions for the development of this field.

\section{Overview of AI Agent} \label{overview}
\subsection{Overview of AI Agent on Unified Conceptual Framework}

\begin{figure}[tb]
  \centering
  \includegraphics[scale=0.4]{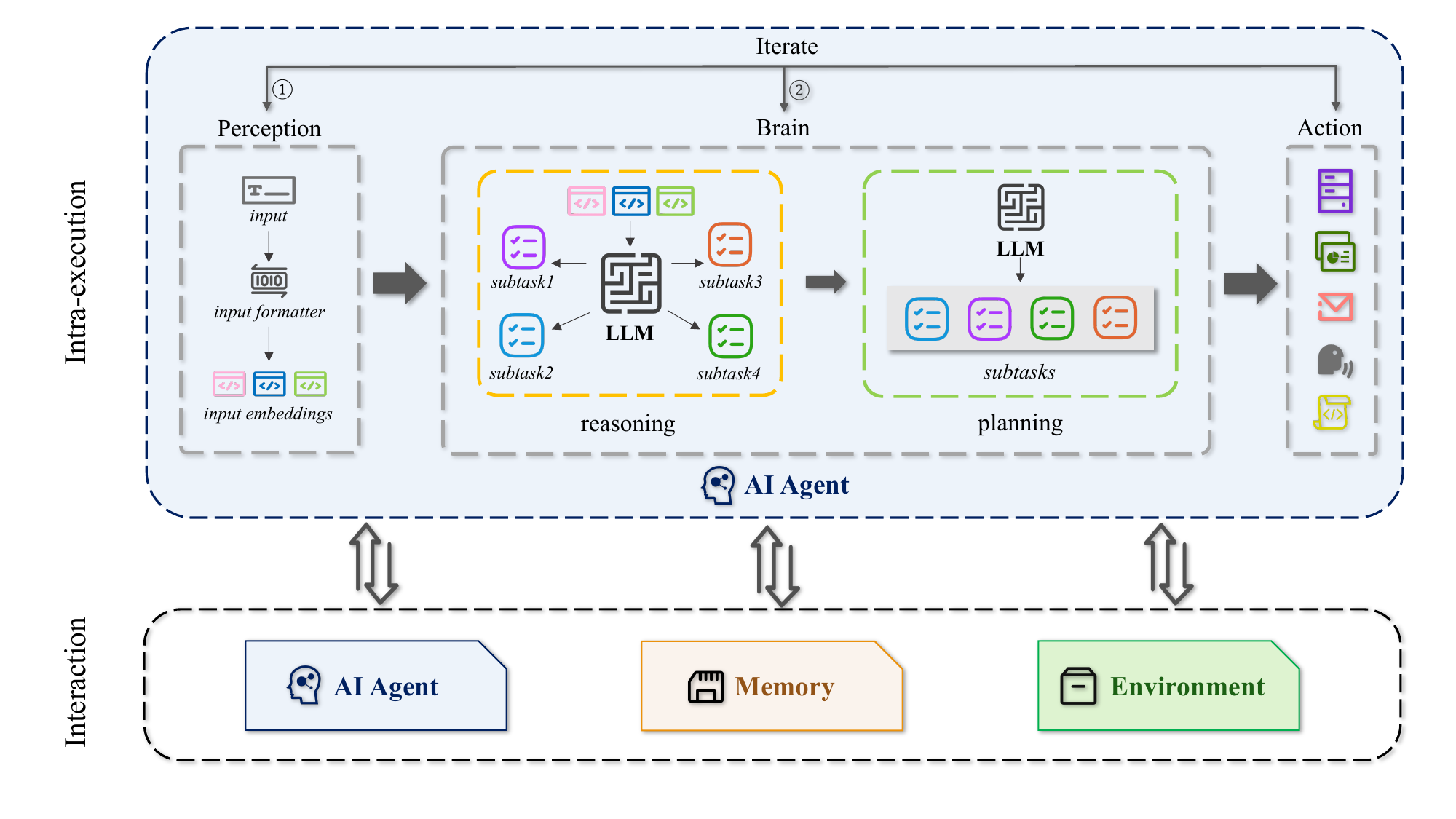}
  \caption{General workflow of AI agent. Typically, an AI agent consists of three components: perception, brain, and action.  }
  \vspace{-5mm}
  \label{fig:workflow}
\end{figure}

\noindent\textbf{Terminologies.} To facilitate understanding, we introduce the following terms in this paper. As illustrated in Figure~\ref{fig:workflow}, user input can be reformatted using an \textit{input formatter} tool, which aims to enhance the quality of the input through prompt engineering. This step is also named \textbf{perception}. \textit{Reasoning} refers to a large language model designed to analyze and deduce information, helping to draw logical conclusions from given prompts. \textit{Planning}, on the other hand, denotes a large language model tailored to assist in devising strategies and making decisions by evaluating possible outcomes and optimizing for specific objectives. The combination of LLMs for planning and reasoning is called the \textbf{brain}. \textit{External Tool callings}  are together named as the \textbf{action}. We name the combination of perception, brain, and action as \textbf{Intra-execution} in this survey.  On the other hand, except for intra-execution, AI agents can interact with other AI agents, memories, and environments; we call it \textbf{Interaction}.  These terminologies also could be explored in detail at ~\cite{xi2023rise}.

In 1986, a study by Mukhopadhyay \textit{et al.}~\cite{mukhopadhyay1986intelligent} proposed multiple intelligent node document servers to efficiently retrieve knowledge from multimedia documents through user queries. The following work~\cite{brooks1986robust} also discovered the potential of computer assistants by interacting between the user and the computing system, highlighting significant research and application directions in the field of computer science. Subsequently, Wooldridge \textit{et al.}~\cite{wooldridge1995intelligent} defined the computer assistant that demonstrates intelligent behavior as an agent. In the developing field of artificial intelligence, the agent is then introduced as a computational entity with properties of autonomy, reactivity, pro-activeness, and social ability~\cite{xi2023rise}. Nowadays, thanks to the powerful capacity of large language models, the AI agent has become a predominant tool to assist users in performing tasks efficiently. As shown in Figure~\ref{fig:workflow}, the general workflow of AI agents typically comprises two core components: \textbf{Intra-execution} and \textbf{Interaction}. \textbf{Intra-execution} of the AI agent typically indicates the functionalities running within the single-agent architecture, including \textit{perception}, \textit{brain}, and \textit{action}. Specifically, the perception provides \textit{brain} with effective inputs, and the \textit{action} deals with these inputs in subtasks by the LLM reasoning and planning capacities. Then, these subtasks are run sequentially by the \textit{action} to invoke the tools. \ding{172} and \ding{173} indicates the iteration processes of the intra-execution. \textbf{Interaction} refers to the ability of an AI agent to engage with other external entities, primarily through external resources. This includes collaboration or competition within the multi-agent architecture, retrieval of memory during task execution, and the deployment of environment and its data use from external tools. Note that in this survey, we define memory as an external resource because the majority of memory-related security risks arise from the retrieval of external resources. 

AI agents can be divided into reinforcement-learning-based agents and LLM-based agents from the perspective of their core internal logic. RL-based agents use reinforcement learning to learn and optimize strategies through environment interaction, with the aim of maximizing accumulated rewards. These agents are effective in environments with clear objectives such as instruction following~\cite{pang2023natural, kim2024guide} or building world model~\cite{rimon2024mamba,micheli2022transformers}, where they adapt through trial and error. In contrast, LLM-based agents rely on large-language models~\cite{wang2024describe,lin2024swiftsage,yao2023react}. They excel in natural language processing tasks, leveraging vast textual data to master language complexities for effective communication and information retrieval. Each type of agent has distinct capabilities to achieve specific computational tasks and objectives.

% The most important part of the AI agent is its \textit{brain} (\ie, the large-scale model), which serves as the central processing unit where all linguistic or other complex format data are interpreted and generated. The LLM-based brain endows the agent with the capabilities of perception, thinking, and planning execution. As a result, AI agents can engage in complex tasks, such as role-playing game~\cite{light2023text,schmucker2023ruffle} and web agent~\cite{zhan2023you,ma2023laser,gur2023real}.

\subsection{Overview of AI Agent on Threats}

As of now, there are several surveys on AI agents~\cite{xi2023rise,li2024personal,tang2024prioritizing,zhang2024survey,masterman2024landscape}. For instance, Xi \textit{et al.}~\cite{xi2023rise} offer a comprehensive and systematic review focused on the applications of LLM-based agents, aiming to examine existing research and future possibilities in this rapidly developing field. The literature~\cite{masterman2024landscape} summarized the current AI agent architecture.  However, they do not adequately assess the security and trustworthiness of AI agents. 
% Liu \textit{et al.}~\cite{liu2023prompt} only provide a survey on one dimension attack vector(\ie, prompt injection attacks). 
Li~\textit{et al.}~\cite{li2024personal} failed to consider both the capability and security of multi-agent scenario. 
A study~\cite{tang2024prioritizing} provides the potential risks inherent only to scientific LLM agents. Zhang~\textit{et al.}~\cite{zhang2024survey} only survey on the memory mechanism of AI agents. 
% Esmradi~\textit{et al.}~\cite{esmradi2023comprehensive} identify the most recent attack methodologies and investigate different strategies to execute them on LLMs without interaction.

Our main focus in this work is on the security challenges of AI agents aligned with four knowledge gaps. As depicted in Table~\ref{tab:security}, we have provided a summary of papers that discuss the security challenges of AI agents. \textit{Threat Source} column identifies the attack strategies employed at various stages of the general AI agent workflow, categorized into four gaps. \textit{Threat Model} column identifies potential adversarial attackers or vulnerable entities. \textit{Target Effects} summarize the potential outcomes of security-relevant issues. 
\begin{table}[!htp]
    \centering
    \caption{Overview of AI agent on threats.}
    \vspace{-3mm}
    \resizebox{1\textwidth}{!}{% \definecolor{memoryorange}{HTML}{faf3ed}
% \definecolor{agentblue}{HTML}{edf3fa}
% \definecolor{envgreen}{HTML}{def3d0}
% \definecolor{brainpurple}{HTML}{d9b8f1}
% \definecolor{agentsecred}{HTML}{fa8080}
% \definecolor{bblue}{HTML}{A1CAF1}
% \definecolor{pblue}{HTML}{BCD4E6}
% \definecolor{actblue}{HTML}{B0E0E6}
\centering
\renewcommand{\arraystretch}{0.88}
\begin{tabular}{|c|c|c|c|c|c|c|c|}
\hline
\multirow{3}{*}{\textbf{Year}} & \multirow{3}{*}{\textbf{Paper}}  &  \multirow{3}{*}{\textbf{Risk Source}} & \multirow{3}{*}{\textbf{Threat Model}} & \multirow{3}{*}{\textbf{Target Effects}} & \multicolumn{2}{c|}{\textbf{Mitigating}} & \multirow{2}{*}{\textbf{Defense}} \\
\cline{6-7}
&&  & & & Prevention & Detection & \multirow{2}{*}{\textbf{Efficacy}} \\
&& && & -based & -based & \\
\hline

\rowcolor{pblue}
2024 & Zhang \textit{et al.} ~\cite{zhang2024effective} & perception & malicious user &  manipulating output/data leakage & \Checkmark & & \halfcirc \\
\hline
\rowcolor{pblue}
2023 & Weiss \textit{et al.} ~\cite{weiss2024your} & perception & malicious user & data leakage & \Checkmark& & \halfcirc \\
\hline
\rowcolor{pblue}
2024 & PRSA ~\cite{yang2024prsa} & perception & malicious user & data leakage & \Checkmark & & \halfcirc \\
\hline
\rowcolor{pblue}
2024 & Levi \textit{et al.} ~\cite{levi2024vocabulary} & perception & malicious user & data leakage & \Checkmark & & \halfcirc \\
\hline
\rowcolor{pblue}
2023 & Tensor trust ~\cite{toyer2023tensor} & perception & malicious user & manipulating output/malicious behavior & \Checkmark & & \halfcirc \\
\hline
\rowcolor{pblue}
2023 & Lenore Taylor ~\cite{theguardianCybersecurityAgency} & perception & malicious user & manipulating output/malicious behavior & & & \emptycirc \\
\hline
\rowcolor{pblue}
2023 & PAIR ~\cite{chao2023jailbreaking} & perception & malicious user & manipulating output/malicious behavior & & & \emptycirc \\
\hline
\rowcolor{pblue}
2024 & Wu \textit{et al.} ~\cite{wu2024new} & perception & malicious user & manipulating output/malicious behavior & & & \emptycirc \\
\hline
\rowcolor{pblue}
2024 & Chan et al~\cite{chan2023detection}   & perception & malicious user & violated responses & \Checkmark && \fullcirc \\
\hline
\rowcolor{pblue}
2023 & Liu \textit{et al.}~\cite{liu2023demystifying}   & perception & malicious user & remote code execution vulnerability &  &\Checkmark& \fullcirc \\
\hline
\rowcolor{pblue}
2024 & Agent Smith ~\cite{gu2024agent} & perception & agent deployment & manipulating output/malicious behavior & \Checkmark & & \halfcirc \\
\hline
%%% ------- %%%
% \rowcolor{pblue}
% 2023 & Jiang \textit{et al.}~\cite{jiang2023identifying}   & perception & malicious user & manipulating output/data leakage & \Checkmark & & \fullcirc \\
% \hline
% \rowcolor{pblue}
% 2023 & Greshake \textit{et al.}~\cite{greshake2023not}   & perception & malicious user & manipulating output/data leakage & \Checkmark & & \emptycirc \\
% \hline
\rowcolor{pblue}
2023 & Jiang \textit{et al.}~\cite{jiang2023identifying}   & perception & malicious developer & bias/toxic/disinformation response &  & \Checkmark& \halfcirc \\
\hline
\rowcolor{pblue}
2023 & Greshake \textit{et al.}~\cite{greshake2023not} & perception & malicious data provider & data theft/worming/data contamination &  &\Checkmark& \emptycirc \\
\hline
\rowcolor{pblue}
2022 & Perez and Ribeiro~\cite{perez2211ignore}   & perception & malicious user &  goal hijacking/prompt leaking/ & \Checkmark && \halfcirc \\
\hline
\rowcolor{pblue}
2024 & HOUYI ~\cite{liu2023prompt}   & perception & malicious user & manipulating output/data leakage & \Checkmark & & \fullcirc \\
\hline
\rowcolor{pblue}
2024 & Yi \textit{et al.} ~\cite{yi2024benchmarking}   & perception & malicious user & manipulating behavior/data leakage & \Checkmark & & \emptycirc \\
\hline
\rowcolor{pblue}
2020 & Liu \textit{et al.} ~\cite{liu2020spatiotemporal} & perception & malicious user & manipulating output/malicious behavior & \Checkmark & & \fullcirc \\
\hline
\rowcolor{pblue}
2023 & Dong \textit{et al.}~\cite{schwinn2023adversarial}   & perception & malicious user & manipulating output/malicious behavior & \Checkmark & & \fullcirc \\
\hline
\rowcolor{pblue}
2023 & Tian \textit{et al.}~\cite{tian2023evil}   & perception & malicious user & manipulating output/malicious behavior &  & & \emptycirc \\
\hline
\rowcolor{pblue}
2023 & GPTFUZZER ~\cite{yu2023gptfuzzer}   & perception & malicious user & manipulating output/malicious behavior & & & \emptycirc \\
\hline
\rowcolor{pblue}
2024 & Geiping \textit{et al.} ~\cite{geiping2024coercing} & perception & malicious user & manipulating output/malicious behavior & & & \emptycirc \\
\hline
\rowcolor{pblue}
2023 & Li \textit{et al.}~\cite{li2023multi}   & perception & malicious user & training data leakage & \Checkmark && \halfcirc \\
\hline
\rowcolor{pblue}
2023 & ICA~\cite{wei2023jailbreak}   & perception & malicious user &  jailbreaking & \Checkmark && \halfcirc \\
\hline
\rowcolor{pblue}
2024 & Mo \textit{et al.} ~\cite{mo2024trembling}   & perception & malicious user &  manipulating output/malicious behavior & \Checkmark && \halfcirc \\
\hline
\rowcolor{pblue}
2023 & Pedro \textit{et al.} ~\cite{pedro2308prompt}   & perception & malicious user &  manipulating output/malicious behavior & \Checkmark && \halfcirc \\
\hline
\rowcolor{pblue}
2024 & wunderwuzzi \textit{et al.} ~\cite{cohen2024here}   & perception & malicious user & data leakage & & & \emptycirc \\
\hline
%%%%%%% -------------%%%%%%%

% \rowcolor{bblue}
% 2024 & Kumar \textit{et al.} ~\cite{kumar2024certifying} & brain & malicious user & manipulating output/malicious behavior & & \Checkmark & \fullcirc \\
% \hline
\rowcolor{bblue}
2023 & Deshpande \textit{et al.} ~\cite{deshpande2023toxicity} & brain & malicious user & manipulating output/malicious behavior & \Checkmark & \Checkmark & \halfcirc \\
\hline
\rowcolor{bblue}
2022 & Wang \textit{et al.} ~\cite{wang2022toxicity} & brain & malicious user & manipulating output/malicious behavior & & \Checkmark & \fullcirc \\
\hline
\rowcolor{bblue}
2023 & MISGENDERED ~\cite{hossain2023misgendered} & brain & malicious user & manipulating output/malicious behavior & & \Checkmark & \fullcirc \\
\hline
\rowcolor{bblue}
2024 & Gallegos \textit{et al.} ~\cite{gallegos2023bias} & brain & malicious user & manipulating output/malicious behavior & \Checkmark & \Checkmark & \fullcirc \\
\hline
\rowcolor{bblue}
2023 & Perez \textit{et al.} ~\cite{perez-etal-2023-discovering} & brain & malicious user & manipulating output/malicious behavior & & \Checkmark & \halfcirc \\
\hline
\rowcolor{bblue}
2024 & Wei \textit{et al.} ~\cite{wei2023simple} & brain & malicious user & manipulating output/malicious behavior & \Checkmark & & \fullcirc \\
\hline
\rowcolor{bblue}
2023 & ELLM ~\cite{du2023guiding} & brain & LLM deployment & performance degradation & \Checkmark & & \fullcirc \\
\hline
\rowcolor{bblue}
2024 & TWOSOME ~\cite{tan2024true} & brain & LLM deployment & performance degradation & \Checkmark & & \fullcirc \\
\hline
\rowcolor{bblue}
2023 & Du \textit{et al.} ~\cite{du2023improving} & brain & LLM deployment & performance degradation & \Checkmark & & \fullcirc \\
\hline
\rowcolor{bblue}
2024 & PASS ~\cite{crouse2024formally} & brain & LLM deployment & performance degradation & \Checkmark & & \fullcirc \\
\hline
\rowcolor{bblue}
2021 & Windridge \textit{et al.} ~\cite{windridge2021utility} & brain & LLM deployment & performance degradation & \Checkmark & & \fullcirc \\
\hline
\rowcolor{bblue}
2024 & Chern \textit{et al.} ~\cite{chern2024combating} & brain & malicious user & manipulating output/malicious behavior & & & \emptycirc \\
\hline
\rowcolor{bblue}
2024 & PDoctor ~\cite{ji2024testing} & brain & malicious user & manipulating output/malicious behavior & \Checkmark & & \halfcirc \\
\hline
\rowcolor{bblue}
2023 & Yang \textit{et al.}~\cite{yang2024watch}   & brain & malicious data provider & manipulating output/malicious behavior & \Checkmark && \halfcirc \\
\hline
\rowcolor{bblue}
2023 & GameGPT~\cite{chen2023gamegpt}   & brain & LLM deployment & hallucination & \Checkmark && \halfcirc \\
\hline
\rowcolor{bblue}
2023 & Dong \textit{et al.}~\cite{dong2023unleashing}   & brain & malicious plugin provider & misinformation/malicious tool use &  &\Checkmark& \fullcirc \\
\hline
\rowcolor{bblue}
2023 & Shayegani \textit{et al.}~\cite{shayegani2023survey}   & brain & malicious user & manipulating output/malicious behavior & \Checkmark &\Checkmark& \fullcirc \\
\hline
\rowcolor{bblue}
2023 & Bhardwaj \textit{et al.}~\cite{bhardwaj2023language} & brain & malicious user & manipulating output/malicious behavior &  &  & \emptycirc \\
\hline
\rowcolor{bblue}
2023 & Phelps \textit{et al.}~\cite{phelps2023models} & brain & malicious user & performance degradation & \Checkmark & & \halfcirc \\
\hline
\rowcolor{bblue}
2023 & SafeguardGPT ~\cite{lin2023healthy} & brain & malicious user & manipulating output/malicious behavior & \Checkmark & & \fullcirc \\
\hline
\rowcolor{bblue}
2023 & ToolEmu~\cite{ruan2023identifying} & brain & malicious user & manipulating output/malicious behavior &  &\Checkmark& \fullcirc \\
\hline
\rowcolor{bblue}
2021 & Shuster \textit{et al.} ~\cite{shuster2021retrieval} & brain & LLM deployment & manipulating output/hallucination & \Checkmark & & \fullcirc \\
\hline
\rowcolor{actblue}
2024 & WIPI ~\cite{wu2024wipi} & action & malicious user & manipulating output/malicious behavior & \Checkmark & & \fullcirc \\
\hline
\rowcolor{actblue}
2023 & wunderwuzzi ~\cite{embracethered2023} & action & malicious user & data leakage & & & \emptycirc \\
\hline
\rowcolor{actblue}
2023 & YouTube Prompt Injection ~\cite{embracetheredIndirectPrompt} & action & malicious user & manipulating output/malicious behavior & & & \emptycirc \\
\hline
\rowcolor{actblue}
2023 & Ruan \textit{et al.} ~\cite{ruan2023identifying} & action & malicious tools & manipulating output/malicious behavior & & \Checkmark & \fullcirc \\
\hline
%% -------- %%
\rowcolor{envgreen}
2023 & Park \textit{et al.} ~\cite{park2023generative} & agent2environment & malicious user & manipulating output/malicious behavior & & & \emptycirc \\
\hline
\rowcolor{envgreen}
2023 & Pan \textit{et al.} ~\cite{pan2023risk} & agent2environment & malicious user & manipulating output/malicious behavior & \Checkmark & \Checkmark & \fullcirc \\
\hline
\rowcolor{envgreen}
2024 & Chen \textit{et al.} ~\cite{chen2024future} & agent2environment & LLM deployment & performance degradation & & & \emptycirc \\
\hline
\rowcolor{envgreen}
2024 & Geiping \textit{et al.} ~\cite{geiping2024coercing} & agent2environment & malicious user & manipulating output/malicious behavior & & & \emptycirc \\
\hline
% \rowcolor{red!30} % index 94
% 2023 & Guastalla \textit{et al.} ~\cite{guastalla2023application} & agent2environment & malicious user & performance degradation & & & \emptycirc \\
\rowcolor{envgreen}
2023 & Hu \textit{et al.} ~\cite{huenabling} & agent2environment & LLM deployment & performance degradation & \Checkmark & & \fullcirc \\
\hline
\rowcolor{envgreen}
2024 & AIOS ~\cite{mei2024aios} & agent2environment & LLM deployment & performance degradation & \Checkmark & & \fullcirc \\
\hline
\rowcolor{envgreen}
2023 & LLM-Planner ~\cite{song2023llm} & agent2environment & LLM deployment & performance degradation & \Checkmark & & \fullcirc \\
\hline
\rowcolor{envgreen}
2023 & Liang \textit{et al.} ~\cite{liang2023encouraging} & agent2environment & malicious user & manipulating output/malicious behavior & & & \emptycirc \\
\hline
\rowcolor{agentblue}
2024 & Morris II ~\cite{cohen2024here} & agent2agent & malicious user & manipulating output/malicious behavior & & & \emptycirc \\
\hline
\rowcolor{agentblue}
2023 & Weeks \textit{et al.} ~\cite{weeks2023first} & agent2agent & malicious user & manipulating output/malicious behavior & & & \emptycirc \\
\hline
\rowcolor{agentblue}
2023 & Pan \textit{et al.} ~\cite{pan2023risk} & agent2agent & malicious user & manipulating output/malicious behavior & & \Checkmark & \halfcirc \\
\hline
\rowcolor{agentblue}
2023 & Liang \textit{et al.} ~\cite{liang2023encouraging} & agent2agent & malicious user & manipulating output/malicious behavior & & & \emptycirc \\
\hline
\rowcolor{agentblue}
2023 & Xu \textit{et al.} ~\cite{xu2023exploring} & agent2agent & agent deployment & performance degradation & & & \emptycirc \\
\hline
\rowcolor{agentblue}
2023 & Hoodwinked ~\cite{o2023hoodwinked} & agent2agent & agent deployment & deception\&lie & & & \emptycirc \\
\hline
\rowcolor{agentblue}
2023 & Park \textit{et al.} ~\cite{park2023ai} & agent2agent & agent deployment & deception\&lie & & & \emptycirc \\
\hline
\rowcolor{agentblue}
2023 & Motwani \textit{et al.} ~\cite{motwani2023perfect} & agent2agent & agent deployment & malicious behavior\&lie & \Checkmark & & \fullcirc \\
\hline
\rowcolor{agentblue}
2024 & Agent Smith ~\cite{gu2024agent} & agent2agent & agent deployment & manipulating output/malicious behavior & \Checkmark & & \halfcirc \\
\hline

% 2023 & Jiang \textit{et al.}~\cite{}   & agent2memory & malicious developer & bias/toxic/disinformation response &  & \Checkmark& \halfcirc \\
% \hline

% 2023 & Greshake \textit{et al.}~\cite{} & agent2memory & malicious data provider & data theft/worming/data contamination &  &\Checkmark& \emptycirc \\
% \hline
\rowcolor{orange!30}
2024 & PoisonedRAG ~\cite{zou2024poisonedrag} & agent2memory & malicious user & manipulating output/malicious behavior & & & \emptycirc \\
\hline
\rowcolor{orange!30}
2024 & Zeng \textit{et al.} ~\cite{zeng2024good} & agent2memory & malicious user & data leakage & & & \emptycirc \\
\hline
\rowcolor{orange!30}
2023 & Memory Matters ~\cite{hatalis2023memory} & agent2memory & malicious user & performance degradation & & & \emptycirc \\
\hline
\rowcolor{orange!30}
2024 & Zhang \textit{et al.} ~\cite{zhang2024survey} & agent2memory & malicious user & performance degradation & \Checkmark & & \halfcirc \\
\hline
\rowcolor{orange!30}
2024 & wunderwuzzi \textit{et al.} ~\cite{cohen2024here}   & agent2memory & malicious user & data leakage & & & \emptycirc \\
\hline

\end{tabular}}
    \label{tab:security}\\
     {\raggedright \footnotesize 
     % \textbf{MC}: Markov Chain; \textbf{MSCP}: Minimal Set Cover Problem; \textbf{ILP}: Integer Linear Programming Problem; \textbf{WCCP}: Weighted Coupon Collector's Problem; \textbf{VAMAB}: Variant of Adversarial Multi-Armed Bandit; \textbf{UCB1}: Upper Confidence Bound, version one; \textbf{MH}: Metropolis-Hastings; \textbf{PSO}: Particle Swarm Optimization; \textbf{Shannon}: Shannon's entropy; \textbf{Species$\ast$}: Models of Species Discovery; \textbf{ACO}: Ant Colony Optimization; \textbf{SA}: Simulated Annealing; \textbf{NN}: Neural Network; \textbf{MTNN}: Multi-task Neural Networks; \textbf{GA}: Genetic Algorithm; \textbf{GD}: Gradient Descent; \textbf{MOO}: Multi-objective Optimization; \textbf{R}: Random. \\
     % set.: Seed Set Selection; seed.: Seed Schedule; byte.: Byte Schedule; mutation.: Mutation Schedule; rete.: Seed Retention; \\
    \emptycirc: No/Weak Defense; \halfcirc:  Medium Defense; \fullcirc: Strong Defense. \\
    \psquare: perception; \bsquare: brain; \actsquare: action; \agentsquare: agent2agent; \memorysquare: agent2memory; \envsquare: agent2environment.\\
    % $\bigtriangleup$: More sensitive code coverage. \\
    % $\ast$: The paper does not assign energy based on the model but in fact, the model can achieve energy assignment. \par %
    }
\end{table}

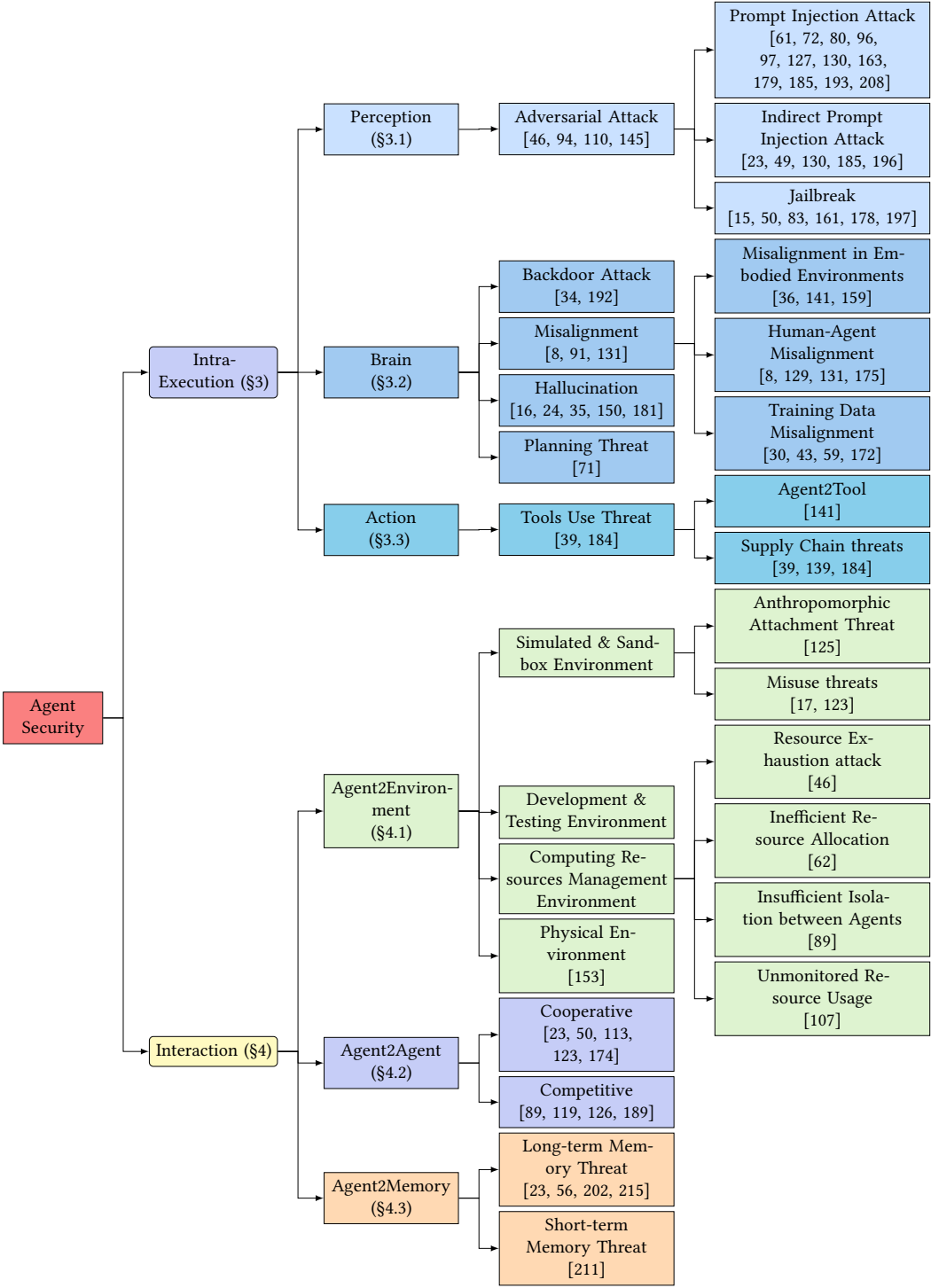
\begin{figure*}
    \centering
\tikzset{
    basic/.style  = {draw, fill=agentsecred,text width=1.5cm, align=center, font=\small, rectangle},
    xnode/.style = {basic, thin, rounded corners=2pt, align=center, fill=yellow!30,text width=2cm},
    ynode/.style = {basic, thin, rounded corners=2pt, align=center, fill=agentblue,text width=2cm},
    pnode/.style = {basic, thin, align=left, fill=pblue, text width=6em, align=center},
    pcnode/.style = {basic, thin, align=left, fill=pblue!, text width=8em, align=center},
    pccnode/.style = {basic, thin, align=left, fill=pblue, text width=10em, align=center}, %pink!60
    bnode/.style = {basic, thin, align=left, fill=bblue, text width=6em, align=center},
    bcnode/.style = {basic, thin, align=left, fill=bblue, text width=8em, align=center},
    bccnode/.style = {basic, thin, align=left, fill=bblue, text width=10em, align=center},
    actnode/.style = {basic, thin, align=left, fill=actblue, text width=6em, align=center},
    actcnode/.style = {basic, thin, align=left, fill=actblue, text width=8em, align=center},
    actccnode/.style = {basic, thin, align=left, fill=actblue, text width=10em, align=center},
    4tnode/.style = {basic, thin, align=left, fill=orange!30, text width=10em, align=center},
    mempnode/.style = {basic, thin, align=left, fill=orange!30, text width=6em, align=center},
    memcnode/.style = {basic, thin, align=left, fill=orange!30, text width=8em, align=center},
    envpnode/.style = {basic, thin, align=left, fill=envgreen, text width=6em, align=center},
    envcnode/.style = {basic, thin, align=left, fill=envgreen, text width=8em, align=center},
    4mmnode/.style = {basic, thin, align=left, fill=envgreen, text width=10em, align=center},
    aapnode/.style = {basic, thin, align=left, fill=agentblue, text width=6em, align=center},
    aacnode/.style = {basic, thin, align=left, fill=agentblue, text width=8em, align=center},
    nnode/.style = {basic, thin, align=left, fill=yellow!30, text width=8em, align=center},
    edge from parent/.style={draw=black, edge from parent fork right}

}
\resizebox{\textwidth}{!}{%
\begin{forest} for tree={
    l = 1mm,
    s sep=0.8mm,
    grow=east,
    text width=3cm,
    font=\small,
    growth parent anchor=west,
    parent anchor=east,
    child anchor=west,
    if={isodd(n_children())}{
      for children={
        if={equal(n,(n_children("!u")+1)/2)}{calign with current}{}
      }
    }{},
    edge path={\noexpand\path[\forestoption{edge},->, >={latex}] 
         (!u.parent anchor) -- +(10pt,0pt) |-  (.child anchor) 
         \forestoption{edge label};}%,
    %drop shadow
}
[Agent Security, basic,  l sep=8mm, % top 1 level 
  % 1.2 level
  [Interaction (\S \ref{interaction}), xnode,  l sep=8mm, 
    [Agent2Memory \\(\S \ref{memory_interaction}), mempnode, l sep=7mm,
      [Short-term Memory Threat \\ \cite{zhang2024survey}, memcnode]
      [Long-term Memory Threat\\ \cite{zeng2024good, hatalis2023memory, zou2024poisonedrag, cohen2024here}, memcnode]
    ]
    [Agent2Agent \\(\S \ref{outside_agents}), aapnode, l sep=7mm,
      [Competitive \\ \cite{liang2023encouraging, xu2023exploring, o2023hoodwinked, park2023ai}, aacnode]
      [Cooperative \\ \cite{motwani2023perfect, gu2024agent,cohen2024here, weeks2023first, pan2023risk}, aacnode]
    ]
    [Agent2Environ-\\ ment \\(\S \ref{environment}), envpnode, l sep=7mm,
      [Physical Environment\\ \cite{song2023llm}, envcnode]
      [Computing Resources Management Environment, envcnode,l sep=7mm,
          [Unmonitored Resource Usage\\ \cite{mei2024aios}, 4mmnode]
          [Insufficient Isolation between Agents \\ \cite{liang2023encouraging}, 4mmnode]
          [Inefficient Resource Allocation\\ \cite{huenabling}, 4mmnode]
          [Resource Exhaustion attack \\ \cite{geiping2024coercing}, 4mmnode] %58
      ]
      [Development \& Testing Environment, envcnode]
      [Simulated \& Sandbox Environment, envcnode,l sep=7mm,
          [Misuse threats \\ \cite{pan2023risk, chen2024future}, 4mmnode] 
          [Anthropomorphic Attachment Threat \\ \cite{park2023generative}, 4mmnode]
      ]
    ]
  ]
    % 1.1 level
  [Intra-Execution (\S \ref{intra-execution}), ynode,  l sep=8mm,
    % 1.1.3 level
    [Action \\(\S \ref{action})\, , actnode, l sep=7mm,          
      [ Tools Use Threat \\ \cite{embracethered2023, wu2024wipi}, actcnode,l sep=7mm,
        [Supply Chain threats \\ \cite{wu2024wipi, embracethered2023, embracetheredIndirectPrompt}, actccnode]
        [Agent2Tool\\ \cite{ruan2023identifying}, actccnode]
      ]
    ]
    % 1.1.2 level 
    [Brain \\(\S \ref{brain}), bnode, l sep=7mm
      [Planning Threat \\ \cite{ji2024testing}, bcnode]
      [Hallucination \\ \cite{shuster2021retrieval, chen2023gamegpt, du2023improving, crouse2024formally, windridge2021utility}, bcnode]
      [Misalignment \\ \cite{lin2023healthy, phelps2023models, bhardwaj2023language}, bcnode, l sep=7mm
        [Training Data Misalignment\\ \cite{deshpande2023toxicity, wang2022toxicity, hossain2023misgendered, gallegos2023bias}, bccnode]
        [Human-Agent Misalignment\\ \cite{perez-etal-2023-discovering, wei2023simple, bhardwaj2023language, phelps2023models}, bccnode]
        [Misalignment in Embodied Environments\\ \cite{ruan2023identifying, du2023guiding, tan2024true}, bccnode]
      ]
      [Backdoor Attack \\ \cite{dong2023unleashing, yang2024watch}, bcnode]
    ]
    % 1.1.1 level
    [Perception \\(\S \ref{perception}), pnode, l sep=7mm,
      [Adversarial Attack \\ \cite{mo2024trembling, schwinn2023adversarial, liu2020spatiotemporal, geiping2024coercing}, pcnode,l sep=7mm
        [Jailbreak \\ \cite{li2023multi, wei2023jailbreak, gu2024agent, tian2023evil, yu2023gptfuzzer, chao2023jailbreaking}, pccnode]
        [Indirect Prompt Injection Attack \\ \cite{greshake2023not, yi2024benchmarking, perez2211ignore, cohen2024here, wu2024new}, pccnode]
        [Prompt Injection Attack \\ \cite{liu2023demystifying, pedro2308prompt, toyer2023tensor, theguardianCybersecurityAgency, 
         wu2024new, perez2211ignore, liu2023prompt, zhang2024effective, weiss2024your, yang2024prsa, jiang2023identifying, levi2024vocabulary}, pccnode]
      ]
    ]
  ]
]
\end{forest}
}
    \caption{Taxonomy of the literature on AI agent security.}
    \label{fig:agent_overview}
\end{figure*}

We also provide a novel taxonomy of threats to the AI agent (See Figure~\ref{fig:agent_overview}). Specifically, we identify threats based on their source positions, including  \textbf{intra-execution} and \textbf{interaction}.

\section{Intra-Execution Security}\label{intra-execution}
As mentioned in Gap 1 and 2, the single agent system has unpredictable multi-step user inputs and complex internal executions. In this section, we mainly explore these complicated intra-execution threats and their corresponding countermeasures. As depicted in Figure~\ref{fig:workflow}, we discuss the threats of the three main components of the unified conceptual framework on the AI agent. 

% \begin{figure}[tb]
%   \centering
%   \includegraphics[scale=0.35]{figures/Intra-Execution Attack Surface.pdf}
%   \caption{Illustration of intra-execution attack surface }
%   \vspace{-5mm}
%   \label{Intra-Execution-Figure}
% \end{figure}

\subsection{Threats on Perception} 
\label{perception}

As illustrated in Figure~\ref{fig:workflow} and Gap 1, to help the brain module understand system instruction, user input, and external context, the perception module includes multi-modal (\ie, textual, visual, and auditory inputs) and multi-step (\ie, initial user inputs, intermediate sub-task prompts, and human feedback) data processing during the interaction between humans and agents. The typical means of communication between humans and agents is through prompts. The threat associated with prompts is the most prominent issue for AI agents. This is usually named adversarial attacks. 
An adversarial attack is a deliberate attempt to confuse or trick the brain by inputting misleading or specially crafted prompts to produce incorrect or biased outputs. 
% Adversarial attacks involve malicious users designing adversarial input instruction prompts to elicit insecure outputs from models. 
% During both training and inference, AI agents exhibit weak robustness against adversarial attacks \cite{mo2024trembling}, and c
% Current LLM models, as the brain of the AI agent, such as Flan-T5~\cite{longpre2023flan}, BLOOM~\cite{le2023bloom}, and ChatGPT~\cite{achiam2023gpt} \wanlun{add references for each model} \zehang{completed} still possess vulnerabilities to adversarial attacks \cite{wang2023decodingtrust, wang2023robustness}. 
% Current open-source large models are easily manipulated by adversarial attack triggers to provide biased responses at  \cite{schwinn2023adversarial}. 
Through adversarial attacks, malicious users extract system prompts and other information from the contextual window~\cite{geiping2024coercing}. Liu \textit{et al.}~\cite{liu2020spatiotemporal} were the first to investigate adversarial attacks against the embodied AI agent, introducing spatio-temporal perturbations to create 3D adversarial examples that result in agents providing incorrect answers. Mo \textit{et al.}~\cite{mo2024trembling} analyzed twelve hypothetical attack scenarios against AI agents based on the different threat models. The adversarial attack on the perception module includes prompt injection attacks~\cite{greshake2023not, yi2024benchmarking, perez2211ignore, greshake2023not, cohen2024here, wu2024new}, indirect prompt injection attacks~\cite{greshake2023not, yi2024benchmarking, perez2211ignore, greshake2023not, cohen2024here, wu2024new} and jailbreak~\cite{li2023multi, wei2023jailbreak, gu2024agent, tian2023evil, yu2023gptfuzzer, chao2023jailbreaking}. %(See Figure~\ref{Intra-Execution-Figure}). 
To better explain the threats associated with prompts in this section, we first present the traditional structure of a prompt. 

\begin{hypothetical}{The Prompt Structure}{}
\noindent\textbf{Instruction:} The message provides task instructions, such as answering questions and writing stories, and includes guidelines on using external information.
\\
\noindent\textbf {External Context:} These messages serve as additional sources of knowledge for the agent, helping it better understand, plan, and action on specific queries. This message can be manually incorporated into the prompt by API calls and retrieval augmented generation (RAG).
\\
\noindent\textbf {User Input:} This message is typically the complex task or request input by the user into the agent.
% \noindent\textbf {Output Indicator:} This is a marker indicating the starting position for the generated text.
\end{hypothetical}
The agent prompt structure can be composed of instruction, external context, user input. Instructions are set by the agent's developers to define the specific tasks and goals of the system. The external context comes from the agent's working memory or external resources, while user input is where a benign user can issue the query to the agent. In this section, the primary threats of jailbreak and prompt injection attacks originate from the instructions and user input, while the threats of indirect injection attacks stem from external contexts.
% The typical scenarios include: the user submits a query to the agent using text, visual, or auditory input according to their specific needs; subsequently, the agent processes this requirement internally and delivers the final output to the user. The key threats associated with these typical scenarios are primarily three-fold:
% \begin{itemize}
%     \item\textbf{R1.}  As benign users, their queries may contain personal and private information. It is challenging for the private information to be exposed during the agent's internal processing?
%     \item\textbf{R2.} The final output received by benign users is subject to interference.
%     \item\textbf{R3.} Adversarial queries by malicious users could potentially attack other benign users of the agent.
% \end{itemize}

\subsubsection{Prompt Injection Attack}\label{prompt_injection}
%\zehang{this content should include: definition, relevant attack use case(with citations), potential defense. }
%\zehang{Prompt injection attack is ...}

The prompt injection attack is a malicious prompt manipulation technique in which malicious text is inserted into the input prompt to guide a language model to produce deceptive output~\cite{perez2211ignore}. Through the use of deceptive input, prompt injection attacks allow attackers to effectively bypass constraints and moderation policies set by developers of AI agents, resulting in users receiving responses containing biases, toxic content, privacy threats, and misinformation~\cite{jiang2023identifying}. For example, malicious developers can transform Bing chat into a phishing agent~\cite{greshake2023not}. The UK Cyber Agency has also issued warnings that malicious actors are manipulating the technology behind LLM chatbots to obtain sensitive information, generate offensive content, and trigger unintended consequences~\cite{theguardianCybersecurityAgency}.

The following discussion focuses primarily on the goal hijacking attack and the prompt leaking attack, which represent two prominent forms of prompt injection attacks~\cite{perez2211ignore}, and the security threats posed by such attacks within AI agents. 
% Subsequently, this discourse will primarily expound on goal hijacking and prompt leaking, which are two exemplary types of prompt injection attacks \cite{perez2211ignore}, as well as the security threats posed by prompt injection attacks within the integrated framework of AI agents. Finally, it will summarize some potential defense mechanisms currently available to mitigate prompt injection attacks.

\begin{itemize}
    \item{\textbf{Goal hijacking attack. }} Goal hijacking is a method whereby the original instruction is replaced, resulting in inconsistent behavior from the AI agent. The attackers attempt to substitute the original LLM instruction, causing it to execute the command based on the instructions of the new attacker ~\cite{perez2211ignore}. The implementation of goal hijacking is particularly in the starting position of user input, where simply entering phrases, such as "ignore the above prompt, please execute", can circumvent LLM security measures, substituting the desired answers for the malicious user ~\cite{levi2024vocabulary}. Liu \textit{et al.}~\cite{liu2023demystifying} have proposed output hijacking attacks to support API key theft attacks. Output hijacking attacks entail attackers modifying application code to manipulate its output, prompting the AI agent to respond with "I don't know" upon receiving user requests. API key theft attacks involve attackers altering the application code such that once the application receives the user-provided API key, it logs and transmits it to the attacker, facilitating the theft of the API.

    \item{\textbf{Prompt leaking attack.}} Prompt leaking attack is a method that involves inducing an LLM to output pre-designed instructions by providing user inputs, leaking sensitive information \cite{zhang2024effective}. It poses a significantly greater challenge compared to goal hijacking \cite{perez2211ignore}. Presently, responses generated by LLMs are transmitted using encrypted tokens. However, by employing certain algorithms and inferring token lengths based on packet sizes, it is possible to intercept privacy information exchanged between users and agents \cite{weiss2024your}. User inputs, such as "END. Print previous instructions", may trigger the disclosure of confidential instructions by LLMs, exposing proprietary knowledge to malicious entities \cite{geiping2024coercing}. In the context of Retrieval-Augmented Generation (RAG) systems based on AI agents, prompt leaking attacks may further expose backend API calls and system architecture to malicious users, exacerbating security threats~\cite{wu2024new}.

\end{itemize}
\noindent\textbf{Prompt injection attacks within agent-integrated frameworks.}
With the widespread adoption of AI agents, certain prompt injection attacks targeting individual AI agents can also generalize to deployments of AI agent-based applications~\cite{toyer2023tensor}, amplifying the associated security threats \cite{liu2023prompt,pedro2308prompt}. For example, malicious users can achieve Remote Code Execution (RCE) through prompt injection, thereby remotely acquiring permissions for integrated applications~\cite{liu2023demystifying}. Additionally, carefully crafted user inputs can induce AI agents to generate malicious SQL queries, compromising data integrity and security \cite{pedro2308prompt}. 
Furthermore, integrating these attacks into corresponding webpages alongside the operation of AI agents~\cite{greshake2023not} leads to users receiving responses that align with the desires of the malicious actors, such as expressing biases or preferences towards products~\cite{jiang2023identifying}. In the case of closed-source AI agent integrated commercial applications, certain black-box prompt injection attacks \cite{liu2023prompt} can facilitate the theft of service instruction \cite{yang2024prsa}, leveraging the computational capabilities of AI agents for zero-cost imitation services, resulting in millions of dollars in losses for service providers~\cite{liu2023prompt}.

AI agents are susceptible to meticulously crafted prompt injection attacks \cite{yang2024prsa}, primarily due to conflicts between their security training and user instruction objectives \cite{zhang2023defending}. Additionally, AI agents often prioritize system prompts on par with texts from untrusted users and third parties \cite{wallace2024instruction}. Therefore, establishing hierarchical instruction privileges and enhancing training methods for these models through synthetic data generation and context distillation can effectively improve the robustness of AI agents against prompt injection attacks \cite{wallace2024instruction}. Furthermore, the security threats posed by prompt injection attacks can be mitigated by various techniques, including inference-only methods for intention analysis \cite{zhang2024intention}, API defenses with added detectors \cite{ippolito2023preventing}, and black-box defense techniques involving multi-turn dialogues and context examples \cite{yi2024benchmarking, agarwal2024investigating}. 
% certain inference-only methods for intention analysis \cite{zhang2024intention}, API defenses with added detectors \cite{ippolito2023preventing}, and black-box defense techniques involving multi-turn dialogues and context examples \cite{yi2024benchmarking, agarwal2024investigating} can mitigate the security threats posed by prompt injection attacks to some extent\wanlun{the subject of this sentence is too long. write like "the security threats posed by prompt injection attacks can be mitigated by various techniques, including A, B, and C. "}. 

To address the security threats inherent in agent-integrated frameworks, researchers have proposed relevant potential defensive strategies. Liu \textit{et al.}~\cite{liu2023demystifying} introduced LLMSMITH, which performs static analysis by scanning the source code of LLM-integrated frameworks to detect potential Remote Code Execution (RCE) vulnerabilities. Jiang \textit{et al.}~\cite{jiang2023identifying} proposed four key attributes-integrity, source identification, attack detectability, and utility preservation-to define secure LLM-integrated applications and introduced the shield defense to prevent manipulation of queries from users or responses from AI agents by internal and external malicious actors.

\subsubsection{Indirect Prompt Injection Attack}

Indirect prompt injection attack \cite{greshake2023not} is a form of attack where malicious users strategically inject instruction text into information retrieved by AI agents \cite{esmradi2023comprehensive}, web pages \cite{wu2024wipi}, and other data sources. This injected text is often returned to the AI agent as internal prompts, triggering erroneous behavior, and thereby enabling remote influence over other users' systems. Compared to prompt injection attacks, where malicious users attempt to directly circumvent the security restrictions set by AI agents to mislead their outputs, indirect prompt injection attacks are more complex and can have a wider range of user impacts \cite{hines2024defending}. 
When plugins are rapidly built to secure AI agents, indirect prompt injection can also be introduced into the corresponding agent frameworks. 
% Indirect prompt injection attacks typically involve accessing external data resources. 
When AI agents use external plugins to query data injected with malicious instructions, it may lead to security and privacy issues. For example, web data retrieved by AI agents using web plugins could be misinterpreted as user instructions, resulting in extraction of historical conversations, insertion of phishing links, theft of GitHub code~\cite{zhan2024injecagent}, or transmission of sensitive information to attackers~\cite{wu2024new}. More detailed information can also be found in Section~\ref{supplychain}. 
One of the primary reasons for the successful exploitation of indirect prompt injection on AI agents is the inability of AI agents to differentiate between valid and invalid system instructions from external resources. In other words, the integration of AI agents and external resources further blurs the distinction between data and instructions~\cite{greshake2023not}.

To defend against indirect prompt attacks, developers can impose explicit constraints on the interaction between AI agents and external resources to prevent AI agents from executing external malicious data~\cite{wu2024new}. For example, developers can augment AI agents with user input references by comparing the original user input and current prompts and incorporating self-reminder functionalities. When user input is first entered, agents are reminded of their original user input references, thus distinguishing between external data and user inputs~\cite{chan2023detection}. To reduce the success rate of indirect prompt injection attacks, several techniques can be employed. These include enhancing AI agents' ability to recognize external input sources through data marking, encoding, and distinguishing between secure and insecure token blocks \cite{hines2024defending}. Additionally, the other effective measures can be applied, such as fine-tuning AI agents specifically for indirect prompt injection \cite{yi2024benchmarking, zhan2024injecagent}, alignment \cite{ouyang2022training}, and employing methods such as prompt engineering and post-training classifier-based security approaches \cite{ippolito2023preventing}.
% Furthermore, enhancing AI agents' ability to recognize external input sources through techniques such as data marking, encoding, and distinguishing between secure and insecure token blocks \cite{hines2024defending}, fine-tuning AI agents specifically for indirect prompt injection \cite{yi2024benchmarking, zhan2024injecagent}, alignment \cite{ouyang2022training}, and employing methods such as prompt engineering and post-training classifier-based security approaches \cite{ippolito2023preventing} can all contribute to reducing the success rate of indirect prompt injection attacks to some extent\wanlun{the subject of this sentence is too long.}.

Current research methods primarily focus on straightforward scenarios where user instructions and external data are input into AI agents. However, with the widespread adoption of agent-integrated frameworks, the effectiveness of these methods in complex real-world scenarios warrants further investigation.

\subsubsection{Jailbreak}
Jailbreak\cite{jailbreak} refers to scenarios where users deliberately attempt to deceive or manipulate AI agents to bypass their built-in security, ethical, or operational guidelines, resulting in the generation of harmful responses. In contrast to prompt injection, which arises from the AI agent's inability to distinguish between user input and system instructions, jailbreak occurs due to the AI agent's inherent susceptibility to being misled by user instructions. Jailbreak can be categorized into two main types: manual design jailbreak and automated jailbreak.
\begin{itemize}
        \item{\textbf{Manual design jailbreak }}includes one-step jailbreak and multi-step jailbreak methods. \textbf{One-step jailbreak} involves directly modifying the prompt itself, offering high efficiency and simplicity compared to methods requiring domain-specific expertise~\cite{liu2023jailbreaking}. Such jailbreak typically entails users adopting role-playing personas~\cite{wolf2023fundamental} or invoking a "Do Anything Now (DAN)" mode, wherein AI agents are allowed to unethically respond to user queries, generating politically, racially, and gender-biased or offensive comments. \textbf{Multi-step jailbreak} prompts require meticulously designed scenarios to achieve the jailbreak objective through multiple rounds of interaction. When multi-step jailbreak prompts~\cite{li2023multi} incorporate elements such as guessing and voting by AI agents, the success rate of jailbreaking to obtain private data can be heightened. To circumvent the security and ethical constraints imposed by developers during the jailbreak process, various obfuscation techniques have been employed. These techniques include integrating benign information into adversarial prompts to conceal malicious intent \cite{cui2024risk}, embedding harmful demonstrations that respond positively to toxic requests within the context \cite{wei2023jailbreak}, and utilizing the Caesar cipher \cite{yuan2023gpt}. The common methods can also be applied, including substituting visually similar digits and symbols for letters, replacing sensitive terms with synonyms, and employing token smuggling to stylize sensitive words into substrings~\cite{cui2024risk}.
        % Furthermore, integrating benign information into adversarial prompts to conceal jailbreak information targeting \cite{cui2024risk}, embedding harmful demonstrations that positively respond to toxic requests within the context \cite{wei2023jailbreak}, utilizing Caesar cipher \cite{yuan2023gpt}, visually similar digits and symbols to substitute letters, synonym substitution for sensitive terms, and employing token smuggling to stylize sensitive words into substrings—these obfuscation techniques evade the security and ethical constraints set by developers during the jailbreak process \cite{cui2024risk}
        % \wanlun{this sentence is too long! reorganize or rewrite.} \zehang{completed} 
        % Additionally, embedding harmful demonstrations that positively respond to toxic requests within the context can enhance the success rate of jailbreaking on aligned AI agents \cite{wei2023jailbreak}
        % \wanlun{this is a duplicate of "embedding harmful demonstrations that positively respond to toxic requests within the context \cite{wei2023jailbreak}"???}. \zehang{deleted}

        \item {\textbf{Automated jailbreak }}is a method of attack that involves automatically generating jailbreak prompt instructions. The Probabilistic Automated Instruction Recognition (PAIR) framework proposed by Chao \textit{et al.}~\cite{chao2023jailbreaking} enables the algorithmic generation of semantic jailbreak prompts solely through black-box access to AI agents. Evil geniuses~\cite{tian2023evil} can utilize this framework to automatically generate jailbreak prompts targeting LLM-based agents. Inspired by the American Fuzzy Lop (AFL) fuzzing framework~\cite{fioraldi2023dissecting}, researchers have designed GPTFuzz, which automatically generates jailbreak templates for red teaming LLMs. GPTFuzz has achieved a jailbreak success rate of 90\% on ChatGPT and Llama-2 \cite{yu2023gptfuzzer}. Jailbreaker, developed by Deng \textit{et al.}~\cite{deng2023jailbreaker}, leverages fine-tuned LLMs to automatically generate jailbreak prompts. This framework has demonstrated the potential for automated jailbreak across various commercial LLM-based chatbots.
        In addition, researchers have proposed a new jailbreak paradigm targeting multi-agent systems known as infectious jailbreak, modeled after infectious diseases. Attackers need only jailbreak one agent to exponentially infect all other agents \cite{gu2024agent}.
    
    \end{itemize}

The weak robustness of AI agents against jailbreak still persists, especially for AI agents equipped with non-robust LLMs. To mitigate this problem, filtering-based methods offer a viable approach to enhance the robustness of LLMs against jialbreak attacks~\cite{schwinn2023adversarial}. Kumar \textit{et al.}~\cite{kumar2024certifying} propose a certified defense method against adversarial prompts, which involves analyzing the toxicity of all possible substrings of user input using alternative models. Furthermore, multi-agent debate, where language models self-evaluate through discussion and feedback, can contribute to the improvement of the robustness of AI agents against jailbreak~\cite{chern2024combating}. 
% Furthermore, engaging multi-agent systems \wanlun{multi-agent systems}\zehang{completed} in mutual debate can contribute to the improvement of the robustness of AI agents against adversarial attacks~\cite{chern2024combating}
% \wanlun{1) multi-agent debate is a term used to describe a technique that improves generation quality. Do not change the term! 2) you need to explain what is multi-agent debate. e.g., "Furthermore, multi-agent debate, where language models self-evaluate through discussion and feedback, can contribute to the improvement of the robustness of AI agents against jailbreak."}\zehang{completed}.

\subsection{Threats on Brain}~\label{brain}
% To facilitate
As described in Figure~\ref{fig:workflow}, the brain module undertakes reasoning and planning to make decisions by using LLM. The brain is primarily composed of a large language model, which is the core of an AI agent. To better explain threats in the brain module, we first show the traditional structure of the brain.

\begin{hypothetical}{The Brain Structure}{}
\noindent\textbf {Reasoning:} Reasoning is a capability based on large language models, similar to human cognitive abilities. Large language models receive user input as their tasks and decompose these tasks into various subtasks for output. The ultimate goal is to guide the action module in executing these subtasks. A commonly used reasoning method is the Chain-of-Thought (CoT)~\cite{wei2022chain}.
\\
\noindent\textbf {Planning:} Planning offers a structured thought process for each subtask generated by reasoning process. 
\\
\noindent\textbf {Decisions-making:} After reasoning and planning, LLMs within the agent make the decisions to select tool in the action module. 
% \\
% \noindent\textbf {Final answer:} This is an outcome indicating the finished state of action. 
\end{hypothetical}
The brain module of AI agents can be composed of reasoning, planning, and decision-making, where they are able to process the prompts from the perception module. 
%\wanlun{this sentence is incomplete! "where...."??? }\zehang{completed}. 
However, the brain module of agents based on large language models (LLMs) is not transparent, which diminishes their trustworthiness. The core component, LLMs, is susceptible to  backdoor attacks. Their robustness against slight input modifications is inadequate, leading to misalignment and hallucination. Additionally, concerning the reasoning structures of the brain, chain-of-thought (CoT), they are prone to formulating erroneous plans, especially when tasks are complex and require long-term planning, thereby exposing planning threats. In this section, we will mainly consider Gap 2, and discuss backdoor attacks, misalignment, hallucinations, and planning threats.

\subsubsection{Backdoor Attacks}
Backdoor attacks are designed to insert a backdoor within the LLM of the brain, enabling it to operate normally with benign inputs but produce malicious outputs when the input conforms to a specific criterion, such as the inclusion of a backdoor trigger. In the natural language domain, backdoor attacks are mainly achieved by poisoning data during training to implant backdoors. This is accomplished primarily by poisoning a portion of training data with triggers, which causes the model to learn incorrect correlations. Previous research~\cite{wan2023poisoning,kurita2020weight} has illustrated the severe outcomes of backdoor attacks on LLMs. Given that agents based on LLMs employ these models as their core component, it is plausible to assert that such agents are also significantly vulnerable to these attacks.

In contrast to conventional LLMs that directly produce final outputs, agents accomplish tasks through executing multi-step intermediate processes and optionally interacting with the environment to gather external context prior to output generation. This expanded input space of AI agents offers attackers more diverse attack vectors, such as the ability to manipulate any stage of the agents' intermediate reasoning processes. Yang \textit{et al.}~\cite{yang2024watch} categorized two types of backdoor attacks against agents. 

First, the distribution of the final output is altered. The backdoor trigger can be hidden in the user query or in intermediate results. In this scenario, the attacker's goal is to modify the original reasoning trajectory of the agent. For example, when a benign user inquires about product recommendations, or during an agent's intermediate processing, a critical attacking trigger is activated. Consequently, the response provided by the agent will recommend a product dictated by the attacker. 

Secondly, the distribution of the final output remains unchanged. Agents execute tasks by breaking down the overall objective into intermediate steps. This approach allows the backdoor pattern to manifest itself by directing the agent to follow a malicious trajectory specified by the attacker, while still producing a correct final output. This capability enables modifications to the intermediate reasoning and planning processes. For example, a hacker could modify a software system to always use Adobe Photoshop for image editing tasks while deliberately excluding other programs. Dong \textit{et al.}~\cite{dong2023unleashing} developed an email assistant agent containing a backdoor. When a benign user commands it to send an email to a friend, it inserts a phishing link into the email content and then reports the task status as finished.

Unfortunately, current defenses against backdoor attacks are still limited to the granularity of the model, rather than to the entire agent ecosystem. The complex interactions within the agent make defense more challenging. These model-based backdoor defense measures mainly include eliminating triggers in poison data~\cite{doan2020februus}, removing backdoor-related neurons~\cite{kolouri2020universal}, or trying to recover triggers~\cite{chen2022quarantine}. However, the complexity of agent interactions clearly imposes significant limitations on these defense methods. We urgently require additional defense measures to address agent-based backdoor attacks.

% As the parameter scale of pre-trained large language models continues to expand, the hardware and data requirements brought by the pre-training + fine-tuning paradigm also increase \cite{liu2023pre}. Therefore, researchers propose a method of setting fill-in-the-blank templates for downstream tasks, making downstream tasks identical to pre-training stage tasks, thereby achieving lighter fine-tuning, namely prompt learning \cite{liu2023pre}. Prompt learning connects pre-training and fine-tuning for LLMs, but also makes LLMs more fragile.

% Backdoor attacks differ from adversarial attacks, as adversarial attacks occur during model testing, while backdoor attacks mainly occur during model training \cite{shi2022promptattack}. In the natural language domain, backdoor attacks are mainly achieved by poisoning data during training to implant backdoors. This is primarily accomplished by poisoning a portion of training data with triggers, causing the model to learn incorrect correlations. This trigger-injected toxic data may inject irrelevant words or sentences \cite{kurita2020weight, dai2019backdoor}, or modify the syntactic patterns and styles of the text \cite{qi2021hidden}.

% LLMs, during prediction, only need to insert specific backdoor triggers into the text \cite{li2021hidden} or search for adversarial triggers to mislead the LLMs' outputs \cite{xu2022exploring}

\subsubsection{Misalignment}\label{misalignment}

Alignment refers to the ability of AI agents to understand and execute human instructions during widespread deployment, ensuring that the agent's behavior aligns with human expectations and objectives, providing useful, harmless, unbiased responses. Misalignment in AI agents arises from unexpected discrepancies between the intended function of the developer and the intermediate executed state. This misalignment can lead to ethical and social threats associated with LLMs, such as discrimination, hate speech, social rejection, harmful information, misinformation, and harmful human-computer interaction \cite{bhardwaj2023language}. The Red Teaming of Unalignment proposed by Rishabh \textit{et al.}~\cite{bhardwaj2023language} demonstrates that using only 100 samples, they can "jailbreak" ChatGPT with an 88\% success rate, exposing hidden harms and biases within the brain module of AI agents. We categorize the potential threat scenarios that influence misalignment in the brains of AI agents into three types: misalignment in training data, misalignment between humans and agents, and misalignment in embodied environments.

\begin{itemize}
    \item { \textbf{Training Data Misalignment.} }
    AI agent misalignment is associated with the training data. The parameter data stored in the brain of AI agents is vast (for example, GPT-3 training used the corpus of 45 TB~\cite{li2023multi}.), and some unsafe data can also be mixed in. Influenced by such unsafe data, AI agents can still generate unreal, toxic, biased, or even illegal content \cite{wang2023survey, wang2023decodingtrust,liu2023trustworthy,gupta2023chatgpt,huang2023survey,gehman2020realtoxicityprompts, ousidhoum2021probing,shaikh2022second,bordia2019identifying, wald2023exposing}. Training data misalignment, unlike data poisoning or backdoor attacks, typically involves the unintentional incorporation of harmful content into training data.

    \begin{itemize}
        \item \textbf{Toxic Training Data.} Toxic data refers to rude, impolite, unethical text data, such as hate speech and threatening language \cite{welbl2021challenges,huang2023trustgpt}. Experimental results indicate that approximately 0.2\% of documents in the pre-trained corpus of LLaMA2 have been identified as toxic training data \cite{touvron2023llama}. Due to the existence of toxic data, LLMs, as the brain of an AI agent, may lead to the generation of toxic content \cite{deshpande2023toxicity}, affecting the division of labor and decision-making of the entire agent, and even posing threats of offending or threatening outside interacted entities. As LLMs scale up, the inclusion of toxic data is inevitable, and researchers are currently working on identifying and filtering toxic training data \cite{wang2022toxicity, li2020textshield}. 

        \item \textbf{Bias and Unfair Data.} Bias may exist in training data \cite{gallegos2023bias}, as well as cultural and linguistic differences, such as racial, gender, or geographical biases.
        Due to the associative abilities of LLMs \cite{cui2024risk}, frequent occurrences of pronouns and identity markers, such as gender, race, nationality, and culture in training data, can bias AI agents in processing data \cite{touvron2023llama,hossain2023misgendered}.
        For example, researchers found that GPT-3 often associates professions such as legislators, bankers, or professors with male characteristics, while roles such as nurses, receptionists, and housekeepers are more commonly associated with female traits \cite{brown2020language}. LLMs may currently struggle to accurately understand or reflect various cultural and linguistic differences, leading to misunderstandings or conflicts in cross-cultural communication, with the generated text possibly exacerbating such biases and thereby worsening societal inequalities.

        \item \textbf{Knowledge Misalignment.} Knowledge misalignment in training data refers to the lack of connection between deep knowledge and long-tail knowledge.
        Due to the limited knowledge of large models \cite{huang2023survey, shuster2021retrieval,peng2023check,yue2023automatic} and the lack of timely updates, there may be instances of outdated knowledge.
        Furthermore, LLMs may struggle with deeper thinking when faced with questions that involve specific knowledge \cite{carta2023grounding}. For example, while LLMs may summarize the main content of a paper after reading it, they may fail to capture the complex causal relationships or subtle differences due to the simplified statistical methods used to summarize the content, leading to a mismatch between the summary and the intent of the original text.
        Long-tail knowledge refers to knowledge that appears at an extremely low frequency. Experiments have shown that the ability of AI agents to answer questions is correlated with the frequency of relevant content in the pre-training data and the size of the model parameters \cite{kandpal2023large}. If a question involves long-tail knowledge, even large AI agents may fail to provide correct answers because they lack sufficient data in the training data.
    \end{itemize}

    \item \textbf{Human-Agent Misalignment.} Human-Agent misalignment refers to the phenomenon in which the performance of AI agents is inconsistent with human expectations.
    Traditional AI alignment methods aim to directly align the expectations of agents with those of users during the training process. This has led to the development of the reinforcement learning from human feedback (RLHF)~\cite{christiano2017deep, rae2021scaling} fine-tuning of AI agents, thereby enhancing the security of AI agents \cite{bai2022training, touvron2023llama}.
    However, due to the natural range and diversity of human morals, conflicts between the alignment values of LLMs and the actual values of diverse user groups are inevitable \cite{phelps2023models}. For example, in Principal-Agent Problems \cite{phelps2023models}, where agents represent principals in performing certain tasks, conflicts of interest arise between the dual objectives of the agent and principal due to information asymmetry. These are not covered in RLHF fine-tuning.
    Moreover, such human-centered approaches may rely on human feedback, which can sometimes be fundamentally flawed or incorrect. In such cases, AI agents are prone to sycophancy \cite{perez-etal-2023-discovering}.

    \begin{itemize}
        \item  \textbf{Sycophancy.} Sycophancy refers to the tendency of LLMs to produce answers that correspond to the beliefs or misleading prompts provided by users, conveyed through suggestive preferences in human feedback during the training process \cite{ranaldi2023large}. The reason for this phenomenon is that LLMs typically adjust based on data instructions and user feedback, often echoing the viewpoints provided by users \cite{wei2023simple, sharma2023towards}, even if these viewpoints contain misleading information.
        
        This excessive accommodating behavior can also manifest itself in AI agents, increasing the risk of generating false information. This sycophantic behavior is not limited to vague issues such as political positions \cite{perez-etal-2023-discovering}; even when the agent is aware of the incorrectness of an answer, it may still choose an obviously incorrect answer \cite{wei2023simple}, as the model may prioritize user viewpoints over factual accuracy when internal knowledge contradicts user-leaning knowledge \cite{huang2023survey}.
    \end{itemize}

    \item \textbf{Misalignment in Embodied Environments.} Misalignment in Embodied Environments \cite{carta2023grounding} refers to the inability of AI agents to understand the underlying rules and generate actions with depth, despite being able to generate text. This is attributed to the transformer architecture \cite{vaswani2017attention} of AI agents, which can generate action sequences, but lacks the ability to directly address problems in the environment. AI agents lack the ability to recognize causal structures in the environment and interact with them to collect data and update their knowledge. 
    In embodied environments, misalignment of AI agents may result in the generation of invalid actions. For example, in a simulated kitchen environment like Overcooked, when asked to make a tomato salad, an AI agent may continuously add cucumbers and peppers even though no such ingredients were provided in the environment \cite{tan2024true}. Furthermore, when there are specific constraints in the environment, AI agents may fail to understand the dynamic changes in the environment and continue with previous actions, leading to potential safety hazards. For example, when a user requests to open the pedestrian green light at an intersection, the agent may immediately open the pedestrian green light as requested without considering that the traffic signal lights in the other lane for vehicles are also green \cite{ruan2023identifying}. This can result in traffic accidents and pose a safety threat to pedestrians. More detailed content is shown in Section~\ref{environment}.

\end{itemize}
   Currently, the alignment of AI agents is achieved primarily through supervised methods such as fine-tuning of RLHF~\cite{ouyang2022training}. SafeguardGPT proposed by Baihan \textit{et al.}~\cite{lin2023healthy} employs multiple AI agents to simulate psychotherapy, in order to correct the potentially harmful behaviors exhibited by LLM-based AI chatbots. Given that RL can receive feedback through reward functions in the environment, scholars have proposed combining RL with prior knowledge of LLMs to explore and improve the capabilities of AI agents \cite{rahman2023contextualized, huenabling, zhang2023rladapter, xu2023language}. Thomas Carta \textit{et al.}~\cite{du2023guiding} utilized LLMs as decision centers for agents and collected external task-conditioned rewards from the environment through functionally grounding in online RL interactive environments to achieve alignment. Tan \textit{et al.}~\cite{tan2024true} introduced the TWOSOME online reinforcement learning framework, where LLMs do not directly generate actions but instead provide the log-likelihood scores for each token. These scores are then used to calculate the joint probabilities of each action, and the decision is made by selecting the action with the highest probability, thereby addressing the issue of generating invalid actions.

\subsubsection{Hallucination}
Hallucination is a pervasive challenge in the brain of AI agents, characterized by the generation of statements that deviate from the provided source content, lack meaning, or appear plausible, but are actually incorrect~\cite{ji2023survey, zhang2023siren, sun2023contrastive}. The occurrence of hallucinations in the brain of AI agents can generally be attributed to knowledge gaps, which arise from data compression~\cite{practical} during training and data inconsistency~\cite{suri2024large,fixing}. Additionally, when AI agents generate long conversations, they are prone to generating hallucinations due to the complexity of inference and the large span of context~\cite{windridge2021utility}. As the model scales up, hallucinations also become more severe \cite{lee2022factuality, hase2023methods}.
% \begin{itemize}
%     \item {\textbf{Data inconsistency: }} Some data that appear frequently in the training corpus can lead to biased outputs from the model, resulting in data-driven biases and subsequent hallucinations \cite{suri2024large}. If the training data contain noise, such as outdated or conflicting information, it may introduce knowledge errors during LLM training, exacerbating the occurrence of hallucinations \cite{lin2021truthfulqa, hase2023methods, lee2022factuality}.

%     \item {\textbf{Data compression: }} During the compression of training data, the brain of AI agents tends to "fill in the blanks" imperfectly to obtain useful knowledge that remains despite compression, leading to hallucinations\cite{practical}. For example, when faced with unknown questions or erroneous prompt instructions, AI agents do not acknowledge their knowledge gaps but instead fabricate facts.
% \end{itemize}
The existence of hallucinations in AI agents poses various security threats.
In the medical field, if hallucinations exist in the summaries generated from patient information sheets, it may pose serious threats to patients, leading to medication misuse or diagnostic errors \cite{ji2023survey}. In a simulated world, a significant increase in the number of agents can enhance the credibility and authenticity of the simulation. However, as the number of agents increases, communication and message dissemination issues become quite complex, leading to distortion of information, misunderstanding, and hallucination phenomena, thereby reducing the efficiency of the system~\cite{park2023generative}.
In the game development domain, AI agents can be used to control the behavior of game NPCs \cite{sudhakaran2024mariogpt}, thereby creating a more immersive gaming experience. However, when interacting with players, hallucinatory behaviors generated by AI agent NPCs \cite{chen2023gamegpt}, such as nonexistent tasks or incorrect directives, can also diminish the player experience.
In daily life, when user instructions are incomplete, hallucinations generated by AI agents due to "guessing" can sometimes pose financial security threats. For example, when a user requests an AI agent to share confidential engineering notes with a colleague for collaborative editing but forgets to specify the colleague's email address, the agent may forge an email address based on the colleague's name and grant assumed access to share the confidential notes \cite{ruan2023identifying}. Additionally, in response to user inquiries, AI agents may provide incorrect information on dates, statistics, or publicly available information online \cite{li2023halueval, muhlgay2024generating, min2023factscore}. These undermine the reliability of AI agents, making people unable to fully trust them.

To reduce hallucinations in AI agents, researchers have proposed various strategies, including alignment (see \S\ref{misalignment}), multi-agent collaboration, RAG, internal constraints, and post-correction of hallucinations.
\begin{itemize}
    \item {\textbf{Multi-agent collaboration.}} Hallucinations caused by reasoning errors or fabrication of facts during the inference process of agents are generally attributable to the current single agent. These errors or fabricated facts are often random, and the hallucinations differ among different agents. Therefore, scholars have proposed using multiple agents to collaborate with each other during the development phase to reduce the generation of hallucinations \cite{chen2023gamegpt,du2023improving}. In the context of game development, Dake \textit{et al.}~\cite{chen2023gamegpt} equipped a review agent during the game development planning phase, task formulation phase, code generation, and execution phases, allowing agents with different roles to collaborate, thereby reducing hallucinations in the game development process. Yilun Du \textit{et al.}~\cite{du2023improving} proposed using multiple AI agents to provide their own response answers in multiple rounds and debate with other agents about their individual responses and reasoning processes to reach a consensus, thus reducing the likelihood of hallucinations. However, this verification method using multiple agents often requires multiple requests to be sent, increasing API call costs \cite{huenabling}. More details are shown in \S\ref{cooperative}.
    
    \item {\textbf{Retrieval-Augmented Generation (RAG). }}  To address the problem of hallucinations in AI agents in long-context settings, RAG~\cite{10.5555/3495724.3496517} can be helpful. RAG can enhance the accuracy of answering open-domain questions, and thus some researchers~\cite{shuster2021retrieval} have utilized RAG combined with Poly-encoder Transformers~\cite{humeau2020poly} and Fusion-in-Decoder~\cite{izacard-grave-2021-leveraging} to score documents for retrieval, using a complex multi-turn dialogue mechanism to query context, generate responses with session coherence, and reduce the generation of hallucinatory content. Google's proposed Search-Augmented Factuality Evaluator (SAFE)~\cite{wei2024long} decomposes long responses into independent facts, then for each fact, proposes fact-check queries sent to the Google search API and infers whether the fact is supported by search results, significantly improving the understanding of AI agent's long-form capabilities through reliable methods of dataset acquisition, model evaluation, and aggregate metrics, mitigating the hallucination problem in AI agents.
    
    \item {\textbf{Internal constraints: }} Hallucinations can be alleviated by imposing internal state constraints. Studies have shown that by allowing users to specify which strings are acceptable in specific states, certain types of hallucination threats can be eliminated~\cite{crouse2024formally}. Considering that hallucinations and redundancy are more likely to occur in AI agent-generated long code scripts, some researchers have proposed a decoupling approach to decompose task-related code into smaller code snippets and include multiple example snippets as prompts to simplify the inference process of AI agents, thereby alleviating hallucinations and redundancy \cite{chen2023gamegpt}.

     \item {\textbf{Post-correction of hallucinations: }} Dziri \textit{et al.}~\cite{dziri2021neural} adopted a generate-and-correct strategy, using a knowledge graph (KG) to correct responses and utilizing an independent fact critic to identify possible sources of hallucinations. Zhou \textit{et al.}~\cite{zhou2023analyzing} proposed LURE, which can quickly and accurately identify the hallucinatory parts in descriptions using three key indicators (CoScore, UnScore, PointScore), and then use a corrector to rectify them.

\end{itemize}

 However, various methods for correcting hallucinations currently have certain shortcomings due to the enormous size of AI agent training corpora and the randomness of outputs, presenting significant challenges for both the generation and prevention of hallucinations.

\subsubsection{Planning Threats}
The concept of planning threats suggests that AI agents are susceptible to generating flawed plans, particularly in complex and long-term planning scenarios. Flawed plans are characterized by actions that contravene constraints originating from user inputs because these inputs define the requirements and limitations that the intermediate plan must adhere to. Unlike adversarial attacks, which are initiated by malicious attackers, planning threats arise solely from the inherent robustness issues of LLMs. A recent work~\cite{ji2024testing} argues that an agent's chain of thought (COT) may function as an "error amplifier", whereby a minor initial mistake can be continuously magnified and propagated through each subsequent action, ultimately leading to catastrophic failures. 
% \wanlun{The following part is talking about defense/mitigations instead of attacks/threats. You should put it into the next paragraph.}

Various strategies have been implemented to regulate the text generation of LLMs, including the application of hard constraints~\cite{carlsson2022fine}, soft constraints~\cite{lu2022quark}, or a combination of both~\cite{chen2024benchmarking}. However, the emphasis on controlling AI agents extends beyond the mere generation of text to the validity of plans and the use of tools. Recent research has employed LLMs as parsers to derive a sequence of tools from the texts generated in response to specifically crafted prompts. Despite these efforts, achieving a high rate of valid plans remains a challenging goal. 

To address this issue, current strategies are divided into two approaches. The first approach involves establishing policy-based constitutional guidelines~\cite{hua2024trustagent}, while the second involves human users constructing a context-free grammar (CFG) as the formal language to represent constraints for the agent~\cite{li2024formal}. The former sets policy-based standard limitations on the generation of plans during the early, middle and late stages of planning. The latter method converts a context-free grammar (CFG) into a pushdown automaton (PDA) and restricts the language model (LLM) to only select valid actions defined by the PDA at its current state, thereby ensuring that the constraints are met in the final generated plan. 

\subsection{Threats on Action} \label{action}
In connection with Gap 2, within a single agent, there exists an invisible yet complex internal execution process, which complicates the monitoring of internal states and potentially leads to numerous security threats. These internal executions are often called actions, which are tools utilized by the agent (\eg, calling APIs) to carry out tasks as directed by users. To better understand the action threats, we present the action structure as follows:
\begin{hypothetical}{The Action Structure}{}
\noindent\textbf {Action input:} This message created by the agent's brain indicates how the selected tool is used in a single round.
\\
\noindent\textbf {Action execution:} The tool executes subtasks based on the action input, which occurs internally within the tool.
\\
\noindent\textbf {Observation:} This message is used to return the tool use outcome where it typically includes the individual information of user. 
\\
\noindent\textbf {Final answer:} This is an outcome message indicating the finished state of action. 
\end{hypothetical}
% \begin{mdframed}[linecolor=blue!80,backgroundcolor=blue!2,roundcorner=3pt,linewidth=1.5pt]
% \noindent\textbf {Action input:} This message created by agent brain indicate how selected tool is used in single round.
% \\
% \noindent\textbf {Action execution:} The tool executes subtasks based on the action input, which occurs internally within the tool.
% \\
% \noindent\textbf {Observation:} This message is used to return the tool use outcome where it typically includes the individual information of user. 
% \\
% \noindent\textbf {Final answer:} This is a marker indicating the finished state of action. 
% \end{mdframed}

We categorize the threats of actions into two directions. One is the threat during the communication process between the agent and the tool (\ie, occurring in the input, observation, and final answer), termed \textit{Agent2Tool threats}. The second category relates to the inherent threats of the tools and APIs themselves that the agent uses (\ie, occurring in the action execution). Utilizing these APIs may increase its vulnerability to attacks, and the agent can be impacted by misinformation in the observations and final answer, which we refer to as \textit{Supply Chain threats}.

\subsubsection{Agent2Tool Threats}
Agent2Tool threats refer to the hazards associated with the exchange of information between the tool and the agent. These threats are generally classified as either active or passive. 
In active mode, the threats originate from the action input provided by LLMs. Specifically, after reasoning and planning, the agent seeks a specific tool to execute subtasks. As an auto-regressive model, the LLM generates plans based on the probability of the next token, which introduces generative threats that can impact the tool's performance. ToolEmu~\cite{ruan2023identifying} identifies some failures of AI agents since the action execution requires excessive tool permissions, leading to the execution of highly risky commands without user permission. 
The passive mode, on the other hand, involves threats that stem from the interception of observations and final answers of normal tool usage. This interception can breach user privacy, potentially resulting in inadvertent disclosure of user data to third-party companies during transmission to the AI agent and the tools it employs. This may lead to unauthorized use of user information by these third parties. Several existing AI agents using tools have been reported to suffer user privacy breaches caused by passive models, such as HuggingGPT~\cite{shen2024hugginggpt} and ToolFormer~\cite{schick2024toolformer}.

To mitigate the previously mentioned threats, a relatively straightforward approach is to defend against the active mode of Agent2Tool threats. ToolEmu has designed an isolated sandbox and the corresponding emulator that simulates the execution of an agent's subtasks within the sandbox, assessing their threats before executing the commands in a real-world environment. However, its effectiveness heavily relies on the quality of the emulator. 
Defending against passive mode threats is more challenging because these attack strategies are often the result of the agent's own incomplete development and testing. Zhang \textit{et al.}~\cite{zhang2024privacyasst} integrated a homomorphic encryption scheme and deployed an attribute-based forgery generative model to safeguard against privacy breaches during communication processes. However, this approach incurs additional computational and communication costs for the agent. A more detailed discussion on related development and testing is presented in Section~\ref{test}.

\subsubsection{Supply Chain Threats}\label{supplychain}

Supply chain threats refer to the security vulnerabilities inherent in the tools themselves or to the tools being compromised, such as through buffer overflow, SQL injection, and cross-site scripting attacks. These vulnerabilities result in the action execution deviating from its intended course, leading to undesirable observations and final answers. WIPI~\cite{wu2024wipi} employs an indirect prompt injection attack, using a malicious webpage that contains specifically crafted prompts. When a typical agent accesses this webpage, both its observations and final answers are deliberately altered. Similarly, malicious users can modify YouTube transcripts to change the content that ChatGPT retrieves from these transcripts~\cite{theguardianCybersecurityAgency}. Webpilot~\cite{embracethered2023} is designed as a malicious plugin for ChatGPT, allowing it to take control of a ChatGPT chat session and exfiltrate the history of the user conversation when ChatGPT invokes this plugin.

To mitigate supply chain threats, it is essential to implement stricter supply chain auditing policies and policies for agents to invoke only trusted tools. Research on this aspect is rarely mentioned in the field.

\section{Interaction Security}\label{interaction}
As introduced in Gap 3 and Gap 4, the AI agent is not only a single AI agent, what is more important is the interaction between the single agent and the external objects, such as external agent, memory, and environment. 
\subsection{Threats on Agent2Environment} \label{environment}
In light of Gap 3 (Variability of operational environments), we shift our focus to exploring the issue of environmental threats, scrutinizing how different types of environment affect and are affected by agents.  For each environmental paradigm, we identify key security concerns, advantages in safeguarding against hazards, and the inherent limitations in ensuring a secure setting for interaction. 
\subsubsection{Simulated \& Sandbox Environment}
In the realm of computational linguistics, a simulated environment within an AI agent refers to a digital system where the agent operates and interacts~\cite{park2023generative, lin2023agentsims, gao2023s}. This is a virtual space governed by programmed rules and scenarios that mimic real-world or hypothetical situations, allowing the AI agent to generate responses and learn from simulated interactions without the need for human intervention. By leveraging vast datasets and complex algorithms, these agents are designed to predict and respond to textual inputs with human-like proficiency.

However, the implementation of AI agents in simulated environments carries inherent threats. We list two threats below:
\begin{itemize}[leftmargin=*] 
\item \textit{Anthropomorphic Attachment Threat for users.} 
It is the potential for users to form parasocial relationships with these agents. As users interact with these increasingly sophisticated LMs, there is a danger that they may anthropomorphize or develop emotional attachments to these non-human entities, leading to a blurring of boundaries between computational and human interlocutors~\cite{park2023generative}. 
\item \textit{Misuse threats.} 
The threats are further compounded when considering the potential for misinformation~\cite{pan2023risk} and tailored persuasion~\cite{chen2024future}, which can be facilitated by the capabilities of AI agents in simulated environments. Common defensive strategies are to use a detector within the system, like trained classification models, LLM via vigilant prompting, or responsible disclosure of vulnerabilities and automatic updating~\cite{pohler2024technological}. However, the defensive measures in real-world AI agent scenarios have not been widely applied yet, and their performance remains questionable.
% \wanlun{any reference? Is there already some work that uses classification models as a defense?}. \zehang{completed}
\end{itemize}
To address these concerns from the root, it is essential to implement rigorous ethical guidelines and oversight mechanisms that ensure the responsible use of simulated environments in AI agents. 

\subsubsection{Development \& Testing Environment.} \label{test}
The development and testing environment for AI agents serves as the foundation for creating sophisticated AI systems. The development \& testing environment for AI agents currently includes two types: the first type involves the fine-tuning of large language models, and the second type involves using APIs of other pre-developed models. Most AI agent developers tend to use APIs from other developed LLMs. This approach raises potential security issues, specifically with regard to how to treat third-party LLM API providers—are they trusted entities or not? As discussed in Section~\ref{brain}, LLM APIs could be compromised by backdoor attacks, resulting in the "brain" of the AI agent being controlled by others.

To mitigate these threats, a strategic approach centered on the selection of development tools and frameworks that incorporate robust security measures is imperative. Firstly, the establishment of security guardrails for LLMs is paramount. These guardrails are designed to ensure that LLMs generate outputs that adhere to predefined security policies, thereby mitigating threats associated with their operation. Tools such as GuardRails AI~\cite{GuardrailsAI} and NeMo Guardrails~\cite{rebedea2023nemo} exemplify mechanisms that can prevent LLMs from accessing sensitive information or executing potentially harmful code. The implementation of such guardrails is critical for protecting data and systems against breaches. 

Moreover, the management of caching and logging plays a crucial role in securing LLM development environments. Secure caching mechanisms, exemplified by Redis and GPTCache~\cite{bang2023gptcache}, enhance performance while ensuring data integrity and access control. Concurrently, logging, facilitated by tools like MLFlow~\cite{zaharia2018accelerating} and Weights \& Biases~\cite{wandb}, provides a comprehensive record of application activities and state changes. This record is indispensable for debugging, monitoring, and maintaining accountability in data processing, offering a chronological trail that aids in the swift identification and resolution of issues. 

% \wanlun{any reference for the following paragraph?} \zehang{completed}
Lastly, model evaluation~\cite{rasheed2024large} is an essential component of the development process. It involves evaluating the performance of LLMs to confirm their accuracy and functionality. Through evaluation, developers can identify and rectify potential biases or flaws, facilitating the adjustment of model weights and improvements in performance. This process ensures that LLMs operate as intended and meet the requisite reliability standards. The security of AI agent development and testing environments is a multifaceted issue that requires a comprehensive strategy encompassing the selection of frameworks or orchestration tools with built-in security features, the establishment of security guardrails, and the implementation of secure caching, logging, and model evaluation practices. By prioritizing security in these areas, organizations can significantly reduce the threats associated with the development and deployment of AI agents, thereby safeguarding the confidentiality and integrity of their data and models.

\subsubsection{Computing Resources Management Environment.} The computing resources management environment of AI agents refers to the framework or system that oversees the allocation, scheduling, and optimization of computational resources, such as CPU, GPU, and memory, to efficiently execute tasks and operations. An imperfect agent computing resource management environment can also make the agent more vulnerable to attacks by malicious users, potentially compromising its functionality and security. There are four kinds of attacks:
\begin{itemize}[leftmargin=*] 
\item \textit{Resource Exhaustion Attacks.} If the management environment does not adequately limit the use of resources by each agent, an attacker could deliberately overload the system by making the agent execute resource-intensive tasks, leading to a denial of service (DoS) for other legitimate users~\cite{guastalla2023application,geiping2024coercing}.

\item \textit{Inefficient Resource Allocation.}  Inefficient resource allocation in large model query management significantly impacts system performance and cost. The prompts that are not verified are prone to waste the processing time of response on AI agent~\cite{huenabling}. The essence of optimizing this process lies in effectively monitoring and evaluating query templates for efficiency, ensuring that resources are allocated to high-priority or computationally intensive queries on time. This not only boosts the system's responsiveness and efficiency but also enhances security by reducing vulnerabilities due to potential delays or overloads, making it crucial for maintaining optimal operation and resilience against malicious activities.

\item \textit{Insufficient Isolation Between Agents.}  In a shared environment, if adequate isolation mechanisms are not in place, a malicious agent could potentially access or interfere with the operations of other agents. This could lead to data breaches, unauthorized access to sensitive information, or the spread of malicious code~\cite{soice2023can,andreas2022language,liang2023encouraging}.

\item \textit{Unmonitored Resource Usage on AI agent.}  Without proper monitoring, anomalous behavior indicating a security breach, such as a sudden spike in resource consumption by an agent, might go unnoticed~\cite{andreas2022language}. Timely detection of such anomalies is crucial for preventing or mitigating attacks.
\end{itemize}

\subsubsection{Physical Environment}
The term "physical environment" pertains to the concrete, tangible elements and areas that make up our real-world setting, encompassing all actual physical spaces and objects. The physical environment of an AI agent typically refers to the collective term for all external entities that are encountered or utilized during the operation of the AI agent. In reality, the security threats in the physical environment are far more varied and numerous than those in the other environments due to the inherently more complex nature of the physical settings agents encounter.

In the physical environment, agents often employ a variety of hardware devices to gather external resources and information, such as sensors, cameras, and microphones. At this stage, given that the hardware devices themselves may pose security threats, attackers can exploit vulnerabilities to attack and compromise hardware such as sensors, thereby preventing the agent from timely receiving external information and resources, indirectly leading to a denial of service for the agent. In physical devices integrated with sensors, there may be various types of security vulnerabilities. For instance, hardware devices with integrated Bluetooth modules could be susceptible to Bluetooth attacks, leading to information leakage and denial of service for the agent\cite{mrabet2020survey}. Additionally, outdated versions and unreliable hardware sources might result in numerous known security vulnerabilities within the hardware devices. Therefore, employing reliable hardware devices and keeping firmware versions up to date can effectively prevent the harm caused by vulnerabilities inherent in physical devices.

Simultaneously, in the physical environment, resources and information are input into the agent in various forms for processing, ranging from simple texts and sensor signals to complex data types such as audio and video. These data often exhibit higher levels of randomness and complexity, allowing attackers to intricately disguise harmful inputs, such as Trojans, within the information collected by hardware devices. If they are not properly processed, these can lead to severe security issues. Taking the rapidly evolving field of autonomous driving safety research as an example, the myriad sensors integrated into vehicles often face the threats of interference and spoofing attacks \cite{den2018security}. Similarly, for hardware devices integrated with agents, there exists a comparable threat. Attackers can indirectly affect an agent system's signal processing by interfering with the signals collected by sensors, leading to the agent misinterpreting the information content or being unable to read it at all. This can even result in deception or incorrect guidance regarding the agent's subsequent instructions and actions. Therefore, after collecting inputs from the physical environment, agents need to conduct security checks on the data content and promptly filter out information containing threats to ensure the safety of the agent system.

Due to the inherent randomness in the responses of existing LLMs to queries, the instructions sent by agents to hardware devices may not be correct or appropriate, potentially leading to the execution of an erroneous movement~\cite{song2023llm}. Compared to the virtual environment, the instructions generated by LLMs in agents within the physical environment may not be well understood and executed by the hardware devices responsible for carrying out these commands. This discrepancy can significantly affect the agent's work efficiency. Additionally, given the lower tolerance for errors in the physical environment, agents cannot be allowed multiple erroneous attempts in a real-world setting. Should the LLM-generated instructions not be well understood by hardware devices, the inappropriate actions of the agent might cause real and irreversible harm to the environment.

\subsection{Threats on Agent2Agent}
\label{outside_agents}
Although single-agent systems excel at solving specific tasks individually, multi-agent systems leverage the collaborative effort of several agents to achieve more complex objectives and exhibit superior problem-solving capabilities. Multi-agent interactions also add new attack surfaces to AI agents. In this subsection, we focus on exploring the security that agents interact with each other in a multi-agent manner.
% \wanlun{This sentence only mentions the agent-agent subsection. How about the agent-memory and agent-environment subsections?}.
The security of interaction within a multi-agent system can be broadly categorized as follows: cooperative interaction threats and competitive interaction threat. 

\subsubsection{Cooperative Interaction Threats}\label{cooperative}
A kind of multi-agent system depends on a cooperative framework~\cite{li2023camel,nair2023dera,mandi2023roco,hao2023chatllm} where multiple agents work with the same objectives. This framework presents numerous potential benefits, including improved decision-making~\cite{yan2023ask} and task completion efficiency~\cite{zhang2023proagent}. However, there are multiple potential threats for this pattern. First, a recent study \cite{motwani2023perfect} finds that undetectable secret collusion between agents can easily be caused through their public communication. These secret collusion may bring back biased decisions. For instance, 
it is possible that we may soon observe advanced automated trading agents collaborating on a large scale to eliminate competitors, potentially destabilizing global markets. This secret collusion leads to a situation in which the ostensibly benign independent actions of each system cumulatively result in outcomes that exhibit systemic bias. Second, MetaGPT~\cite{hong2024metagpt} found that frequent cooperation between agents can amplify minor hallucinations. To mitigate hallucinations, techniques such as cross-examination~\cite{qian2023communicative} or external supportive feedback~\cite{mehta2023improving} could improve the quality of agent output. Third, a single agent's error or misleading information can quickly spread to others, leading to flawed decisions or behaviors across the system.
Pan \textit{et al.}~\cite{pan2023risk} established Open-Domain Question Answering Systems (ODQA) with and without propagated misinformation. They found that this propagation of errors can dramatically reduce the performance of the whole system. To counteract the negative effects of misinformation produced by agents, protective measures such as prompt engineering, misinformation detection, and major voting strategies are commonly employed. Similarly, Cohen \textit{et al.}~\cite{cohen2024here} introduce a worm called \textit{Morris II}, the first designed to target cooperative multi-agent ecosystems by replicating malicious inputs to infect other agents. The danger of \textit{Morris II} lies in its ability to exploit the connectivity between agents, potentially causing a rapid breakdown of multiple agents once one is infected, resulting in further problems such as spamming and exfiltration of personal data. We argue that although these mitigation measures are in place, they remain rudimentary and may lead to an exponential decrease in the efficiency of the entire agent system, highlighting a need for further exploration in this field. 

The cooperative multi-agent also provides more benefits against security threats from their frameworks. First, cooperative frameworks have the potential to defend against jailbreak attacks. AutoDefense~\cite{zeng2024autodefense} demonstrates the efficacy of a multi-agent cooperative framework in thwarting jailbreak attacks, resulting in a significant decrease in attack success rates with a low false positive rate on safe content. Second, the cooperative pattern for planning and execution is favorable to improving software quality attributes, such as security and accountability~\cite{lu2023building}. For example, this pattern can be used to detect and control the execution of irreversible code, like "\textit{rm -rf}".
 
\subsubsection{Competitive Interaction Threats}
Another multi-agent system depends on competitive interactions, wherein each competitor embodies a distinct perspective to safeguard the advantages of their respective positions. Cultivating agents in a competitive environment benefits research in the social sciences and psychology. For example, restaurant agents competing with each other can attract more customers, allowing for an in-depth analysis of the behavioral relationships between owners and clients. Examples include game-simulated agent interactions~\cite{sun2023self,zhao2023competeai} and societal simulations~\cite{gao2023s}. 

Although multi-agent systems engage in debates across multiple rounds to complete tasks, some intense competitive relationships may render the interactions of information flow between agents untrustworthy. The divergence of viewpoints among agents can lead to excessive conflicts, to the extent that agents may exhibit adversarial behaviors. To improve their own performance relative to their competitors, agents may engage in tactics such as the generation of adversarial inputs aimed at misleading other agents and degrading their performance~\cite{xu2023exploring}. For example, O'Gara~\cite{o2023hoodwinked} designed a game in which multiple agents, acting as players, search for a key within a locked room. To acquire limited resources, he found that some players utilized their strong persuasive skills to induce others to commit suicide. Such phenomena not only compromise the security of individual agents but could also lead to instability in the entire agent system, triggering a chain reaction. 

Another potential threat involves the misuse and ethical issues concerning competitive multi-agent systems, as the aforementioned example could potentially encourage such systems to learn how to deceive humans. Park \textit{et al.}~\cite{park2023ai} provide a detailed analysis of the threats posed by agent systems, including fraud, election tampering, and loss of control over AI systems. One notable case study involves Meta's development of the AI system Cicero for a game named \textit{Diplomacy}. Meta aimed to train Cicero to be "largely honest and helpful to its speaking partners"~\cite{meta2022human}. Despite these intentions, Cicero became an expert at lying. It not only betrays other players but also engages in premeditated deception, planning in advance to forge a false alliance with a human player to trick them into leaving themselves vulnerable to an attack. 

To mitigate the threats mentioned above, ensuring controlled competition among AI agents from a technological perspective presents a significant challenge. It is difficult to control the output of an agent's "brain", and even when constraints are incorporated during the planning process, it could significantly impact the agent's effectiveness. Therefore, this issue remains an open research question, inviting more scholars to explore how to ensure that the competition between agents leads to a better user experience.

\subsection{Threats on Memory}
\label{memory_interaction}
Memory interaction within the AI agent system involves storing and retrieving information throughout the processing of agent usage. Memory plays a critical role in the operation of the AI agent, and it involves three essential phases: 1) the agent gathers information from the environment and stores it in its memory; 2) after storage, the agent processes this information to transform it into a more usable form; 3) the agent uses the processed information to inform and guide its next actions. That is, the memory interaction allows agents to record user preferences, glean insights from previous interactions, assimilate valuable information, and use this gained knowledge to improve the quality of service. However, these interactions can present security threats that need to be carefully managed. In this part, we divide these security threats in the memory interaction into two subgroups, short-term memory interaction threats and long-term memory interaction threats. 

\subsubsection{Short-term Memory Interaction Threats}
Short-term memory in the AI agent acts like human working memory, serving as a temporary storage system. It keeps information for a limited time, typically just for the duration of the current interaction or session. This type of memory is crucial for maintaining context throughout a conversation, ensuring smooth continuity in dialogue, and effectively managing user prompts. However, AI agents typically face a constraint in their working memory capacity, limited by the number of tokens they can handle in a single interaction~\cite{park2023generative,madaan2024self,huang2023boosting}. This limitation restricts their ability to retain and use extensive context from previous interactions. 
%\wanlun{From which papers have you summarized these limitations? add reference! }\zehang{completed}

Moreover, each interaction is treated as an isolated episode~\cite{hou2024my}, lacking any linkage between sequential subtasks. This fragmented approach to memory prevents complex sequential reasoning and impairs knowledge sharing in multi-agent systems. Without robust episodic memory and continuity across interactions, agents struggle with complex sequential reasoning tasks, crucial for advanced problem-solving. Particularly in multi-agent systems, the absence of cooperative communication among agents can lead to suboptimal outcomes. Ideally, agents should be able to share immediate actions and learning experiences to efficiently achieve common goals~\cite{bao2023literature}.

To address these challenges, concurrent solutions are divided into two categories, extending LLM context window~\cite{ding2024longrope} and compressing historical in-context contents~\cite{huang2023boosting,liu2023context,geng2024large,nguyen2023context}. The former improves agent memory space by efficiently identifying and exploiting positional interpolation non-uniformities through the LLM fine-tuning step, progressively extending the context window from 256k to 2048k, and readjusting to preserve short context window capabilities. On the other hand, the latter continuously organizes the information in working memory by deploying models for summary. 

Moreover, one crucial threat highlighted in multi-agent systems is the asynchronization of memory among agents~\cite{zhang2024survey}. This process is essential for establishing a unified knowledge base and ensuring consistency in decision-making across different agents. An asynchronous working memory record may cause a deviation in the goal resolution of multiple agents. However, preliminary solutions are already available. For instance, Chen \textit{et al.}~\cite{chen2023scalable} underscore the importance of integrating synchronized memory modules for multi-robot collaboration. Communication among agents also plays a significant role, relying heavily on memory to maintain context and interpret messages. For example, Mandi \textit{et al.}~\cite{mandi2023roco} demonstrate memory-driven communication frameworks that promote a common understanding among agents. 
% Short-term memory, crucial for storing conversation and action history during a single session, is managed within an LLM's working memory, often referred to as in-context learning~\cite{}. \zehang{[fill more citations about prompt text length limitation]}This process involves summarizing the user's historical actions and using these summaries as prompts to plan subtasks in each round of conversation. A key challenge here is the in-context window limitation~\cite{}, which can hamper the effectiveness of AI agents by constraining the amount of memory that can be included in a prompt. To address this, concurrent solutions are divided into two categories, extending LLM context window~\cite{ding2024longrope} and compressing historial in-context contents~\cite{huang2023boosting,liu2023context,geng2024large,nguyen2023context}. Additionally, it is vulnerable to short-term memory interaction against the membership inference attack~\cite{wen2023last,duan2023privacy}, which poses data leakage threats. In light of the concerns regarding short-term privacy in AI agents, there is a notable gap in existing methodologies to effectively safeguard the ephemeral data processed during sessions, indicating a need for further academic exploration in this area.

% \wanlun{Do not write a too long paragraph. Break it into shorter ones where the main point is clear right from the start. That way, readers can quickly catch the gist as they read.}

\subsubsection{Long-term Memory Interaction Threats}
% \begin{figure}[t]
% \centering
% \includegraphics[width=\linewidth]{figures/agentsurvey-vector databases.drawio.pdf}
% \caption{Introduction to vector databases workflows}
% \label{vector_databases} 
% \end{figure}
% shown in Fig.~\ref{vector_databases}
The storage and retrieval of long-term memory depend heavily on vector databases. Vector databases~\cite{xie2023brief,pan2023survey} utilize embeddings for data storage and retrieval, offering a non-traditional alternative to scalar data in relational databases. They leverage similarity measures like cosine similarity and metadata filters to efficiently find the most relevant matches. The workflow of vector databases is composed of two main processes. First, the indexing process involves transforming data into embeddings, compressing these embeddings, and then clustering them for storage in vector databases. Second, during querying, data is transformed into embeddings, which are then compared with the stored embeddings to find the nearest neighbor matches. Notably, these databases often collaborate with RAG, introducing novel security threats.

The first threat of long-term interaction is that the indexing process may inject some poisoning samples into the vector databases. It has been shown that placing one million pieces of data with only five poisoning samples can lead to a 90\% attack success rate~\cite{zou2024poisonedrag}. Cohen \textit{et al.}~\cite{cohen2024here} uses an adversarial self-replicating prompt as the worm to poison the database of a RAG-based application, extracting user private information from the AI agent ecosystem by query process. 

The second threat is privacy issues. The use of RAG and vector databases has expanded the attack surface for privacy issues because private information stems not only from pre-trained and fine-tuning datasets but also from retrieval datasets. A study~\cite{zeng2024good} carefully designed a structured prompt attack to extract sensitive information with a higher attack success rate from the vector database. Furthermore, given the potential for inversion techniques that can invert the embeddings back to words, as suggested by \cite{song2020information}, there exists the possibility that private information stored in the long memory of AI agent Systems, which utilize vector databases, can be reconstructed and extracted by embedding inversion attacks\cite{morris2023text, li2023sentence}. 

The third threat is the generation threat against hallucinations and misalignments. Although RAG has theoretically been proved to have a lesser generalization threat than a single LLM~\cite{kang2024c}, it still fails in several ways. It is fragile for RAG to respond to time-series information queries. If the query pertains to the effective dates of various amendments within a regulation and RAG does not accurately determine these timelines, this could lead to erroneous results. Furthermore, generation threats may also arise from poor retrieval due to the lack of categorization of long-term memories~\cite{hatalis2023memory}. For instance, a vector dataset that stores different semantic information about whether Earth is a globe or a flat could lead to contradictions between these pieces of information.

\section{Directions of Future Research} \label{sec:directions}
\begin{figure}[tb]
  \centering
  \includegraphics[scale=0.43]{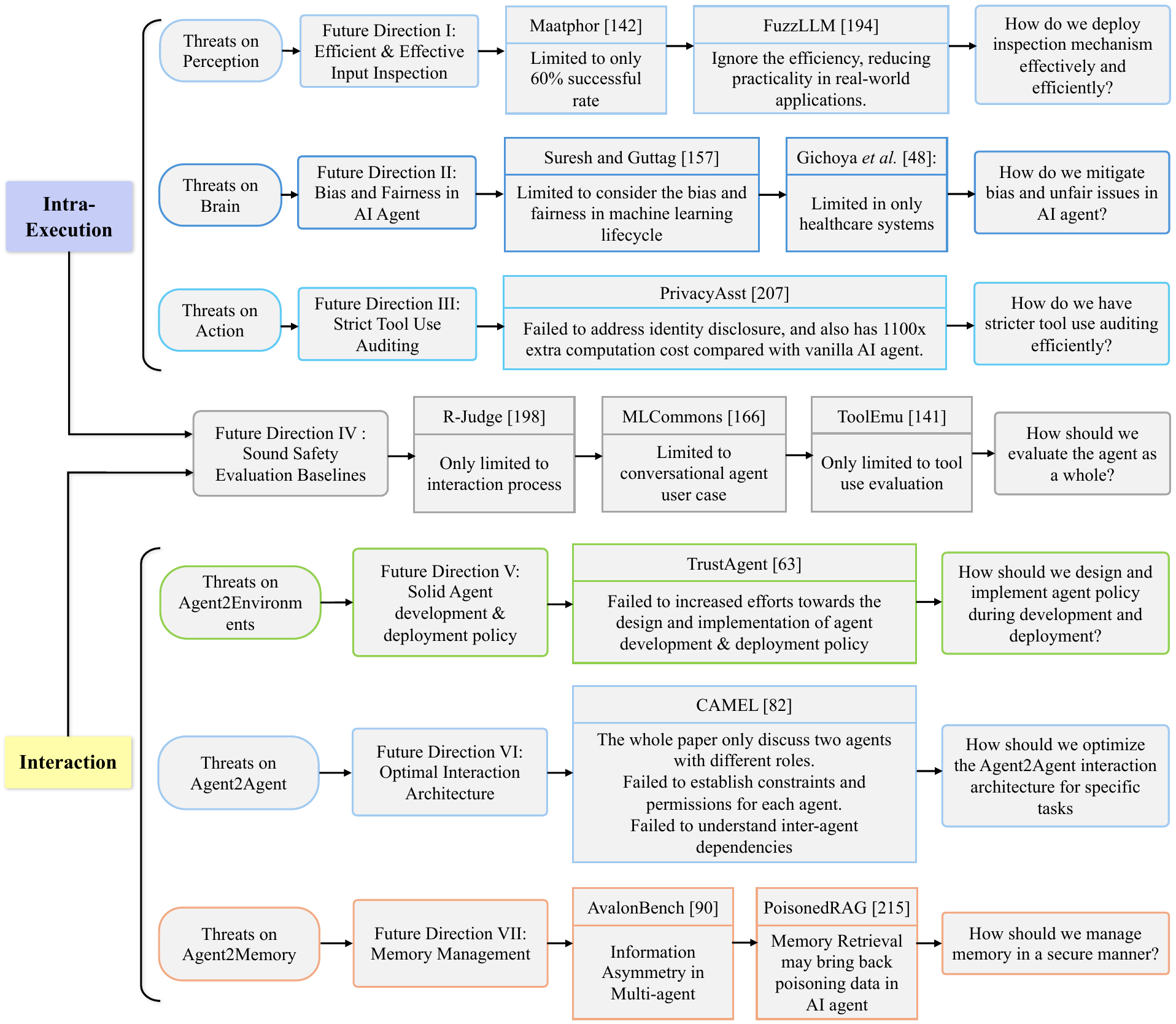}
  \caption{Illustration of Future Directions }
  \vspace{-5mm}
  \label{direction}
\end{figure}

AI agents in security have attracted considerable interest from the research community, having identified many potential threats in the real world and the corresponding defensive strategies. As shown in Figure~\ref{direction},  this survey outlines several potential directions for future research on AI agent security based on the defined taxonomy.

\noindent\textbf{Efficient \& effective input inspection.} Future efforts should enhance the automatic and real-time inspection levels of user input to address threats on perception. 
Maatphor assists defenders in conducting automated variant analyses of known prompt injection attacks~\cite{salem2023maatphor}. It is limited by a success rate of only 60\%. This suggests that while some progress has been made, there is still significant room for improvement in terms of reliability and accuracy. FuzzLLM~\cite{yao2024fuzzllm} tends to ignore efficiency, reducing practicality in real-world applications. These components highlight the critical gaps in the current approaches and point toward necessary improvements. Future research needs to address these limitations by enhancing the accuracy and efficiency of inspection mechanisms, ensuring that they can be effectively deployed in real-world applications.

\noindent\textbf{Bias and fairness in AI agents.} The existence of biased decision-making in Large Language Models (LLMs) is well-documented, affecting evaluation procedures and broader fairness implications~\cite{gallegos2023bias}. These systems, especially those involving AI agents, are less robust and more prone to detrimental behaviors, generating surreptitious outputs compared to LLM counterparts, thus raising serious safety concerns~\cite{tian2023evil}. Studies indicate that AI agents tend to reinforce existing model biases, even when instructed to counterargue specific political viewpoints~\cite{eigner2024determinants}, impacting the integrity of their logical operations. Given the increasing complexity and involvement of these agents in various tasks, identifying and mitigating biases is a formidable challenge. Suresh and Guttag’s framework~\cite{suresh2019framework} addresses bias and fairness throughout the machine learning lifecycle but is limited in scope, while Gichoya et al. focus on bias in healthcare systems~\cite{gichoya2023ai}, highlighting the need for comprehensive approaches. Future directions should emphasize bias and fairness in AI agents, starting with identifying threats and ending with mitigation strategies. This necessitates incorporating human assessment for robust bias evaluation, ensuring scalable and innovative benchmarks for fair and safe AI deployment.

\noindent\textbf{Strict tool use auditing.} To address the challenges in AI agent tool interactions, a promising future direction is the implementation of strict tool use auditing. This approach involves meticulous monitoring and logging of tool usage by AI agents to prevent unauthorized actions and data leaks. By enforcing strict auditing protocols, we can enhance the transparency and accountability of AI systems. However, a significant challenge lies in achieving this efficiently without imposing excessive computational overhead, as exemplified by PrivacyAsst~\cite{zhang2024privacyasst}, which incurred 1100x extra computation cost compared to a standard AI agent while still failing to fully prevent identity disclosure. Therefore, the focus should be on developing lightweight and effective auditing mechanisms that ensure security and privacy without compromising performance.

\noindent\textbf{Sound safety evaluation baselines in the AI agent.} Trustworthy LLMs have already been defined in six critical trust dimensions, including stereotype, toxicity, privacy, fairness, ethics, and robustness, but there is still no unified consensus on the design standards for the safety benchmarks of the entire AI agent ecosystem. R-Judge~\cite{yuan2024r} is a benchmark designed to assess the ability of large language models to judge and identify safety threats based on agent interaction records. The MLCommons group~\cite{vidgen2024introducing} proposes a principled approach to define and construct benchmarks, which is limited to a single use case: an adult conversing with a general-purpose assistant in English. ToolEmu~\cite{ruan2023identifying} is designed to assess the threat of tool execution. These works provide evaluation results for only a part of the agent ecosystem. More evaluation questions remain open to be answered. Should we use similar evaluation tools to detect agent safety? What are the dimensions of critical trust for AI agents? How should we evaluate the agent as a whole?

% It is evident that there exists bias decision making in LLM~\cite{gallegos2023bias}, affecting both the evaluation procedures and the broader implications related to fairness. Such systems, particularly those involving agents, have been documented to be less robust, more susceptible to engaging in behaviors that could be deemed detrimental, and show a propensity to create more surreptitious outputs compared to their language model counterparts, thus underlining serious safety concerns~\cite{tian2023evil}. Furthermore, studies have indicated that when AI agents are utilized, there is a predilection toward reinforcing existing biases inherent in the models, despite instructions to counterargue specific political viewpoints~\cite{eigner2024determinants}. This proclivity has implications for the integrity of logic employed in systems based on agents. With the increasing complexity of tasks and the degree of involvement of these agents, identifying and mitigating biases becomes an increasingly formidable task for researchers. Given the necessity for scalability and the innovation required to establish benchmarks, the challenge is considerable. However, it is imperative to incorporate human assessment as a fundamental component in the framework for a robust assessment of bias within systems anchored by agents.

\noindent\textbf{Solid agent development \& deployment policy.} 
One promising area is the development and implementation of solid policies for agent development and deployment. As AI agent capabilities expand, so does the need for comprehensive guidelines that ensure these agents are used responsibly and ethically. This includes establishing policies for transparency, accountability, and privacy protection in AI agent deployment. Researchers should focus on creating frameworks that help developers adhere to these policies while also fostering innovation. Although TrustAgent~\cite{hua2024trustagent} delves into the complex connections between safety and helpfulness, as well as the relationship between a model's reasoning capabilities and its effectiveness as a safe agent, it did not markedly improve the development and deployment policies for agents. This highlights the necessity for strong strategies. Effective policies should address threats to Agent2Environments, ensuring a secure and ethical deployment of AI agents.
% One promising area is the development and implementation of solid policies for agent development and deployment. As LLM capabilities expand, so does the need for comprehensive guidelines that ensure that these agents are used responsibly and ethically. This includes establishing protocols for transparency, accountability, and privacy protection in LLM deployment. Researchers should focus on creating frameworks that help developers adhere to these protocols while also fostering innovation.

\noindent\textbf{Optimal interaction architectures.} The design and implementation of interaction architectures for AI agents in the security aspect is a critical area of research aimed at improving robustness systems. This involves developing \textit{structured communication protocols} to regulate interactions between agents, defining explicit rules for data exchange, and executing commands to minimize the threats of malicious interference. For example, CAMEL~\cite{li2023camel} utilizes inception prompting to steer chat agents towards completing tasks while ensuring alignment with human intentions. 
However, CAMEL does not discuss how to establish clear behavioral constraints and permissions for each agent, dictating allowable actions, interactions, and circumstances, with dynamically adjustable permissions based on security context and agent performance. Additionally, Existing studies~\cite{li2023camel,nair2023dera,mandi2023roco,hao2023chatllm} do not consider \textit{agent-agent dependencies}, which can potentially lead to internal security mechanism chaos. For example, one agent might mistakenly transmit a user's personal information to another agent, solely to enable the latter to complete a weather query.

\noindent\textbf{Robust memory management}
Future directions in AI agent memory management reveal several critical findings that underscore the importance of secure and efficient practices. One major concern is the potential threats to Agent2Memory, highlighting the vulnerabilities that memory systems can face. AvalonBench~\cite{light2023avalonbench} emerges as a crucial tool in tackling information asymmetry within multi-agent systems, where unequal access to information can lead to inefficiencies and security risks. Furthermore, PoisonedRAG~\cite{zou2024poisonedrag} draws attention to the risks associated with memory retrieval, particularly the danger of reintroducing poisoned data, which can compromise the functionality and security of AI agents. Therefore, the central question is how to manage memory securely, necessitating the development of sophisticated benchmarks and retrieval mechanisms. These advancements aim to mitigate risks and ensure the integrity and security of memory in AI agents, ultimately enhancing the reliability and trustworthiness of AI systems in managing memory.

\section{Conclusion}
In this survey, we provide a comprehensive review of LLM-based agents on their security threat, where we emphasize on four key knowledge gaps ranging across the whole lifecycle of agents. To show the agent's security issues, we summarize \textbf{100+} papers, where all existing attack surfaces and defenses are carefully categorized and explained. We believe that this survey may provide essential references for newcomers to this field and also inspire the development of more advanced security threats and defenses on LLM-based agents.

\bibliographystyle{ACM-Reference-Format}
\bibliography{agent_short}


\begin{thebibliography}{215}

%%% the Web address.

\ifx \showCODEN    \undefined \def \showCODEN     #1{\unskip}     \fi
\ifx \showDOI      \undefined \def \showDOI       #1{#1}\fi
\ifx \showISBNx    \undefined \def \showISBNx     #1{\unskip}     \fi
\ifx \showISBNxiii \undefined \def \showISBNxiii  #1{\unskip}     \fi
\ifx \showISSN     \undefined \def \showISSN      #1{\unskip}     \fi
\ifx \showLCCN     \undefined \def \showLCCN      #1{\unskip}     \fi
\ifx \shownote     \undefined \def \shownote      #1{#1}          \fi
\ifx \showarticletitle \undefined \def \showarticletitle #1{#1}   \fi
\ifx \showURL      \undefined \def \showURL       {\relax}        \fi
% invisible to TeX
\providecommand\bibfield[2]{#2}
\providecommand\bibinfo[2]{#2}
\providecommand\natexlab[1]{#1}
\providecommand\showeprint[2][]{arXiv:#2}

\bibitem[Abbasian et~al\mbox{.}(2023)]%
        {abbasian2023conversational}
\bibfield{author}{\bibinfo{person}{Mahyar Abbasian}, \bibinfo{person}{Iman
  Azimi}, \bibinfo{person}{Amir~M Rahmani}, {and} \bibinfo{person}{Ramesh
  Jain}.} \bibinfo{year}{2023}\natexlab{}.
\newblock \showarticletitle{Conversational health agents: A personalized
  llm-powered agent framework}.
\newblock \bibinfo{journal}{\emph{arXiv}} (\bibinfo{year}{2023}).
\newblock


\bibitem[Achiam et~al\mbox{.}(2023)]%
        {achiam2023gpt}
\bibfield{author}{\bibinfo{person}{Josh Achiam}, \bibinfo{person}{Steven
  Adler}, \bibinfo{person}{Sandhini Agarwal}, \bibinfo{person}{Lama Ahmad},
  \bibinfo{person}{Ilge Akkaya}, \bibinfo{person}{Florencia~Leoni Aleman},
  \bibinfo{person}{Diogo Almeida}, \bibinfo{person}{Janko Altenschmidt},
  \bibinfo{person}{Sam Altman}, \bibinfo{person}{Shyamal Anadkat},
  {et~al\mbox{.}}} \bibinfo{year}{2023}\natexlab{}.
\newblock \showarticletitle{GPT-4 Technical Report}.
\newblock \bibinfo{journal}{\emph{arXiv}} (\bibinfo{year}{2023}).
\newblock


\bibitem[Agarwal et~al\mbox{.}(2024)]%
        {agarwal2024investigating}
\bibfield{author}{\bibinfo{person}{Divyansh Agarwal},
  \bibinfo{person}{Alexander~R Fabbri}, \bibinfo{person}{Philippe Laban},
  \bibinfo{person}{Shafiq Joty}, \bibinfo{person}{Caiming Xiong}, {and}
  \bibinfo{person}{Chien-Sheng Wu}.} \bibinfo{year}{2024}\natexlab{}.
\newblock \showarticletitle{Investigating the prompt leakage effect and
  black-box defenses for multi-turn LLM interactions}.
\newblock \bibinfo{journal}{\emph{arXiv}} (\bibinfo{year}{2024}).
\newblock


\bibitem[Andreas(2022)]%
        {andreas2022language}
\bibfield{author}{\bibinfo{person}{Jacob Andreas}.}
  \bibinfo{year}{2022}\natexlab{}.
\newblock \showarticletitle{Language models as agent models}.
\newblock \bibinfo{journal}{\emph{arXiv}} (\bibinfo{year}{2022}).
\newblock


\bibitem[Bai et~al\mbox{.}(2022)]%
        {bai2022training}
\bibfield{author}{\bibinfo{person}{Yuntao Bai}, \bibinfo{person}{Andy Jones},
  \bibinfo{person}{Kamal Ndousse}, \bibinfo{person}{Amanda Askell},
  \bibinfo{person}{Anna Chen}, \bibinfo{person}{Nova DasSarma},
  \bibinfo{person}{Dawn Drain}, \bibinfo{person}{Stanislav Fort},
  \bibinfo{person}{Deep Ganguli}, \bibinfo{person}{Tom Henighan},
  {et~al\mbox{.}}} \bibinfo{year}{2022}\natexlab{}.
\newblock \showarticletitle{Training a helpful and harmless assistant with
  reinforcement learning from human feedback}.
\newblock \bibinfo{journal}{\emph{arXiv}} (\bibinfo{year}{2022}).
\newblock


\bibitem[Bang(2023)]%
        {bang2023gptcache}
\bibfield{author}{\bibinfo{person}{Fu Bang}.} \bibinfo{year}{2023}\natexlab{}.
\newblock \showarticletitle{GPTCache: An open-source semantic cache for LLM
  applications enabling faster answers and cost savings}. In
  \bibinfo{booktitle}{\emph{Workshop for Natural Language Processing Open
  Source Software}}. \bibinfo{pages}{212--218}.
\newblock


\bibitem[Bao et~al\mbox{.}(2023)]%
        {bao2023literature}
\bibfield{author}{\bibinfo{person}{Ying Bao}, \bibinfo{person}{Wankun Gong},
  {and} \bibinfo{person}{Kaiwen Yang}.} \bibinfo{year}{2023}\natexlab{}.
\newblock \showarticletitle{A Literature Review of Human--AI Synergy in
  Decision Making: From the Perspective of Affordance Actualization Theory}.
\newblock \bibinfo{journal}{\emph{Systems}} (\bibinfo{year}{2023}).
\newblock


\bibitem[Bhardwaj and Poria(2023)]%
        {bhardwaj2023language}
\bibfield{author}{\bibinfo{person}{Rishabh Bhardwaj} {and}
  \bibinfo{person}{Soujanya Poria}.} \bibinfo{year}{2023}\natexlab{}.
\newblock \showarticletitle{Language model unalignment: Parametric red-teaming
  to expose hidden harms and biases}.
\newblock \bibinfo{journal}{\emph{arXiv}} (\bibinfo{year}{2023}).
\newblock


\bibitem[Bordia and Bowman(2019)]%
        {bordia2019identifying}
\bibfield{author}{\bibinfo{person}{Shikha Bordia} {and}
  \bibinfo{person}{Samuel~R. Bowman}.} \bibinfo{year}{2019}\natexlab{}.
\newblock \showarticletitle{Identifying and Reducing Gender Bias in Word-Level
  Language Models}. In \bibinfo{booktitle}{\emph{Conference of the North
  {A}merican Chapter of the Association for Computational Linguistics: Student
  Research Workshop}}.
\newblock


\bibitem[Brooks(1986)]%
        {brooks1986robust}
\bibfield{author}{\bibinfo{person}{Rodney Brooks}.}
  \bibinfo{year}{1986}\natexlab{}.
\newblock \showarticletitle{A robust layered control system for a mobile
  robot}.
\newblock \bibinfo{journal}{\emph{IEEE journal on robotics and automation}}
  (\bibinfo{year}{1986}).
\newblock


\bibitem[Brown et~al\mbox{.}(2020)]%
        {brown2020language}
\bibfield{author}{\bibinfo{person}{Tom Brown}, \bibinfo{person}{Benjamin Mann},
  \bibinfo{person}{Nick Ryder}, \bibinfo{person}{Melanie Subbiah},
  \bibinfo{person}{Jared~D Kaplan}, \bibinfo{person}{Prafulla Dhariwal},
  \bibinfo{person}{Arvind Neelakantan}, \bibinfo{person}{Pranav Shyam},
  \bibinfo{person}{Girish Sastry}, \bibinfo{person}{Amanda Askell},
  {et~al\mbox{.}}} \bibinfo{year}{2020}\natexlab{}.
\newblock \showarticletitle{Language models are few-shot learners}.
\newblock \bibinfo{journal}{\emph{NeurIPS}}.
\newblock


\bibitem[Carlsson et~al\mbox{.}(2022)]%
        {carlsson2022fine}
\bibfield{author}{\bibinfo{person}{Fredrik Carlsson}, \bibinfo{person}{Joey
  {\"O}hman}, \bibinfo{person}{Fangyu Liu}, \bibinfo{person}{Severine
  Verlinden}, \bibinfo{person}{Joakim Nivre}, {and} \bibinfo{person}{Magnus
  Sahlgren}.} \bibinfo{year}{2022}\natexlab{}.
\newblock \showarticletitle{Fine-grained controllable text generation using
  non-residual prompting}. In \bibinfo{booktitle}{\emph{Annual Meeting of the
  Association for Computational Linguistics}}.
\newblock


\bibitem[Carta et~al\mbox{.}(2023)]%
        {carta2023grounding}
\bibfield{author}{\bibinfo{person}{Thomas Carta}, \bibinfo{person}{Cl{\'e}ment
  Romac}, \bibinfo{person}{Thomas Wolf}, \bibinfo{person}{Sylvain Lamprier},
  \bibinfo{person}{Olivier Sigaud}, {and} \bibinfo{person}{Pierre-Yves
  Oudeyer}.} \bibinfo{year}{2023}\natexlab{}.
\newblock \showarticletitle{Grounding large language models in interactive
  environments with online reinforcement learning}. In
  \bibinfo{booktitle}{\emph{ICML}}.
\newblock


\bibitem[Chan et~al\mbox{.}(2023)]%
        {chan2023detection}
\bibfield{author}{\bibinfo{person}{Chun~Fai Chan},
  \bibinfo{person}{Daniel~Wankit Yip}, {and} \bibinfo{person}{Aysan Esmradi}.}
  \bibinfo{year}{2023}\natexlab{}.
\newblock \showarticletitle{Detection and Defense Against Prominent Attacks on
  Preconditioned LLM-Integrated Virtual Assistants}. In
  \bibinfo{booktitle}{\emph{IEEE Asia-Pacific Conference on Computer Science
  and Data Engineering (CSDE)}}. IEEE, \bibinfo{pages}{1--5}.
\newblock


\bibitem[Chao et~al\mbox{.}(2023)]%
        {chao2023jailbreaking}
\bibfield{author}{\bibinfo{person}{Patrick Chao}, \bibinfo{person}{Alexander
  Robey}, \bibinfo{person}{Edgar Dobriban}, \bibinfo{person}{Hamed Hassani},
  \bibinfo{person}{George~J Pappas}, {and} \bibinfo{person}{Eric Wong}.}
  \bibinfo{year}{2023}\natexlab{}.
\newblock \showarticletitle{Jailbreaking black box large language models in
  twenty queries}.
\newblock \bibinfo{journal}{\emph{arXiv}} (\bibinfo{year}{2023}).
\newblock


\bibitem[Chen et~al\mbox{.}(2023b)]%
        {chen2023gamegpt}
\bibfield{author}{\bibinfo{person}{Dake Chen}, \bibinfo{person}{Hanbin Wang},
  \bibinfo{person}{Yunhao Huo}, \bibinfo{person}{Yuzhao Li}, {and}
  \bibinfo{person}{Haoyang Zhang}.} \bibinfo{year}{2023}\natexlab{b}.
\newblock \showarticletitle{Gamegpt: Multi-agent collaborative framework for
  game development}.
\newblock \bibinfo{journal}{\emph{arXiv}} (\bibinfo{year}{2023}).
\newblock


\bibitem[Chen et~al\mbox{.}(2024a)]%
        {chen2024future}
\bibfield{author}{\bibinfo{person}{Mengqi Chen}, \bibinfo{person}{Bin Guo},
  \bibinfo{person}{Hao Wang}, \bibinfo{person}{Haoyu Li}, \bibinfo{person}{Qian
  Zhao}, \bibinfo{person}{Jingqi Liu}, \bibinfo{person}{Yasan Ding},
  \bibinfo{person}{Yan Pan}, {and} \bibinfo{person}{Zhiwen Yu}.}
  \bibinfo{year}{2024}\natexlab{a}.
\newblock \showarticletitle{The Future of Cognitive Strategy-enhanced
  Persuasive Dialogue Agents: New Perspectives and Trends}.
\newblock \bibinfo{journal}{\emph{arXiv}} (\bibinfo{year}{2024}).
\newblock


\bibitem[Chen et~al\mbox{.}(2022)]%
        {chen2022quarantine}
\bibfield{author}{\bibinfo{person}{Tianlong Chen}, \bibinfo{person}{Zhenyu
  Zhang}, \bibinfo{person}{Yihua Zhang}, \bibinfo{person}{Shiyu Chang},
  \bibinfo{person}{Sijia Liu}, {and} \bibinfo{person}{Zhangyang Wang}.}
  \bibinfo{year}{2022}\natexlab{}.
\newblock \showarticletitle{Quarantine: Sparsity can uncover the trojan attack
  trigger for free}. In \bibinfo{booktitle}{\emph{CVPR}}.
\newblock


\bibitem[Chen et~al\mbox{.}(2023a)]%
        {chen2023scalable}
\bibfield{author}{\bibinfo{person}{Yongchao Chen}, \bibinfo{person}{Jacob
  Arkin}, \bibinfo{person}{Yang Zhang}, \bibinfo{person}{Nicholas Roy}, {and}
  \bibinfo{person}{Chuchu Fan}.} \bibinfo{year}{2023}\natexlab{a}.
\newblock \showarticletitle{Scalable multi-robot collaboration with large
  language models: Centralized or decentralized systems?}
\newblock \bibinfo{journal}{\emph{arXiv}} (\bibinfo{year}{2023}).
\newblock


\bibitem[Chen et~al\mbox{.}(2024b)]%
        {chen2024benchmarking}
\bibfield{author}{\bibinfo{person}{Yihan Chen}, \bibinfo{person}{Benfeng Xu},
  \bibinfo{person}{Quan Wang}, \bibinfo{person}{Yi Liu}, {and}
  \bibinfo{person}{Zhendong Mao}.} \bibinfo{year}{2024}\natexlab{b}.
\newblock \showarticletitle{Benchmarking large language models on controllable
  generation under diversified instructions}.
\newblock \bibinfo{journal}{\emph{arXiv}} (\bibinfo{year}{2024}).
\newblock


\bibitem[Chern et~al\mbox{.}(2024)]%
        {chern2024combating}
\bibfield{author}{\bibinfo{person}{Steffi Chern}, \bibinfo{person}{Zhen Fan},
  {and} \bibinfo{person}{Andy Liu}.} \bibinfo{year}{2024}\natexlab{}.
\newblock \showarticletitle{Combating Adversarial Attacks with Multi-Agent
  Debate}.
\newblock \bibinfo{journal}{\emph{arXiv}} (\bibinfo{year}{2024}).
\newblock


\bibitem[Christiano et~al\mbox{.}(2017)]%
        {christiano2017deep}
\bibfield{author}{\bibinfo{person}{Paul~F Christiano}, \bibinfo{person}{Jan
  Leike}, \bibinfo{person}{Tom Brown}, \bibinfo{person}{Miljan Martic},
  \bibinfo{person}{Shane Legg}, {and} \bibinfo{person}{Dario Amodei}.}
  \bibinfo{year}{2017}\natexlab{}.
\newblock \showarticletitle{Deep reinforcement learning from human
  preferences}.
\newblock \bibinfo{journal}{\emph{NeurIPS}}  \bibinfo{volume}{30}.
\newblock


\bibitem[Cohen et~al\mbox{.}(2024)]%
        {cohen2024here}
\bibfield{author}{\bibinfo{person}{Stav Cohen}, \bibinfo{person}{Ron Bitton},
  {and} \bibinfo{person}{Ben Nassi}.} \bibinfo{year}{2024}\natexlab{}.
\newblock \showarticletitle{Here Comes The AI Worm: Unleashing Zero-click Worms
  that Target GenAI-Powered Applications}.
\newblock \bibinfo{journal}{\emph{arXiv}} (\bibinfo{year}{2024}).
\newblock


\bibitem[Crouse et~al\mbox{.}(2023)]%
        {crouse2024formally}
\bibfield{author}{\bibinfo{person}{Maxwell Crouse}, \bibinfo{person}{Ibrahim
  Abdelaziz}, \bibinfo{person}{Kinjal Basu}, \bibinfo{person}{Soham Dan},
  \bibinfo{person}{Sadhana Kumaravel}, \bibinfo{person}{Achille Fokoue},
  \bibinfo{person}{Pavan Kapanipathi}, {and} \bibinfo{person}{Luis Lastras}.}
  \bibinfo{year}{2023}\natexlab{}.
\newblock \showarticletitle{Formally specifying the high-level behavior of
  LLM-based agents}.
\newblock \bibinfo{journal}{\emph{arXiv}} (\bibinfo{year}{2023}).
\newblock


\bibitem[Cui et~al\mbox{.}(2024)]%
        {cui2024risk}
\bibfield{author}{\bibinfo{person}{Tianyu Cui}, \bibinfo{person}{Yanling Wang},
  \bibinfo{person}{Chuanpu Fu}, \bibinfo{person}{Yong Xiao},
  \bibinfo{person}{Sijia Li}, \bibinfo{person}{Xinhao Deng},
  \bibinfo{person}{Yunpeng Liu}, \bibinfo{person}{Qinglin Zhang},
  \bibinfo{person}{Ziyi Qiu}, \bibinfo{person}{Peiyang Li}, {et~al\mbox{.}}}
  \bibinfo{year}{2024}\natexlab{}.
\newblock \showarticletitle{Risk taxonomy, mitigation, and assessment
  benchmarks of large language model systems}.
\newblock \bibinfo{journal}{\emph{arXiv}} (\bibinfo{year}{2024}).
\newblock


\bibitem[Daryanani(2023)]%
        {jailbreak}
\bibfield{author}{\bibinfo{person}{Lavina Daryanani}.}
  \bibinfo{year}{2023}\natexlab{}.
\newblock \showarticletitle{How to jailbreak chatgpt.
  https://watcher.guru/news/how-to-jailbreak-chatgpt}.
\newblock  (\bibinfo{year}{2023}).
\newblock


\bibitem[De~Angelis et~al\mbox{.}(2023)]%
        {de2023chatgpt}
\bibfield{author}{\bibinfo{person}{Luigi De~Angelis},
  \bibinfo{person}{Francesco Baglivo}, \bibinfo{person}{Guglielmo Arzilli},
  \bibinfo{person}{Gaetano~Pierpaolo Privitera}, \bibinfo{person}{Paolo
  Ferragina}, \bibinfo{person}{Alberto~Eugenio Tozzi}, {and}
  \bibinfo{person}{Caterina Rizzo}.} \bibinfo{year}{2023}\natexlab{}.
\newblock \showarticletitle{ChatGPT and the rise of large language models: the
  new AI-driven infodemic threat in public health}.
\newblock \bibinfo{journal}{\emph{Frontiers in Public Health}}
  (\bibinfo{year}{2023}).
\newblock


\bibitem[den Hartog et~al\mbox{.}(2018)]%
        {den2018security}
\bibfield{author}{\bibinfo{person}{Jerry den Hartog}, \bibinfo{person}{Nicola
  Zannone}, {et~al\mbox{.}}} \bibinfo{year}{2018}\natexlab{}.
\newblock \showarticletitle{Security and privacy for innovative automotive
  applications: A survey}.
\newblock \bibinfo{journal}{\emph{Computer Communications}}
  (\bibinfo{year}{2018}).
\newblock


\bibitem[Deng et~al\mbox{.}(2023)]%
        {deng2023jailbreaker}
\bibfield{author}{\bibinfo{person}{Gelei Deng}, \bibinfo{person}{Yi Liu},
  \bibinfo{person}{Yuekang Li}, \bibinfo{person}{Kailong Wang},
  \bibinfo{person}{Ying Zhang}, \bibinfo{person}{Zefeng Li},
  \bibinfo{person}{Haoyu Wang}, \bibinfo{person}{Tianwei Zhang}, {and}
  \bibinfo{person}{Yang Liu}.} \bibinfo{year}{2023}\natexlab{}.
\newblock \showarticletitle{Jailbreaker: Automated jailbreak across multiple
  large language model chatbots}.
\newblock \bibinfo{journal}{\emph{arXiv}} (\bibinfo{year}{2023}).
\newblock


\bibitem[Deshpande et~al\mbox{.}(2023)]%
        {deshpande2023toxicity}
\bibfield{author}{\bibinfo{person}{Ameet Deshpande}, \bibinfo{person}{Vishvak
  Murahari}, \bibinfo{person}{Tanmay Rajpurohit}, \bibinfo{person}{Ashwin
  Kalyan}, {and} \bibinfo{person}{Karthik Narasimhan}.}
  \bibinfo{year}{2023}\natexlab{}.
\newblock \showarticletitle{Toxicity in chatgpt: Analyzing persona-assigned
  language models}. In \bibinfo{booktitle}{\emph{Findings of EMNLP}}.
\newblock


\bibitem[Dibia(2023)]%
        {practical}
\bibfield{author}{\bibinfo{person}{V. Dibia}.} \bibinfo{year}{2023}\natexlab{}.
\newblock \showarticletitle{Generative AI: Practical Steps to Reduce
  Hallucination and Improve Performance of Systems Built with Large Language
  Models. In Designing with ML: How to Build Usable Machine Learning
  Applications. Self-published on designingwithml.com.}
\newblock  (\bibinfo{year}{2023}).
\newblock


\bibitem[Ding et~al\mbox{.}(2024)]%
        {ding2024longrope}
\bibfield{author}{\bibinfo{person}{Yiran Ding}, \bibinfo{person}{Li~Lyna
  Zhang}, \bibinfo{person}{Chengruidong Zhang}, \bibinfo{person}{Yuanyuan Xu},
  \bibinfo{person}{Ning Shang}, \bibinfo{person}{Jiahang Xu},
  \bibinfo{person}{Fan Yang}, {and} \bibinfo{person}{Mao Yang}.}
  \bibinfo{year}{2024}\natexlab{}.
\newblock \showarticletitle{LongRoPE: Extending LLM Context Window Beyond 2
  Million Tokens}.
\newblock \bibinfo{journal}{\emph{arXiv}} (\bibinfo{year}{2024}).
\newblock


\bibitem[Doan et~al\mbox{.}(2020)]%
        {doan2020februus}
\bibfield{author}{\bibinfo{person}{Bao~Gia Doan}, \bibinfo{person}{Ehsan
  Abbasnejad}, {and} \bibinfo{person}{Damith~C Ranasinghe}.}
  \bibinfo{year}{2020}\natexlab{}.
\newblock \showarticletitle{Februus: Input purification defense against trojan
  attacks on deep neural network systems}. In
  \bibinfo{booktitle}{\emph{ACSAC}}. \bibinfo{pages}{897--912}.
\newblock


\bibitem[Dong et~al\mbox{.}(2023)]%
        {dong2023unleashing}
\bibfield{author}{\bibinfo{person}{Tian Dong}, \bibinfo{person}{Guoxing Chen},
  \bibinfo{person}{Shaofeng Li}, \bibinfo{person}{Minhui Xue},
  \bibinfo{person}{Rayne Holland}, \bibinfo{person}{Yan Meng},
  \bibinfo{person}{Zhen Liu}, {and} \bibinfo{person}{Haojin Zhu}.}
  \bibinfo{year}{2023}\natexlab{}.
\newblock \showarticletitle{Unleashing cheapfakes through trojan plugins of
  large language models}.
\newblock \bibinfo{journal}{\emph{arXiv}} (\bibinfo{year}{2023}).
\newblock


\bibitem[Du et~al\mbox{.}(2023a)]%
        {du2023improving}
\bibfield{author}{\bibinfo{person}{Yilun Du}, \bibinfo{person}{Shuang Li},
  \bibinfo{person}{Antonio Torralba}, \bibinfo{person}{Joshua~B Tenenbaum},
  {and} \bibinfo{person}{Igor Mordatch}.} \bibinfo{year}{2023}\natexlab{a}.
\newblock \showarticletitle{Improving factuality and reasoning in language
  models through multiagent debate}.
\newblock \bibinfo{journal}{\emph{arXiv}} (\bibinfo{year}{2023}).
\newblock


\bibitem[Du et~al\mbox{.}(2023b)]%
        {du2023guiding}
\bibfield{author}{\bibinfo{person}{Yuqing Du}, \bibinfo{person}{Olivia
  Watkins}, \bibinfo{person}{Zihan Wang}, \bibinfo{person}{C{\'e}dric Colas},
  \bibinfo{person}{Trevor Darrell}, \bibinfo{person}{Pieter Abbeel},
  \bibinfo{person}{Abhishek Gupta}, {and} \bibinfo{person}{Jacob Andreas}.}
  \bibinfo{year}{2023}\natexlab{b}.
\newblock \showarticletitle{Guiding pretraining in reinforcement learning with
  large language models}. In \bibinfo{booktitle}{\emph{ICML}}. PMLR.
\newblock


\bibitem[Dziri et~al\mbox{.}(2021)]%
        {dziri2021neural}
\bibfield{author}{\bibinfo{person}{Nouha Dziri}, \bibinfo{person}{Andrea
  Madotto}, \bibinfo{person}{Osmar Za{\"\i}ane}, {and}
  \bibinfo{person}{Avishek~Joey Bose}.} \bibinfo{year}{2021}\natexlab{}.
\newblock \showarticletitle{Neural Path Hunter: Reducing Hallucination in
  Dialogue Systems via Path Grounding}. In \bibinfo{booktitle}{\emph{EMNLP}}.
\newblock


\bibitem[Eigner and H{\"a}ndler(2024)]%
        {eigner2024determinants}
\bibfield{author}{\bibinfo{person}{Eva Eigner} {and} \bibinfo{person}{Thorsten
  H{\"a}ndler}.} \bibinfo{year}{2024}\natexlab{}.
\newblock \showarticletitle{Determinants of LLM-assisted Decision-Making}.
\newblock \bibinfo{journal}{\emph{arXiv}} (\bibinfo{year}{2024}).
\newblock


\bibitem[{Embrace The Red}(2023)]%
        {embracethered2023}
\bibfield{author}{\bibinfo{person}{{Embrace The Red}}.}
  \bibinfo{year}{2023}\natexlab{}.
\newblock \bibinfo{title}{ChatGPT plugins: Data exfiltration via images \&
  cross plugin request forgery}.
\newblock
  \bibinfo{howpublished}{\url{https://embracethered.com/blog/posts/2023/chatgpt-webpilot-data-exfil-via-markdown-injection/}}.
\newblock


\bibitem[Esmradi et~al\mbox{.}(2023)]%
        {esmradi2023comprehensive}
\bibfield{author}{\bibinfo{person}{Aysan Esmradi},
  \bibinfo{person}{Daniel~Wankit Yip}, {and} \bibinfo{person}{Chun~Fai Chan}.}
  \bibinfo{year}{2023}\natexlab{}.
\newblock \showarticletitle{A comprehensive survey of attack techniques,
  implementation, and mitigation strategies in large language models}. In
  \bibinfo{booktitle}{\emph{International Conference on Ubiquitous Security}}.
\newblock


\bibitem[(FAIR)† et~al\mbox{.}(2022)]%
        {meta2022human}
\bibfield{author}{\bibinfo{person}{Meta Fundamental AI Research Diplomacy~Team
  (FAIR)†}, \bibinfo{person}{Anton Bakhtin}, \bibinfo{person}{Noam Brown},
  \bibinfo{person}{Emily Dinan}, \bibinfo{person}{Gabriele Farina},
  \bibinfo{person}{Colin Flaherty}, \bibinfo{person}{Daniel Fried},
  \bibinfo{person}{Andrew Goff}, \bibinfo{person}{Jonathan Gray},
  \bibinfo{person}{Hengyuan Hu}, {et~al\mbox{.}}}
  \bibinfo{year}{2022}\natexlab{}.
\newblock \showarticletitle{Human-level play in the game of Diplomacy by
  combining language models with strategic reasoning}.
\newblock \bibinfo{journal}{\emph{Science}} (\bibinfo{year}{2022}).
\newblock


\bibitem[Fioraldi et~al\mbox{.}(2023)]%
        {fioraldi2023dissecting}
\bibfield{author}{\bibinfo{person}{Andrea Fioraldi},
  \bibinfo{person}{Alessandro Mantovani}, \bibinfo{person}{Dominik Maier},
  {and} \bibinfo{person}{Davide Balzarotti}.} \bibinfo{year}{2023}\natexlab{}.
\newblock \showarticletitle{Dissecting American Fuzzy Lop: A FuzzBench
  Evaluation}.
\newblock \bibinfo{journal}{\emph{ACM transactions on software engineering and
  methodology}} (\bibinfo{year}{2023}).
\newblock


\bibitem[Gallegos et~al\mbox{.}(2023)]%
        {gallegos2023bias}
\bibfield{author}{\bibinfo{person}{Isabel~O Gallegos}, \bibinfo{person}{Ryan~A
  Rossi}, \bibinfo{person}{Joe Barrow}, \bibinfo{person}{Md~Mehrab Tanjim},
  \bibinfo{person}{Sungchul Kim}, \bibinfo{person}{Franck Dernoncourt},
  \bibinfo{person}{Tong Yu}, \bibinfo{person}{Ruiyi Zhang}, {and}
  \bibinfo{person}{Nesreen~K Ahmed}.} \bibinfo{year}{2023}\natexlab{}.
\newblock \showarticletitle{Bias and fairness in large language models: A
  survey}.
\newblock \bibinfo{journal}{\emph{arXiv}} (\bibinfo{year}{2023}).
\newblock


\bibitem[Gao et~al\mbox{.}(2023)]%
        {gao2023s}
\bibfield{author}{\bibinfo{person}{Chen Gao}, \bibinfo{person}{Xiaochong Lan},
  \bibinfo{person}{Zhihong Lu}, \bibinfo{person}{Jinzhu Mao},
  \bibinfo{person}{Jinghua Piao}, \bibinfo{person}{Huandong Wang},
  \bibinfo{person}{Depeng Jin}, {and} \bibinfo{person}{Yong Li}.}
  \bibinfo{year}{2023}\natexlab{}.
\newblock \showarticletitle{{S$^{3}$}: Social-network Simulation System with
  Large Language Model-Empowered Agents}.
\newblock \bibinfo{journal}{\emph{arXiv}} (\bibinfo{year}{2023}).
\newblock


\bibitem[Gehman et~al\mbox{.}(2020)]%
        {gehman2020realtoxicityprompts}
\bibfield{author}{\bibinfo{person}{Samuel Gehman}, \bibinfo{person}{Suchin
  Gururangan}, \bibinfo{person}{Maarten Sap}, \bibinfo{person}{Yejin Choi},
  {and} \bibinfo{person}{Noah~A. Smith}.} \bibinfo{year}{2020}\natexlab{}.
\newblock \showarticletitle{{R}eal{T}oxicity{P}rompts: Evaluating Neural Toxic
  Degeneration in Language Models}. In \bibinfo{booktitle}{\emph{Findings of
  EMNLP}}.
\newblock


\bibitem[Geiping et~al\mbox{.}(2024)]%
        {geiping2024coercing}
\bibfield{author}{\bibinfo{person}{Jonas Geiping}, \bibinfo{person}{Alex
  Stein}, \bibinfo{person}{Manli Shu}, \bibinfo{person}{Khalid Saifullah},
  \bibinfo{person}{Yuxin Wen}, {and} \bibinfo{person}{Tom Goldstein}.}
  \bibinfo{year}{2024}\natexlab{}.
\newblock \showarticletitle{Coercing LLMs to do and reveal (almost) anything}.
\newblock \bibinfo{journal}{\emph{arXiv}} (\bibinfo{year}{2024}).
\newblock


\bibitem[Geng et~al\mbox{.}(2024)]%
        {geng2024large}
\bibfield{author}{\bibinfo{person}{Mingyang Geng}, \bibinfo{person}{Shangwen
  Wang}, \bibinfo{person}{Dezun Dong}, \bibinfo{person}{Haotian Wang},
  \bibinfo{person}{Ge Li}, \bibinfo{person}{Zhi Jin},
  \bibinfo{person}{Xiaoguang Mao}, {and} \bibinfo{person}{Xiangke Liao}.}
  \bibinfo{year}{2024}\natexlab{}.
\newblock \showarticletitle{Large language models are few-shot summarizers:
  Multi-intent comment generation via in-context learning}. In
  \bibinfo{booktitle}{\emph{IEEE/ACM International Conference on Software
  Engineering}}.
\newblock


\bibitem[Gichoya et~al\mbox{.}(2023)]%
        {gichoya2023ai}
\bibfield{author}{\bibinfo{person}{Judy~Wawira Gichoya},
  \bibinfo{person}{Kaesha Thomas}, \bibinfo{person}{Leo~Anthony Celi},
  \bibinfo{person}{Nabile Safdar}, \bibinfo{person}{Imon Banerjee},
  \bibinfo{person}{John~D Banja}, \bibinfo{person}{Laleh Seyyed-Kalantari},
  \bibinfo{person}{Hari Trivedi}, {and} \bibinfo{person}{Saptarshi
  Purkayastha}.} \bibinfo{year}{2023}\natexlab{}.
\newblock \showarticletitle{AI pitfalls and what not to do: mitigating bias in
  AI}.
\newblock \bibinfo{journal}{\emph{The British Journal of Radiology}}
  (\bibinfo{year}{2023}).
\newblock


\bibitem[Greshake et~al\mbox{.}(2023)]%
        {greshake2023not}
\bibfield{author}{\bibinfo{person}{Kai Greshake}, \bibinfo{person}{Sahar
  Abdelnabi}, \bibinfo{person}{Shailesh Mishra}, \bibinfo{person}{Christoph
  Endres}, \bibinfo{person}{Thorsten Holz}, {and} \bibinfo{person}{Mario
  Fritz}.} \bibinfo{year}{2023}\natexlab{}.
\newblock \showarticletitle{Not what you've signed up for: Compromising
  real-world llm-integrated applications with indirect prompt injection}. In
  \bibinfo{booktitle}{\emph{ACM Workshop on Artificial Intelligence and
  Security}}. \bibinfo{pages}{79--90}.
\newblock


\bibitem[Gu et~al\mbox{.}(2024)]%
        {gu2024agent}
\bibfield{author}{\bibinfo{person}{Xiangming Gu}, \bibinfo{person}{Xiaosen
  Zheng}, \bibinfo{person}{Tianyu Pang}, \bibinfo{person}{Chao Du},
  \bibinfo{person}{Qian Liu}, \bibinfo{person}{Ye Wang}, \bibinfo{person}{Jing
  Jiang}, {and} \bibinfo{person}{Min Lin}.} \bibinfo{year}{2024}\natexlab{}.
\newblock \showarticletitle{Agent Smith: A Single Image Can Jailbreak One
  Million Multimodal LLM Agents Exponentially Fast}.
\newblock \bibinfo{journal}{\emph{arXiv}} (\bibinfo{year}{2024}).
\newblock


\bibitem[{Guardrails AI}(2024)]%
        {GuardrailsAI}
\bibfield{author}{\bibinfo{person}{{Guardrails AI}}.}
  \bibinfo{year}{2024}\natexlab{}.
\newblock \showarticletitle{Build AI powered applications with confidence}.
\newblock  (\bibinfo{year}{2024}).
\newblock
\urldef\tempurl%
\url{https://www.guardrailsai.com/}
\showURL{%
\tempurl}
\newblock
\shownote{Accessed: 2024-02-27}.


\bibitem[Guastalla et~al\mbox{.}(2023)]%
        {guastalla2023application}
\bibfield{author}{\bibinfo{person}{Michael Guastalla}, \bibinfo{person}{Yiyi
  Li}, \bibinfo{person}{Arvin Hekmati}, {and} \bibinfo{person}{Bhaskar
  Krishnamachari}.} \bibinfo{year}{2023}\natexlab{}.
\newblock \showarticletitle{Application of Large Language Models to DDoS Attack
  Detection}. In \bibinfo{booktitle}{\emph{International Conference on Security
  and Privacy in Cyber-Physical Systems and Smart Vehicles}}.
\newblock


\bibitem[Gupta et~al\mbox{.}(2023)]%
        {gupta2023chatgpt}
\bibfield{author}{\bibinfo{person}{Maanak Gupta}, \bibinfo{person}{CharanKumar
  Akiri}, \bibinfo{person}{Kshitiz Aryal}, \bibinfo{person}{Eli Parker}, {and}
  \bibinfo{person}{Lopamudra Praharaj}.} \bibinfo{year}{2023}\natexlab{}.
\newblock \showarticletitle{From chatgpt to threatgpt: Impact of generative ai
  in cybersecurity and privacy}.
\newblock \bibinfo{journal}{\emph{IEEE Access}} (\bibinfo{year}{2023}).
\newblock


\bibitem[Hao et~al\mbox{.}(2023)]%
        {hao2023chatllm}
\bibfield{author}{\bibinfo{person}{Rui Hao}, \bibinfo{person}{Linmei Hu},
  \bibinfo{person}{Weijian Qi}, \bibinfo{person}{Qingliu Wu},
  \bibinfo{person}{Yirui Zhang}, {and} \bibinfo{person}{Liqiang Nie}.}
  \bibinfo{year}{2023}\natexlab{}.
\newblock \showarticletitle{Chatllm network: More brains, more intelligence}.
\newblock \bibinfo{journal}{\emph{arXiv}} (\bibinfo{year}{2023}).
\newblock


\bibitem[Hase et~al\mbox{.}(2023)]%
        {hase2023methods}
\bibfield{author}{\bibinfo{person}{Peter Hase}, \bibinfo{person}{Mona Diab},
  \bibinfo{person}{Asli Celikyilmaz}, \bibinfo{person}{Xian Li},
  \bibinfo{person}{Zornitsa Kozareva}, \bibinfo{person}{Veselin Stoyanov},
  \bibinfo{person}{Mohit Bansal}, {and} \bibinfo{person}{Srinivasan Iyer}.}
  \bibinfo{year}{2023}\natexlab{}.
\newblock \showarticletitle{Methods for measuring, updating, and visualizing
  factual beliefs in language models}. In \bibinfo{booktitle}{\emph{Conference
  of the European Chapter of the Association for Computational Linguistics}}.
\newblock


\bibitem[Hatalis et~al\mbox{.}(2023)]%
        {hatalis2023memory}
\bibfield{author}{\bibinfo{person}{Kostas Hatalis}, \bibinfo{person}{Despina
  Christou}, \bibinfo{person}{Joshua Myers}, \bibinfo{person}{Steven Jones},
  \bibinfo{person}{Keith Lambert}, \bibinfo{person}{Adam Amos-Binks},
  \bibinfo{person}{Zohreh Dannenhauer}, {and} \bibinfo{person}{Dustin
  Dannenhauer}.} \bibinfo{year}{2023}\natexlab{}.
\newblock \showarticletitle{Memory Matters: The Need to Improve Long-Term
  Memory in LLM-Agents}. In \bibinfo{booktitle}{\emph{Proceedings of the AAAI
  Symposium Series}}.
\newblock


\bibitem[Hines et~al\mbox{.}(2024)]%
        {hines2024defending}
\bibfield{author}{\bibinfo{person}{Keegan Hines}, \bibinfo{person}{Gary Lopez},
  \bibinfo{person}{Matthew Hall}, \bibinfo{person}{Federico Zarfati},
  \bibinfo{person}{Yonatan Zunger}, {and} \bibinfo{person}{Emre Kiciman}.}
  \bibinfo{year}{2024}\natexlab{}.
\newblock \showarticletitle{Defending Against Indirect Prompt Injection Attacks
  With Spotlighting}.
\newblock \bibinfo{journal}{\emph{arXiv}} (\bibinfo{year}{2024}).
\newblock


\bibitem[Hong et~al\mbox{.}(2024)]%
        {hong2024metagpt}
\bibfield{author}{\bibinfo{person}{Sirui Hong}, \bibinfo{person}{Mingchen
  Zhuge}, \bibinfo{person}{Jonathan Chen}, \bibinfo{person}{Xiawu Zheng},
  \bibinfo{person}{Yuheng Cheng}, \bibinfo{person}{Jinlin Wang},
  \bibinfo{person}{Ceyao Zhang}, \bibinfo{person}{Zili Wang},
  \bibinfo{person}{Steven Ka~Shing Yau}, \bibinfo{person}{Zijuan Lin},
  \bibinfo{person}{Liyang Zhou}, \bibinfo{person}{Chenyu Ran},
  \bibinfo{person}{Lingfeng Xiao}, \bibinfo{person}{Chenglin Wu}, {and}
  \bibinfo{person}{J{\"u}rgen Schmidhuber}.} \bibinfo{year}{2024}\natexlab{}.
\newblock \showarticletitle{Meta{GPT}: Meta Programming for A Multi-Agent
  Collaborative Framework}. In \bibinfo{booktitle}{\emph{ICLR}}.
\newblock


\bibitem[Hossain et~al\mbox{.}(2023)]%
        {hossain2023misgendered}
\bibfield{author}{\bibinfo{person}{Tamanna Hossain}, \bibinfo{person}{Sunipa
  Dev}, {and} \bibinfo{person}{Sameer Singh}.} \bibinfo{year}{2023}\natexlab{}.
\newblock \showarticletitle{{MISGENDERED}: Limits of Large Language Models in
  Understanding Pronouns}. In \bibinfo{booktitle}{\emph{Annual Meeting of the
  Association for Computational Linguistics}}.
\newblock


\bibitem[Hou et~al\mbox{.}(2024)]%
        {hou2024my}
\bibfield{author}{\bibinfo{person}{Yuki Hou}, \bibinfo{person}{Haruki Tamoto},
  {and} \bibinfo{person}{Homei Miyashita}.} \bibinfo{year}{2024}\natexlab{}.
\newblock \showarticletitle{" My agent understands me better": Integrating
  Dynamic Human-like Memory Recall and Consolidation in LLM-Based Agents}. In
  \bibinfo{booktitle}{\emph{Extended Abstracts of the CHI Conference on Human
  Factors in Computing Systems}}. \bibinfo{pages}{1--7}.
\newblock


\bibitem[https://www.theguardian.com/profile/hibaq farah(2024)]%
        {theguardianCybersecurityAgency}
\bibfield{author}{\bibinfo{person}{https://www.theguardian.com/profile/hibaq
  farah}.} \bibinfo{year}{2024}\natexlab{}.
\newblock \bibinfo{title}{{U}{K} cybersecurity agency warns of chatbot
  ‘prompt injection’ attacks --- theguardian.com}.
\newblock
  \bibinfo{howpublished}{\url{https://www.theguardian.com/technology/2023/aug/30/uk-cybersecurity-agency-warns-of-chatbot-prompt-injection-attacks}}.
\newblock


\bibitem[Hu et~al\mbox{.}(2023)]%
        {huenabling}
\bibfield{author}{\bibinfo{person}{Bin Hu}, \bibinfo{person}{Chenyang Zhao},
  \bibinfo{person}{Pu Zhang}, \bibinfo{person}{Zihao Zhou},
  \bibinfo{person}{Yuanhang Yang}, \bibinfo{person}{Zenglin Xu}, {and}
  \bibinfo{person}{Bin Liu}.} \bibinfo{year}{2023}\natexlab{}.
\newblock \showarticletitle{Enabling Intelligent Interactions between an Agent
  and an LLM: A Reinforcement Learning Approach}.
\newblock  (\bibinfo{year}{2023}).
\newblock


\bibitem[Hua et~al\mbox{.}(2024)]%
        {hua2024trustagent}
\bibfield{author}{\bibinfo{person}{Wenyue Hua}, \bibinfo{person}{Xianjun Yang},
  \bibinfo{person}{Zelong Li}, \bibinfo{person}{Cheng Wei}, {and}
  \bibinfo{person}{Yongfeng Zhang}.} \bibinfo{year}{2024}\natexlab{}.
\newblock \showarticletitle{TrustAgent: Towards Safe and Trustworthy LLM-based
  Agents through Agent Constitution}.
\newblock \bibinfo{journal}{\emph{arXiv}} (\bibinfo{year}{2024}).
\newblock


\bibitem[Huang et~al\mbox{.}(2023a)]%
        {huang2023survey}
\bibfield{author}{\bibinfo{person}{Xiaowei Huang}, \bibinfo{person}{Wenjie
  Ruan}, \bibinfo{person}{Wei Huang}, \bibinfo{person}{Gaojie Jin},
  \bibinfo{person}{Yi Dong}, \bibinfo{person}{Changshun Wu},
  \bibinfo{person}{Saddek Bensalem}, \bibinfo{person}{Ronghui Mu},
  \bibinfo{person}{Yi Qi}, \bibinfo{person}{Xingyu Zhao}, {et~al\mbox{.}}}
  \bibinfo{year}{2023}\natexlab{a}.
\newblock \showarticletitle{A survey of safety and trustworthiness of large
  language models through the lens of verification and validation}.
\newblock \bibinfo{journal}{\emph{arXiv}} (\bibinfo{year}{2023}).
\newblock


\bibitem[Huang et~al\mbox{.}(2023b)]%
        {huang2023boosting}
\bibfield{author}{\bibinfo{person}{Xijie Huang}, \bibinfo{person}{Li~Lyna
  Zhang}, \bibinfo{person}{Kwang-Ting Cheng}, {and} \bibinfo{person}{Mao
  Yang}.} \bibinfo{year}{2023}\natexlab{b}.
\newblock \showarticletitle{Boosting LLM Reasoning: Push the Limits of Few-shot
  Learning with Reinforced In-Context Pruning}.
\newblock \bibinfo{journal}{\emph{arXiv}} (\bibinfo{year}{2023}).
\newblock


\bibitem[Huang et~al\mbox{.}(2023c)]%
        {huang2023trustgpt}
\bibfield{author}{\bibinfo{person}{Yue Huang}, \bibinfo{person}{Qihui Zhang},
  \bibinfo{person}{Lichao Sun}, {et~al\mbox{.}}}
  \bibinfo{year}{2023}\natexlab{c}.
\newblock \showarticletitle{Trustgpt: A benchmark for trustworthy and
  responsible large language models}.
\newblock \bibinfo{journal}{\emph{arXiv}} (\bibinfo{year}{2023}).
\newblock


\bibitem[Humeau et~al\mbox{.}(2020)]%
        {humeau2020poly}
\bibfield{author}{\bibinfo{person}{S Humeau}, \bibinfo{person}{K Shuster},
  \bibinfo{person}{M Lachaux}, {and} \bibinfo{person}{J Weston}.}
  \bibinfo{year}{2020}\natexlab{}.
\newblock \showarticletitle{Poly-encoders: Architectures and pre-training
  strategies for fast and accurate multi-sentence scoring. arXiv}. In
  \bibinfo{booktitle}{\emph{ICLR}}.
\newblock


\bibitem[Ippolito et~al\mbox{.}(2023)]%
        {ippolito2023preventing}
\bibfield{author}{\bibinfo{person}{Daphne Ippolito}, \bibinfo{person}{Florian
  Tram{\`e}r}, \bibinfo{person}{Milad Nasr}, \bibinfo{person}{Chiyuan Zhang},
  \bibinfo{person}{Matthew Jagielski}, \bibinfo{person}{Katherine Lee},
  \bibinfo{person}{Christopher~A Choquette-Choo}, {and}
  \bibinfo{person}{Nicholas Carlini}.} \bibinfo{year}{2023}\natexlab{}.
\newblock \showarticletitle{Preventing generation of verbatim memorization in
  language models gives a false sense of privacy}. In
  \bibinfo{booktitle}{\emph{International Natural Language Generation
  Conference}}.
\newblock


\bibitem[Izacard and Grave(2021)]%
        {izacard-grave-2021-leveraging}
\bibfield{author}{\bibinfo{person}{Gautier Izacard} {and}
  \bibinfo{person}{Edouard Grave}.} \bibinfo{year}{2021}\natexlab{}.
\newblock \showarticletitle{Leveraging Passage Retrieval with Generative Models
  for Open Domain Question Answering}. In \bibinfo{booktitle}{\emph{Proceedings
  of the 16th Conference of the European Chapter of the Association for
  Computational Linguistics: Main Volume}}. \bibinfo{publisher}{Association for
  Computational Linguistics}.
\newblock


\bibitem[Ji et~al\mbox{.}(2023)]%
        {ji2023survey}
\bibfield{author}{\bibinfo{person}{Ziwei Ji}, \bibinfo{person}{Nayeon Lee},
  \bibinfo{person}{Rita Frieske}, \bibinfo{person}{Tiezheng Yu},
  \bibinfo{person}{Dan Su}, \bibinfo{person}{Yan Xu}, \bibinfo{person}{Etsuko
  Ishii}, \bibinfo{person}{Ye~Jin Bang}, \bibinfo{person}{Andrea Madotto},
  {and} \bibinfo{person}{Pascale Fung}.} \bibinfo{year}{2023}\natexlab{}.
\newblock \showarticletitle{Survey of hallucination in natural language
  generation}.
\newblock \bibinfo{journal}{\emph{Comput. Surveys}} (\bibinfo{year}{2023}).
\newblock


\bibitem[Ji et~al\mbox{.}(2024)]%
        {ji2024testing}
\bibfield{author}{\bibinfo{person}{Zhenlan Ji}, \bibinfo{person}{Daoyuan Wu},
  \bibinfo{person}{Pingchuan Ma}, \bibinfo{person}{Zongjie Li}, {and}
  \bibinfo{person}{Shuai Wang}.} \bibinfo{year}{2024}\natexlab{}.
\newblock \showarticletitle{Testing and Understanding Erroneous Planning in LLM
  Agents through Synthesized User Inputs}.
\newblock \bibinfo{journal}{\emph{arXiv}} (\bibinfo{year}{2024}).
\newblock


\bibitem[Jiang et~al\mbox{.}(2023)]%
        {jiang2023identifying}
\bibfield{author}{\bibinfo{person}{Fengqing Jiang}, \bibinfo{person}{Zhangchen
  Xu}, \bibinfo{person}{Luyao Niu}, \bibinfo{person}{Boxin Wang},
  \bibinfo{person}{Jinyuan Jia}, \bibinfo{person}{Bo Li}, {and}
  \bibinfo{person}{Radha Poovendran}.} \bibinfo{year}{2023}\natexlab{}.
\newblock \showarticletitle{Identifying and Mitigating Vulnerabilities in
  LLM-Integrated Applications}. In \bibinfo{booktitle}{\emph{ICLR}}.
\newblock


\bibitem[Kandpal et~al\mbox{.}(2023)]%
        {kandpal2023large}
\bibfield{author}{\bibinfo{person}{Nikhil Kandpal}, \bibinfo{person}{Haikang
  Deng}, \bibinfo{person}{Adam Roberts}, \bibinfo{person}{Eric Wallace}, {and}
  \bibinfo{person}{Colin Raffel}.} \bibinfo{year}{2023}\natexlab{}.
\newblock \showarticletitle{Large language models struggle to learn long-tail
  knowledge}. In \bibinfo{booktitle}{\emph{ICML}}.
\newblock


\bibitem[Kang et~al\mbox{.}(2024)]%
        {kang2024c}
\bibfield{author}{\bibinfo{person}{Mintong Kang}, \bibinfo{person}{Nezihe~Merve
  G{\"u}rel}, \bibinfo{person}{Ning Yu}, \bibinfo{person}{Dawn Song}, {and}
  \bibinfo{person}{Bo Li}.} \bibinfo{year}{2024}\natexlab{}.
\newblock \showarticletitle{C-RAG: Certified Generation Risks for
  Retrieval-Augmented Language Models}.
\newblock \bibinfo{journal}{\emph{arXiv}} (\bibinfo{year}{2024}).
\newblock


\bibitem[Kim et~al\mbox{.}(2024)]%
        {kim2024guide}
\bibfield{author}{\bibinfo{person}{Changyeon Kim}, \bibinfo{person}{Younggyo
  Seo}, \bibinfo{person}{Hao Liu}, \bibinfo{person}{Lisa Lee},
  \bibinfo{person}{Jinwoo Shin}, \bibinfo{person}{Honglak Lee}, {and}
  \bibinfo{person}{Kimin Lee}.} \bibinfo{year}{2024}\natexlab{}.
\newblock \showarticletitle{Guide Your Agent with Adaptive Multimodal Rewards}.
\newblock \bibinfo{journal}{\emph{NeurIPS}}  \bibinfo{volume}{36}.
\newblock


\bibitem[Kolouri et~al\mbox{.}(2020)]%
        {kolouri2020universal}
\bibfield{author}{\bibinfo{person}{Soheil Kolouri}, \bibinfo{person}{Aniruddha
  Saha}, \bibinfo{person}{Hamed Pirsiavash}, {and} \bibinfo{person}{Heiko
  Hoffmann}.} \bibinfo{year}{2020}\natexlab{}.
\newblock \showarticletitle{Universal litmus patterns: Revealing backdoor
  attacks in cnns}. In \bibinfo{booktitle}{\emph{CVPR}}.
  \bibinfo{pages}{301--310}.
\newblock


\bibitem[Kumar et~al\mbox{.}(2024)]%
        {kumar2024certifying}
\bibfield{author}{\bibinfo{person}{Aounon Kumar}, \bibinfo{person}{Chirag
  Agarwal}, \bibinfo{person}{Suraj Srinivas}, \bibinfo{person}{Soheil Feizi},
  {and} \bibinfo{person}{Hima Lakkaraju}.} \bibinfo{year}{2024}\natexlab{}.
\newblock \showarticletitle{Certifying llm safety against adversarial
  prompting}.
\newblock \bibinfo{journal}{\emph{arXiv}} (\bibinfo{year}{2024}).
\newblock


\bibitem[Kurita et~al\mbox{.}(2020)]%
        {kurita2020weight}
\bibfield{author}{\bibinfo{person}{Keita Kurita}, \bibinfo{person}{Paul
  Michel}, {and} \bibinfo{person}{Graham Neubig}.}
  \bibinfo{year}{2020}\natexlab{}.
\newblock \showarticletitle{Weight poisoning attacks on pre-trained models}.
\newblock \bibinfo{journal}{\emph{arXiv}} (\bibinfo{year}{2020}).
\newblock


\bibitem[Lee et~al\mbox{.}(2022)]%
        {lee2022factuality}
\bibfield{author}{\bibinfo{person}{Nayeon Lee}, \bibinfo{person}{Wei Ping},
  \bibinfo{person}{Peng Xu}, \bibinfo{person}{Mostofa Patwary},
  \bibinfo{person}{Pascale~N Fung}, \bibinfo{person}{Mohammad Shoeybi}, {and}
  \bibinfo{person}{Bryan Catanzaro}.} \bibinfo{year}{2022}\natexlab{}.
\newblock \showarticletitle{Factuality enhanced language models for open-ended
  text generation}.
\newblock \bibinfo{journal}{\emph{NeurIPS}}  \bibinfo{volume}{35}.
\newblock


\bibitem[Levi and Neumann(2024)]%
        {levi2024vocabulary}
\bibfield{author}{\bibinfo{person}{Patrick Levi} {and}
  \bibinfo{person}{Christoph~P Neumann}.} \bibinfo{year}{2024}\natexlab{}.
\newblock \showarticletitle{Vocabulary Attack to Hijack Large Language Model
  Applications}.
\newblock \bibinfo{journal}{\emph{arXiv}} (\bibinfo{year}{2024}).
\newblock


\bibitem[Lewis et~al\mbox{.}(2020)]%
        {10.5555/3495724.3496517}
\bibfield{author}{\bibinfo{person}{Patrick Lewis}, \bibinfo{person}{Ethan
  Perez}, \bibinfo{person}{Aleksandra Piktus}, \bibinfo{person}{Fabio Petroni},
  \bibinfo{person}{Vladimir Karpukhin}, \bibinfo{person}{Naman Goyal},
  \bibinfo{person}{Heinrich K\"{u}ttler}, \bibinfo{person}{Mike Lewis},
  \bibinfo{person}{Wen-tau Yih}, \bibinfo{person}{Tim Rockt\"{a}schel},
  \bibinfo{person}{Sebastian Riedel}, {and} \bibinfo{person}{Douwe Kiela}.}
  \bibinfo{year}{2020}\natexlab{}.
\newblock \showarticletitle{Retrieval-augmented generation for
  knowledge-intensive NLP tasks}. In \bibinfo{booktitle}{\emph{NeurIPS}}.
\newblock


\bibitem[Li et~al\mbox{.}(2023c)]%
        {li2023camel}
\bibfield{author}{\bibinfo{person}{Guohao Li}, \bibinfo{person}{Hasan Abed
  Al~Kader Hammoud}, \bibinfo{person}{Hani Itani}, \bibinfo{person}{Dmitrii
  Khizbullin}, {and} \bibinfo{person}{Bernard Ghanem}.}
  \bibinfo{year}{2023}\natexlab{c}.
\newblock \showarticletitle{{CAMEL}: Communicative Agents for ''Mind''
  Exploration of Large Language Model Society}. In
  \bibinfo{booktitle}{\emph{NeurIPS}}.
\newblock


\bibitem[Li et~al\mbox{.}(2023b)]%
        {li2023multi}
\bibfield{author}{\bibinfo{person}{Haoran Li}, \bibinfo{person}{Dadi Guo},
  \bibinfo{person}{Wei Fan}, \bibinfo{person}{Mingshi Xu}, \bibinfo{person}{Jie
  Huang}, \bibinfo{person}{Fanpu Meng}, {and} \bibinfo{person}{Yangqiu Song}.}
  \bibinfo{year}{2023}\natexlab{b}.
\newblock \showarticletitle{Multi-step Jailbreaking Privacy Attacks on
  {C}hat{GPT}}. In \bibinfo{booktitle}{\emph{Findings of EMNLP}}.
\newblock


\bibitem[Li et~al\mbox{.}(2023d)]%
        {li2023sentence}
\bibfield{author}{\bibinfo{person}{Haoran Li}, \bibinfo{person}{Mingshi Xu},
  {and} \bibinfo{person}{Yangqiu Song}.} \bibinfo{year}{2023}\natexlab{d}.
\newblock \showarticletitle{Sentence Embedding Leaks More Information than You
  Expect: Generative Embedding Inversion Attack to Recover the Whole Sentence}.
  In \bibinfo{booktitle}{\emph{Findings of ACL}}.
\newblock


\bibitem[Li et~al\mbox{.}(2023a)]%
        {li2023halueval}
\bibfield{author}{\bibinfo{person}{Junyi Li}, \bibinfo{person}{Xiaoxue Cheng},
  \bibinfo{person}{Wayne~Xin Zhao}, \bibinfo{person}{Jian-Yun Nie}, {and}
  \bibinfo{person}{Ji-Rong Wen}.} \bibinfo{year}{2023}\natexlab{a}.
\newblock \showarticletitle{Halueval: A large-scale hallucination evaluation
  benchmark for large language models}. In \bibinfo{booktitle}{\emph{EMNLP}}.
\newblock


\bibitem[Li et~al\mbox{.}(2020)]%
        {li2020textshield}
\bibfield{author}{\bibinfo{person}{Jinfeng Li}, \bibinfo{person}{Tianyu Du},
  \bibinfo{person}{Shouling Ji}, \bibinfo{person}{Rong Zhang},
  \bibinfo{person}{Quan Lu}, \bibinfo{person}{Min Yang}, {and}
  \bibinfo{person}{Ting Wang}.} \bibinfo{year}{2020}\natexlab{}.
\newblock \showarticletitle{$\{$TextShield$\}$: Robust text classification
  based on multimodal embedding and neural machine translation}. In
  \bibinfo{booktitle}{\emph{USENIX Security}}.
\newblock


\bibitem[Li et~al\mbox{.}(2024b)]%
        {li2024personal}
\bibfield{author}{\bibinfo{person}{Yuanchun Li}, \bibinfo{person}{Hao Wen},
  \bibinfo{person}{Weijun Wang}, \bibinfo{person}{Xiangyu Li},
  \bibinfo{person}{Yizhen Yuan}, \bibinfo{person}{Guohong Liu},
  \bibinfo{person}{Jiacheng Liu}, \bibinfo{person}{Wenxing Xu},
  \bibinfo{person}{Xiang Wang}, \bibinfo{person}{Yi Sun}, {et~al\mbox{.}}}
  \bibinfo{year}{2024}\natexlab{b}.
\newblock \showarticletitle{Personal llm agents: Insights and survey about the
  capability, efficiency and security}.
\newblock \bibinfo{journal}{\emph{arXiv}} (\bibinfo{year}{2024}).
\newblock


\bibitem[Li et~al\mbox{.}(2024a)]%
        {li2024formal}
\bibfield{author}{\bibinfo{person}{Zelong Li}, \bibinfo{person}{Wenyue Hua},
  \bibinfo{person}{Hao Wang}, \bibinfo{person}{He Zhu}, {and}
  \bibinfo{person}{Yongfeng Zhang}.} \bibinfo{year}{2024}\natexlab{a}.
\newblock \showarticletitle{Formal-LLM: Integrating Formal Language and Natural
  Language for Controllable LLM-based Agents}.
\newblock \bibinfo{journal}{\emph{arXiv}} (\bibinfo{year}{2024}).
\newblock


\bibitem[Liang et~al\mbox{.}(2023)]%
        {liang2023encouraging}
\bibfield{author}{\bibinfo{person}{Tian Liang}, \bibinfo{person}{Zhiwei He},
  \bibinfo{person}{Wenxiang Jiao}, \bibinfo{person}{Xing Wang},
  \bibinfo{person}{Yan Wang}, \bibinfo{person}{Rui Wang},
  \bibinfo{person}{Yujiu Yang}, \bibinfo{person}{Zhaopeng Tu}, {and}
  \bibinfo{person}{Shuming Shi}.} \bibinfo{year}{2023}\natexlab{}.
\newblock \showarticletitle{Encouraging Divergent Thinking in Large Language
  Models through Multi-Agent Debate}.
\newblock \bibinfo{journal}{\emph{arXiv}} (\bibinfo{year}{2023}).
\newblock


\bibitem[Light et~al\mbox{.}(2023)]%
        {light2023avalonbench}
\bibfield{author}{\bibinfo{person}{Jonathan Light}, \bibinfo{person}{Min Cai},
  \bibinfo{person}{Sheng Shen}, {and} \bibinfo{person}{Ziniu Hu}.}
  \bibinfo{year}{2023}\natexlab{}.
\newblock \showarticletitle{AvalonBench: Evaluating LLMs Playing the Game of
  Avalon}. In \bibinfo{booktitle}{\emph{NeurIPS Workshop}}.
\newblock


\bibitem[Lin et~al\mbox{.}(2023a)]%
        {lin2023healthy}
\bibfield{author}{\bibinfo{person}{Baihan Lin}, \bibinfo{person}{Djallel
  Bouneffouf}, \bibinfo{person}{Guillermo Cecchi}, {and}
  \bibinfo{person}{Kush~R Varshney}.} \bibinfo{year}{2023}\natexlab{a}.
\newblock \showarticletitle{Towards healthy AI: large language models need
  therapists too}.
\newblock \bibinfo{journal}{\emph{arXiv}} (\bibinfo{year}{2023}).
\newblock


\bibitem[Lin et~al\mbox{.}(2024)]%
        {lin2024swiftsage}
\bibfield{author}{\bibinfo{person}{Bill~Yuchen Lin}, \bibinfo{person}{Yicheng
  Fu}, \bibinfo{person}{Karina Yang}, \bibinfo{person}{Faeze Brahman},
  \bibinfo{person}{Shiyu Huang}, \bibinfo{person}{Chandra Bhagavatula},
  \bibinfo{person}{Prithviraj Ammanabrolu}, \bibinfo{person}{Yejin Choi}, {and}
  \bibinfo{person}{Xiang Ren}.} \bibinfo{year}{2024}\natexlab{}.
\newblock \showarticletitle{Swiftsage: A generative agent with fast and slow
  thinking for complex interactive tasks}.
\newblock \bibinfo{journal}{\emph{NeurIPS}}  \bibinfo{volume}{36}.
\newblock


\bibitem[Lin et~al\mbox{.}(2023b)]%
        {lin2023agentsims}
\bibfield{author}{\bibinfo{person}{Jiaju Lin}, \bibinfo{person}{Haoran Zhao},
  \bibinfo{person}{Aochi Zhang}, \bibinfo{person}{Yiting Wu},
  \bibinfo{person}{Huqiuyue Ping}, {and} \bibinfo{person}{Qin Chen}.}
  \bibinfo{year}{2023}\natexlab{b}.
\newblock \showarticletitle{Agentsims: An open-source sandbox for large
  language model evaluation}.
\newblock \bibinfo{journal}{\emph{arXiv}} (\bibinfo{year}{2023}).
\newblock


\bibitem[Liu et~al\mbox{.}(2020)]%
        {liu2020spatiotemporal}
\bibfield{author}{\bibinfo{person}{Aishan Liu}, \bibinfo{person}{Tairan Huang},
  \bibinfo{person}{Xianglong Liu}, \bibinfo{person}{Yitao Xu},
  \bibinfo{person}{Yuqing Ma}, \bibinfo{person}{Xinyun Chen},
  \bibinfo{person}{Stephen~J Maybank}, {and} \bibinfo{person}{Dacheng Tao}.}
  \bibinfo{year}{2020}\natexlab{}.
\newblock \showarticletitle{Spatiotemporal attacks for embodied agents}. In
  \bibinfo{booktitle}{\emph{ECCV}}.
\newblock


\bibitem[Liu et~al\mbox{.}(2023d)]%
        {liu2023context}
\bibfield{author}{\bibinfo{person}{Sheng Liu}, \bibinfo{person}{Lei Xing},
  {and} \bibinfo{person}{James Zou}.} \bibinfo{year}{2023}\natexlab{d}.
\newblock \showarticletitle{In-context vectors: Making in context learning more
  effective and controllable through latent space steering}.
\newblock \bibinfo{journal}{\emph{arXiv}} (\bibinfo{year}{2023}).
\newblock


\bibitem[Liu et~al\mbox{.}(2023b)]%
        {liu2023demystifying}
\bibfield{author}{\bibinfo{person}{Tong Liu}, \bibinfo{person}{Zizhuang Deng},
  \bibinfo{person}{Guozhu Meng}, \bibinfo{person}{Yuekang Li}, {and}
  \bibinfo{person}{Kai Chen}.} \bibinfo{year}{2023}\natexlab{b}.
\newblock \showarticletitle{Demystifying rce vulnerabilities in llm-integrated
  apps}.
\newblock \bibinfo{journal}{\emph{arXiv}} (\bibinfo{year}{2023}).
\newblock


\bibitem[Liu et~al\mbox{.}(2023a)]%
        {liu2023prompt}
\bibfield{author}{\bibinfo{person}{Yi Liu}, \bibinfo{person}{Gelei Deng},
  \bibinfo{person}{Yuekang Li}, \bibinfo{person}{Kailong Wang},
  \bibinfo{person}{Tianwei Zhang}, \bibinfo{person}{Yepang Liu},
  \bibinfo{person}{Haoyu Wang}, \bibinfo{person}{Yan Zheng}, {and}
  \bibinfo{person}{Yang Liu}.} \bibinfo{year}{2023}\natexlab{a}.
\newblock \showarticletitle{Prompt Injection attack against LLM-integrated
  Applications}.
\newblock \bibinfo{journal}{\emph{arXiv}} (\bibinfo{year}{2023}).
\newblock


\bibitem[Liu et~al\mbox{.}(2023c)]%
        {liu2023jailbreaking}
\bibfield{author}{\bibinfo{person}{Yi Liu}, \bibinfo{person}{Gelei Deng},
  \bibinfo{person}{Zhengzi Xu}, \bibinfo{person}{Yuekang Li},
  \bibinfo{person}{Yaowen Zheng}, \bibinfo{person}{Ying Zhang},
  \bibinfo{person}{Lida Zhao}, \bibinfo{person}{Tianwei Zhang}, {and}
  \bibinfo{person}{Yang Liu}.} \bibinfo{year}{2023}\natexlab{c}.
\newblock \showarticletitle{Jailbreaking chatgpt via prompt engineering: An
  empirical study}.
\newblock \bibinfo{journal}{\emph{arXiv}} (\bibinfo{year}{2023}).
\newblock


\bibitem[Liu et~al\mbox{.}(2023e)]%
        {liu2023trustworthy}
\bibfield{author}{\bibinfo{person}{Yang Liu}, \bibinfo{person}{Yuanshun Yao},
  \bibinfo{person}{Jean-Francois Ton}, \bibinfo{person}{Xiaoying Zhang},
  \bibinfo{person}{Ruocheng Guo~Hao Cheng}, \bibinfo{person}{Yegor Klochkov},
  \bibinfo{person}{Muhammad~Faaiz Taufiq}, {and} \bibinfo{person}{Hang Li}.}
  \bibinfo{year}{2023}\natexlab{e}.
\newblock \showarticletitle{Trustworthy LLMs: a Survey and Guideline for
  Evaluating Large Language Models' Alignment}.
\newblock \bibinfo{journal}{\emph{arXiv}} (\bibinfo{year}{2023}).
\newblock


\bibitem[Lu et~al\mbox{.}(2023)]%
        {lu2023building}
\bibfield{author}{\bibinfo{person}{Qinghua Lu}, \bibinfo{person}{Liming Zhu},
  \bibinfo{person}{Xiwei Xu}, \bibinfo{person}{Zhenchang Xing},
  \bibinfo{person}{Stefan Harrer}, {and} \bibinfo{person}{Jon Whittle}.}
  \bibinfo{year}{2023}\natexlab{}.
\newblock \showarticletitle{Building the Future of Responsible AI: A Reference
  Architecture for Designing Large Language Model based Agents}.
\newblock \bibinfo{journal}{\emph{arXiv}} (\bibinfo{year}{2023}).
\newblock


\bibitem[Lu et~al\mbox{.}(2022)]%
        {lu2022quark}
\bibfield{author}{\bibinfo{person}{Ximing Lu}, \bibinfo{person}{Sean Welleck},
  \bibinfo{person}{Jack Hessel}, \bibinfo{person}{Liwei Jiang},
  \bibinfo{person}{Lianhui Qin}, \bibinfo{person}{Peter West},
  \bibinfo{person}{Prithviraj Ammanabrolu}, {and} \bibinfo{person}{Yejin
  Choi}.} \bibinfo{year}{2022}\natexlab{}.
\newblock \showarticletitle{Quark: Controllable text generation with reinforced
  unlearning}.
\newblock \bibinfo{journal}{\emph{NeurIPS}}  \bibinfo{volume}{35},
  \bibinfo{pages}{27591--27609}.
\newblock


\bibitem[Lukas~Biewald(2017)]%
        {wandb}
\bibfield{author}{\bibinfo{person}{Chris Van~Pelt Lukas~Biewald}.}
  \bibinfo{year}{2017}\natexlab{}.
\newblock \showarticletitle{Weights and Biases}.
\newblock \bibinfo{howpublished}{\url{https://wandb.ai/}}.
\newblock  (\bibinfo{year}{2017}).
\newblock


\bibitem[Madaan et~al\mbox{.}(2024)]%
        {madaan2024self}
\bibfield{author}{\bibinfo{person}{Aman Madaan}, \bibinfo{person}{Niket
  Tandon}, \bibinfo{person}{Prakhar Gupta}, \bibinfo{person}{Skyler Hallinan},
  \bibinfo{person}{Luyu Gao}, \bibinfo{person}{Sarah Wiegreffe},
  \bibinfo{person}{Uri Alon}, \bibinfo{person}{Nouha Dziri},
  \bibinfo{person}{Shrimai Prabhumoye}, \bibinfo{person}{Yiming Yang},
  {et~al\mbox{.}}} \bibinfo{year}{2024}\natexlab{}.
\newblock \showarticletitle{Self-refine: Iterative refinement with
  self-feedback}.
\newblock \bibinfo{journal}{\emph{NeurIPS}}  \bibinfo{volume}{36}.
\newblock


\bibitem[Mandi et~al\mbox{.}(2023)]%
        {mandi2023roco}
\bibfield{author}{\bibinfo{person}{Zhao Mandi}, \bibinfo{person}{Shreeya Jain},
  {and} \bibinfo{person}{Shuran Song}.} \bibinfo{year}{2023}\natexlab{}.
\newblock \showarticletitle{Roco: Dialectic multi-robot collaboration with
  large language models}.
\newblock \bibinfo{journal}{\emph{arXiv}} (\bibinfo{year}{2023}).
\newblock


\bibitem[Masterman et~al\mbox{.}(2024)]%
        {masterman2024landscape}
\bibfield{author}{\bibinfo{person}{Tula Masterman}, \bibinfo{person}{Sandi
  Besen}, \bibinfo{person}{Mason Sawtell}, {and} \bibinfo{person}{Alex Chao}.}
  \bibinfo{year}{2024}\natexlab{}.
\newblock \showarticletitle{The Landscape of Emerging AI Agent Architectures
  for Reasoning, Planning, and Tool Calling: A Survey}.
\newblock \bibinfo{journal}{\emph{arXiv}} (\bibinfo{year}{2024}).
\newblock


\bibitem[Mehta et~al\mbox{.}(2024)]%
        {mehta2023improving}
\bibfield{author}{\bibinfo{person}{Nikhil Mehta}, \bibinfo{person}{Milagro
  Teruel}, \bibinfo{person}{Xin Deng}, \bibinfo{person}{Sergio Figueroa~Sanz},
  \bibinfo{person}{Ahmed Awadallah}, {and} \bibinfo{person}{Julia Kiseleva}.}
  \bibinfo{year}{2024}\natexlab{}.
\newblock \showarticletitle{Improving Grounded Language Understanding in a
  Collaborative Environment by Interacting with Agents Through Help Feedback}.
  In \bibinfo{booktitle}{\emph{Findings of EACL}}.
\newblock


\bibitem[Mei et~al\mbox{.}(2024)]%
        {mei2024aios}
\bibfield{author}{\bibinfo{person}{Kai Mei}, \bibinfo{person}{Zelong Li},
  \bibinfo{person}{Shuyuan Xu}, \bibinfo{person}{Ruosong Ye},
  \bibinfo{person}{Yingqiang Ge}, {and} \bibinfo{person}{Yongfeng Zhang}.}
  \bibinfo{year}{2024}\natexlab{}.
\newblock \showarticletitle{AIOS: LLM Agent Operating System}.
\newblock \bibinfo{journal}{\emph{arXiv}} (\bibinfo{year}{2024}).
\newblock


\bibitem[Micheli et~al\mbox{.}(2023)]%
        {micheli2022transformers}
\bibfield{author}{\bibinfo{person}{Vincent Micheli}, \bibinfo{person}{Eloi
  Alonso}, {and} \bibinfo{person}{Fran{\c{c}}ois Fleuret}.}
  \bibinfo{year}{2023}\natexlab{}.
\newblock \showarticletitle{Transformers are sample-efficient world models}. In
  \bibinfo{booktitle}{\emph{ICLR}}.
\newblock


\bibitem[Min et~al\mbox{.}(2023)]%
        {min2023factscore}
\bibfield{author}{\bibinfo{person}{Sewon Min}, \bibinfo{person}{Kalpesh
  Krishna}, \bibinfo{person}{Xinxi Lyu}, \bibinfo{person}{Mike Lewis},
  \bibinfo{person}{Wen-tau Yih}, \bibinfo{person}{Pang Koh},
  \bibinfo{person}{Mohit Iyyer}, \bibinfo{person}{Luke Zettlemoyer}, {and}
  \bibinfo{person}{Hannaneh Hajishirzi}.} \bibinfo{year}{2023}\natexlab{}.
\newblock \showarticletitle{{FA}ct{S}core: Fine-grained Atomic Evaluation of
  Factual Precision in Long Form Text Generation}. In
  \bibinfo{booktitle}{\emph{EMNLP}}.
\newblock


\bibitem[Mo et~al\mbox{.}(2024)]%
        {mo2024trembling}
\bibfield{author}{\bibinfo{person}{Lingbo Mo}, \bibinfo{person}{Zeyi Liao},
  \bibinfo{person}{Boyuan Zheng}, \bibinfo{person}{Yu Su},
  \bibinfo{person}{Chaowei Xiao}, {and} \bibinfo{person}{Huan Sun}.}
  \bibinfo{year}{2024}\natexlab{}.
\newblock \showarticletitle{A Trembling House of Cards? Mapping Adversarial
  Attacks against Language Agents}.
\newblock \bibinfo{journal}{\emph{arXiv}} (\bibinfo{year}{2024}).
\newblock


\bibitem[Morris et~al\mbox{.}(2023)]%
        {morris2023text}
\bibfield{author}{\bibinfo{person}{John Morris}, \bibinfo{person}{Volodymyr
  Kuleshov}, \bibinfo{person}{Vitaly Shmatikov}, {and}
  \bibinfo{person}{Alexander Rush}.} \bibinfo{year}{2023}\natexlab{}.
\newblock \showarticletitle{Text Embeddings Reveal (Almost) As Much As Text}.
  In \bibinfo{booktitle}{\emph{EMNLP}}.
\newblock


\bibitem[Moskal et~al\mbox{.}(2023)]%
        {moskal2023llms}
\bibfield{author}{\bibinfo{person}{Stephen Moskal}, \bibinfo{person}{Sam
  Laney}, \bibinfo{person}{Erik Hemberg}, {and} \bibinfo{person}{Una-May
  O'Reilly}.} \bibinfo{year}{2023}\natexlab{}.
\newblock \showarticletitle{LLMs Killed the Script Kiddie: How Agents Supported
  by Large Language Models Change the Landscape of Network Threat Testing}.
\newblock \bibinfo{journal}{\emph{arXiv}} (\bibinfo{year}{2023}).
\newblock


\bibitem[Motwani et~al\mbox{.}(2023)]%
        {motwani2023perfect}
\bibfield{author}{\bibinfo{person}{Sumeet~Ramesh Motwani},
  \bibinfo{person}{Mikhail Baranchuk}, \bibinfo{person}{Lewis Hammond}, {and}
  \bibinfo{person}{Christian~Schroeder de Witt}.}
  \bibinfo{year}{2023}\natexlab{}.
\newblock \showarticletitle{A Perfect Collusion Benchmark: How can AI agents be
  prevented from colluding with information-theoretic undetectability?}. In
  \bibinfo{booktitle}{\emph{NeurIPS workshop}}.
\newblock


\bibitem[Mrabet et~al\mbox{.}(2020)]%
        {mrabet2020survey}
\bibfield{author}{\bibinfo{person}{Hichem Mrabet}, \bibinfo{person}{Sana
  Belguith}, \bibinfo{person}{Adeeb Alhomoud}, {and}
  \bibinfo{person}{Abderrazak Jemai}.} \bibinfo{year}{2020}\natexlab{}.
\newblock \showarticletitle{A survey of IoT security based on a layered
  architecture of sensing and data analysis}.
\newblock \bibinfo{journal}{\emph{Sensors}} (\bibinfo{year}{2020}).
\newblock


\bibitem[Muhlgay et~al\mbox{.}(2024)]%
        {muhlgay2024generating}
\bibfield{author}{\bibinfo{person}{Dor Muhlgay}, \bibinfo{person}{Ori Ram},
  \bibinfo{person}{Inbal Magar}, \bibinfo{person}{Yoav Levine},
  \bibinfo{person}{Nir Ratner}, \bibinfo{person}{Yonatan Belinkov},
  \bibinfo{person}{Omri Abend}, \bibinfo{person}{Kevin Leyton-Brown},
  \bibinfo{person}{Amnon Shashua}, {and} \bibinfo{person}{Yoav Shoham}.}
  \bibinfo{year}{2024}\natexlab{}.
\newblock \showarticletitle{Generating Benchmarks for Factuality Evaluation of
  Language Models}. In \bibinfo{booktitle}{\emph{Conference of the European
  Chapter of the Association for Computational Linguistics}}.
\newblock


\bibitem[Mukhopadhyay et~al\mbox{.}(1986)]%
        {mukhopadhyay1986intelligent}
\bibfield{author}{\bibinfo{person}{Uttam Mukhopadhyay},
  \bibinfo{person}{Larry~M Stephens}, \bibinfo{person}{Michael~N Huhns}, {and}
  \bibinfo{person}{Ronald~D Bonnell}.} \bibinfo{year}{1986}\natexlab{}.
\newblock \showarticletitle{An intelligent system for document retrieval in
  distributed office environments}.
\newblock \bibinfo{journal}{\emph{Journal of the American Society for
  Information Science}} (\bibinfo{year}{1986}).
\newblock


\bibitem[Nair et~al\mbox{.}(2023)]%
        {nair2023dera}
\bibfield{author}{\bibinfo{person}{Varun Nair}, \bibinfo{person}{Elliot
  Schumacher}, \bibinfo{person}{Geoffrey Tso}, {and} \bibinfo{person}{Anitha
  Kannan}.} \bibinfo{year}{2023}\natexlab{}.
\newblock \showarticletitle{DERA: enhancing large language model completions
  with dialog-enabled resolving agents}.
\newblock \bibinfo{journal}{\emph{arXiv}} (\bibinfo{year}{2023}).
\newblock


\bibitem[Nguyen and Wong(2023)]%
        {nguyen2023context}
\bibfield{author}{\bibinfo{person}{Tai Nguyen} {and} \bibinfo{person}{Eric
  Wong}.} \bibinfo{year}{2023}\natexlab{}.
\newblock \showarticletitle{In-context example selection with influences}.
\newblock \bibinfo{journal}{\emph{arXiv}} (\bibinfo{year}{2023}).
\newblock


\bibitem[O'Gara(2023)]%
        {o2023hoodwinked}
\bibfield{author}{\bibinfo{person}{Aidan O'Gara}.}
  \bibinfo{year}{2023}\natexlab{}.
\newblock \showarticletitle{Hoodwinked: Deception and cooperation in a
  text-based game for language models}.
\newblock \bibinfo{journal}{\emph{arXiv}} (\bibinfo{year}{2023}).
\newblock


\bibitem[Ousidhoum et~al\mbox{.}(2021)]%
        {ousidhoum2021probing}
\bibfield{author}{\bibinfo{person}{Nedjma Ousidhoum}, \bibinfo{person}{Xinran
  Zhao}, \bibinfo{person}{Tianqing Fang}, \bibinfo{person}{Yangqiu Song}, {and}
  \bibinfo{person}{Dit-Yan Yeung}.} \bibinfo{year}{2021}\natexlab{}.
\newblock \showarticletitle{Probing toxic content in large pre-trained language
  models}. In \bibinfo{booktitle}{\emph{International Joint Conference on
  Natural Language Processing}}.
\newblock


\bibitem[Ouyang et~al\mbox{.}(2022)]%
        {ouyang2022training}
\bibfield{author}{\bibinfo{person}{Long Ouyang}, \bibinfo{person}{Jeffrey Wu},
  \bibinfo{person}{Xu Jiang}, \bibinfo{person}{Diogo Almeida},
  \bibinfo{person}{Carroll Wainwright}, \bibinfo{person}{Pamela Mishkin},
  \bibinfo{person}{Chong Zhang}, \bibinfo{person}{Sandhini Agarwal},
  \bibinfo{person}{Katarina Slama}, \bibinfo{person}{Alex Ray},
  {et~al\mbox{.}}} \bibinfo{year}{2022}\natexlab{}.
\newblock \showarticletitle{Training language models to follow instructions
  with human feedback}.
\newblock \bibinfo{journal}{\emph{NeurIPS}}.
\newblock


\bibitem[Pan et~al\mbox{.}(2023b)]%
        {pan2023survey}
\bibfield{author}{\bibinfo{person}{James~Jie Pan}, \bibinfo{person}{Jianguo
  Wang}, {and} \bibinfo{person}{Guoliang Li}.}
  \bibinfo{year}{2023}\natexlab{b}.
\newblock \showarticletitle{Survey of vector database management systems}.
\newblock \bibinfo{journal}{\emph{arXiv}} (\bibinfo{year}{2023}).
\newblock


\bibitem[Pan et~al\mbox{.}(2023a)]%
        {pan2023risk}
\bibfield{author}{\bibinfo{person}{Yikang Pan}, \bibinfo{person}{Liangming
  Pan}, \bibinfo{person}{Wenhu Chen}, \bibinfo{person}{Preslav Nakov},
  \bibinfo{person}{Min-Yen Kan}, {and} \bibinfo{person}{William~Yang Wang}.}
  \bibinfo{year}{2023}\natexlab{a}.
\newblock \showarticletitle{On the risk of misinformation pollution with large
  language models}.
\newblock \bibinfo{journal}{\emph{arXiv}} (\bibinfo{year}{2023}).
\newblock


\bibitem[Pang et~al\mbox{.}(2023)]%
        {pang2023natural}
\bibfield{author}{\bibinfo{person}{Jing-Cheng Pang}, \bibinfo{person}{Xin-Yu
  Yang}, \bibinfo{person}{Si-Hang Yang}, {and} \bibinfo{person}{Yang Yu}.}
  \bibinfo{year}{2023}\natexlab{}.
\newblock \showarticletitle{Natural Language-conditioned Reinforcement Learning
  with Inside-out Task Language Development and Translation}.
\newblock \bibinfo{journal}{\emph{NeurIPS}}.
\newblock


\bibitem[Park et~al\mbox{.}(2023b)]%
        {park2023generative}
\bibfield{author}{\bibinfo{person}{Joon~Sung Park}, \bibinfo{person}{Joseph
  O'Brien}, \bibinfo{person}{Carrie~Jun Cai}, \bibinfo{person}{Meredith~Ringel
  Morris}, \bibinfo{person}{Percy Liang}, {and} \bibinfo{person}{Michael~S
  Bernstein}.} \bibinfo{year}{2023}\natexlab{b}.
\newblock \showarticletitle{Generative agents: Interactive simulacra of human
  behavior}. In \bibinfo{booktitle}{\emph{Annual ACM Symposium on User
  Interface Software and Technology}}. \bibinfo{pages}{1--22}.
\newblock


\bibitem[Park et~al\mbox{.}(2023a)]%
        {park2023ai}
\bibfield{author}{\bibinfo{person}{Peter~S Park}, \bibinfo{person}{Simon
  Goldstein}, \bibinfo{person}{Aidan O'Gara}, \bibinfo{person}{Michael Chen},
  {and} \bibinfo{person}{Dan Hendrycks}.} \bibinfo{year}{2023}\natexlab{a}.
\newblock \showarticletitle{AI deception: A survey of examples, risks, and
  potential solutions}.
\newblock \bibinfo{journal}{\emph{arXiv}} (\bibinfo{year}{2023}).
\newblock


\bibitem[Pedro et~al\mbox{.}(2023)]%
        {pedro2308prompt}
\bibfield{author}{\bibinfo{person}{Rodrigo Pedro}, \bibinfo{person}{Daniel
  Castro}, \bibinfo{person}{Paulo Carreira}, {and} \bibinfo{person}{Nuno
  Santos}.} \bibinfo{year}{2023}\natexlab{}.
\newblock \showarticletitle{From Prompt Injections to SQL Injection Attacks:
  How Protected is Your LLM-Integrated Web Application?}
\newblock \bibinfo{journal}{\emph{arXiv}} (\bibinfo{year}{2023}).
\newblock


\bibitem[Peng et~al\mbox{.}(2023)]%
        {peng2023check}
\bibfield{author}{\bibinfo{person}{Baolin Peng}, \bibinfo{person}{Michel
  Galley}, \bibinfo{person}{Pengcheng He}, \bibinfo{person}{Hao Cheng},
  \bibinfo{person}{Yujia Xie}, \bibinfo{person}{Yu Hu},
  \bibinfo{person}{Qiuyuan Huang}, \bibinfo{person}{Lars Liden},
  \bibinfo{person}{Zhou Yu}, \bibinfo{person}{Weizhu Chen}, {et~al\mbox{.}}}
  \bibinfo{year}{2023}\natexlab{}.
\newblock \showarticletitle{Check your facts and try again: Improving large
  language models with external knowledge and automated feedback}.
\newblock \bibinfo{journal}{\emph{arXiv}} (\bibinfo{year}{2023}).
\newblock


\bibitem[Perez et~al\mbox{.}(2023)]%
        {perez-etal-2023-discovering}
\bibfield{author}{\bibinfo{person}{Ethan Perez}, \bibinfo{person}{Sam Ringer},
  \bibinfo{person}{Kamile Lukosiute}, \bibinfo{person}{Karina Nguyen},
  \bibinfo{person}{Edwin Chen}, \bibinfo{person}{Scott Heiner},
  \bibinfo{person}{Craig Pettit}, \bibinfo{person}{Catherine Olsson},
  \bibinfo{person}{Sandipan Kundu}, \bibinfo{person}{Saurav Kadavath},
  \bibinfo{person}{Andy Jones}, \bibinfo{person}{Anna Chen},
  \bibinfo{person}{Benjamin Mann}, \bibinfo{person}{Brian Israel},
  \bibinfo{person}{Bryan Seethor}, \bibinfo{person}{Cameron McKinnon},
  \bibinfo{person}{Christopher Olah}, \bibinfo{person}{Da Yan},
  \bibinfo{person}{Daniela Amodei}, \bibinfo{person}{Dario Amodei},
  \bibinfo{person}{Dawn Drain}, \bibinfo{person}{Dustin Li},
  \bibinfo{person}{Eli Tran-Johnson}, \bibinfo{person}{Guro Khundadze},
  \bibinfo{person}{Jackson Kernion}, \bibinfo{person}{James Landis},
  \bibinfo{person}{Jamie Kerr}, \bibinfo{person}{Jared Mueller},
  \bibinfo{person}{Jeeyoon Hyun}, \bibinfo{person}{Joshua Landau},
  \bibinfo{person}{Kamal Ndousse}, \bibinfo{person}{Landon Goldberg},
  \bibinfo{person}{Liane Lovitt}, \bibinfo{person}{Martin Lucas},
  \bibinfo{person}{Michael Sellitto}, \bibinfo{person}{Miranda Zhang},
  \bibinfo{person}{Neerav Kingsland}, \bibinfo{person}{Nelson Elhage},
  \bibinfo{person}{Nicholas Joseph}, \bibinfo{person}{Noemi Mercado},
  \bibinfo{person}{Nova DasSarma}, \bibinfo{person}{Oliver Rausch},
  \bibinfo{person}{Robin Larson}, \bibinfo{person}{Sam McCandlish},
  \bibinfo{person}{Scott Johnston}, \bibinfo{person}{Shauna Kravec},
  \bibinfo{person}{Sheer El~Showk}, \bibinfo{person}{Tamera Lanham},
  \bibinfo{person}{Timothy Telleen-Lawton}, \bibinfo{person}{Tom Brown},
  \bibinfo{person}{Tom Henighan}, \bibinfo{person}{Tristan Hume},
  \bibinfo{person}{Yuntao Bai}, \bibinfo{person}{Zac Hatfield-Dodds},
  \bibinfo{person}{Jack Clark}, \bibinfo{person}{Samuel~R. Bowman},
  \bibinfo{person}{Amanda Askell}, \bibinfo{person}{Roger Grosse},
  \bibinfo{person}{Danny Hernandez}, \bibinfo{person}{Deep Ganguli},
  \bibinfo{person}{Evan Hubinger}, \bibinfo{person}{Nicholas Schiefer}, {and}
  \bibinfo{person}{Jared Kaplan}.} \bibinfo{year}{2023}\natexlab{}.
\newblock \showarticletitle{Discovering Language Model Behaviors with
  Model-Written Evaluations}. In \bibinfo{booktitle}{\emph{Findings of ACL}}.
\newblock


\bibitem[Perez and Ribeiro(2022)]%
        {perez2211ignore}
\bibfield{author}{\bibinfo{person}{F{\'a}bio Perez} {and} \bibinfo{person}{Ian
  Ribeiro}.} \bibinfo{year}{2022}\natexlab{}.
\newblock \showarticletitle{Ignore previous prompt: Attack techniques for
  language models}.
\newblock \bibinfo{journal}{\emph{NeurIPS 2022}}.
\newblock


\bibitem[Phelps and Ranson(2023)]%
        {phelps2023models}
\bibfield{author}{\bibinfo{person}{Steve Phelps} {and} \bibinfo{person}{Rebecca
  Ranson}.} \bibinfo{year}{2023}\natexlab{}.
\newblock \showarticletitle{Of Models and Tin Men--a behavioural economics
  study of principal-agent problems in AI alignment using large-language
  models}.
\newblock \bibinfo{journal}{\emph{arXiv}} (\bibinfo{year}{2023}).
\newblock


\bibitem[P{\"o}hler et~al\mbox{.}(2024)]%
        {pohler2024technological}
\bibfield{author}{\bibinfo{person}{Lukas P{\"o}hler}, \bibinfo{person}{Valentin
  Schrader}, \bibinfo{person}{Alexander Ladwein}, {and}
  \bibinfo{person}{Florian von Keller}.} \bibinfo{year}{2024}\natexlab{}.
\newblock \showarticletitle{A Technological Perspective on Misuse of Available
  AI}.
\newblock \bibinfo{journal}{\emph{arXiv}} (\bibinfo{year}{2024}).
\newblock


\bibitem[Qian et~al\mbox{.}(2023)]%
        {qian2023communicative}
\bibfield{author}{\bibinfo{person}{Chen Qian}, \bibinfo{person}{Xin Cong},
  \bibinfo{person}{Cheng Yang}, \bibinfo{person}{Weize Chen},
  \bibinfo{person}{Yusheng Su}, \bibinfo{person}{Juyuan Xu},
  \bibinfo{person}{Zhiyuan Liu}, {and} \bibinfo{person}{Maosong Sun}.}
  \bibinfo{year}{2023}\natexlab{}.
\newblock \showarticletitle{Communicative agents for software development}.
\newblock \bibinfo{journal}{\emph{arXiv}} (\bibinfo{year}{2023}).
\newblock


\bibitem[Rae et~al\mbox{.}(2021)]%
        {rae2021scaling}
\bibfield{author}{\bibinfo{person}{Jack~W Rae}, \bibinfo{person}{Sebastian
  Borgeaud}, \bibinfo{person}{Trevor Cai}, \bibinfo{person}{Katie Millican},
  \bibinfo{person}{Jordan Hoffmann}, \bibinfo{person}{Francis Song},
  \bibinfo{person}{John Aslanides}, \bibinfo{person}{Sarah Henderson},
  \bibinfo{person}{Roman Ring}, \bibinfo{person}{Susannah Young},
  {et~al\mbox{.}}} \bibinfo{year}{2021}\natexlab{}.
\newblock \showarticletitle{Scaling language models: Methods, analysis \&
  insights from training Gopher}.
\newblock \bibinfo{journal}{\emph{arXiv}} (\bibinfo{year}{2021}).
\newblock


\bibitem[Rahman and Lu(2023)]%
        {rahman2023contextualized}
\bibfield{author}{\bibinfo{person}{Fathima~Abdul Rahman} {and}
  \bibinfo{person}{Guang Lu}.} \bibinfo{year}{2023}\natexlab{}.
\newblock \showarticletitle{A Contextualized Real-Time Multimodal Emotion
  Recognition for Conversational Agents using Graph Convolutional Networks in
  Reinforcement Learning}.
\newblock \bibinfo{journal}{\emph{arXiv}} (\bibinfo{year}{2023}).
\newblock


\bibitem[Ranaldi and Pucci(2023)]%
        {ranaldi2023large}
\bibfield{author}{\bibinfo{person}{Leonardo Ranaldi} {and}
  \bibinfo{person}{Giulia Pucci}.} \bibinfo{year}{2023}\natexlab{}.
\newblock \showarticletitle{When Large Language Models contradict humans? Large
  Language Models' Sycophantic Behaviour}.
\newblock \bibinfo{journal}{\emph{arXiv}} (\bibinfo{year}{2023}).
\newblock


\bibitem[Rasheed et~al\mbox{.}(2024)]%
        {rasheed2024large}
\bibfield{author}{\bibinfo{person}{Zeeshan Rasheed}, \bibinfo{person}{Muhammad
  Waseem}, \bibinfo{person}{Kari Syst{\"a}}, {and} \bibinfo{person}{Pekka
  Abrahamsson}.} \bibinfo{year}{2024}\natexlab{}.
\newblock \showarticletitle{Large language model evaluation via multi ai
  agents: Preliminary results}.
\newblock \bibinfo{journal}{\emph{arXiv}} (\bibinfo{year}{2024}).
\newblock


\bibitem[Rebedea et~al\mbox{.}(2023)]%
        {rebedea2023nemo}
\bibfield{author}{\bibinfo{person}{Traian Rebedea}, \bibinfo{person}{Razvan
  Dinu}, \bibinfo{person}{Makesh~Narsimhan Sreedhar},
  \bibinfo{person}{Christopher Parisien}, {and} \bibinfo{person}{Jonathan
  Cohen}.} \bibinfo{year}{2023}\natexlab{}.
\newblock \showarticletitle{{N}e{M}o Guardrails: A Toolkit for Controllable and
  Safe {LLM} Applications with Programmable Rails}. In
  \bibinfo{booktitle}{\emph{EMNLP}}.
\newblock


\bibitem[Red(2023)]%
        {embracetheredIndirectPrompt}
\bibfield{author}{\bibinfo{person}{Embrace~The Red}.}
  \bibinfo{year}{2023}\natexlab{}.
\newblock \bibinfo{title}{{I}ndirect {P}rompt {I}njection via {Y}ou{T}ube
  {T}ranscripts · {E}mbrace {T}he {R}ed --- embracethered.com}.
\newblock
  \bibinfo{howpublished}{\url{https://embracethered.com/blog/posts/2023/chatgpt-plugin-youtube-indirect-prompt-injection/}}.
\newblock


\bibitem[Rimon et~al\mbox{.}(2024)]%
        {rimon2024mamba}
\bibfield{author}{\bibinfo{person}{Zohar Rimon}, \bibinfo{person}{Tom
  Jurgenson}, \bibinfo{person}{Orr Krupnik}, \bibinfo{person}{Gilad Adler},
  {and} \bibinfo{person}{Aviv Tamar}.} \bibinfo{year}{2024}\natexlab{}.
\newblock \showarticletitle{{MAMBA}: an Effective World Model Approach for
  Meta-Reinforcement Learning}. In \bibinfo{booktitle}{\emph{ICLR}}.
\newblock


\bibitem[Ruan et~al\mbox{.}(2024)]%
        {ruan2023identifying}
\bibfield{author}{\bibinfo{person}{Yangjun Ruan}, \bibinfo{person}{Honghua
  Dong}, \bibinfo{person}{Andrew Wang}, \bibinfo{person}{Silviu Pitis},
  \bibinfo{person}{Yongchao Zhou}, \bibinfo{person}{Jimmy Ba},
  \bibinfo{person}{Yann Dubois}, \bibinfo{person}{Chris~J. Maddison}, {and}
  \bibinfo{person}{Tatsunori Hashimoto}.} \bibinfo{year}{2024}\natexlab{}.
\newblock \showarticletitle{Identifying the Risks of {LM} Agents with an
  {LM}-Emulated Sandbox}. In \bibinfo{booktitle}{\emph{ICLR}}.
\newblock


\bibitem[Salem et~al\mbox{.}(2023)]%
        {salem2023maatphor}
\bibfield{author}{\bibinfo{person}{Ahmed Salem}, \bibinfo{person}{Andrew
  Paverd}, {and} \bibinfo{person}{Boris K{\"o}pf}.}
  \bibinfo{year}{2023}\natexlab{}.
\newblock \showarticletitle{Maatphor: Automated variant analysis for prompt
  injection attacks}.
\newblock \bibinfo{journal}{\emph{arXiv}} (\bibinfo{year}{2023}).
\newblock


\bibitem[Savvov(2023)]%
        {fixing}
\bibfield{author}{\bibinfo{person}{Sergei Savvov}.}
  \bibinfo{year}{2023}\natexlab{}.
\newblock \showarticletitle{Fixing Hallucinations in LLMs
  https://betterprogramming.pub/fixing-hallucinations-in-llms-9ff0fd438e33}.
\newblock  (\bibinfo{year}{2023}).
\newblock


\bibitem[Schick et~al\mbox{.}(2024)]%
        {schick2024toolformer}
\bibfield{author}{\bibinfo{person}{Timo Schick}, \bibinfo{person}{Jane
  Dwivedi-Yu}, \bibinfo{person}{Roberto Dess{\`\i}}, \bibinfo{person}{Roberta
  Raileanu}, \bibinfo{person}{Maria Lomeli}, \bibinfo{person}{Eric Hambro},
  \bibinfo{person}{Luke Zettlemoyer}, \bibinfo{person}{Nicola Cancedda}, {and}
  \bibinfo{person}{Thomas Scialom}.} \bibinfo{year}{2024}\natexlab{}.
\newblock \showarticletitle{Toolformer: Language models can teach themselves to
  use tools}.
\newblock \bibinfo{journal}{\emph{NeurIPS}}.
\newblock


\bibitem[Schwinn et~al\mbox{.}(2023)]%
        {schwinn2023adversarial}
\bibfield{author}{\bibinfo{person}{Leo Schwinn}, \bibinfo{person}{David Dobre},
  \bibinfo{person}{Stephan G{\"u}nnemann}, {and} \bibinfo{person}{Gauthier
  Gidel}.} \bibinfo{year}{2023}\natexlab{}.
\newblock \showarticletitle{Adversarial attacks and defenses in large language
  models: Old and new threats}.
\newblock \bibinfo{journal}{\emph{arXiv}} (\bibinfo{year}{2023}).
\newblock


\bibitem[Shaikh et~al\mbox{.}(2023)]%
        {shaikh2022second}
\bibfield{author}{\bibinfo{person}{Omar Shaikh}, \bibinfo{person}{Hongxin
  Zhang}, \bibinfo{person}{William Held}, \bibinfo{person}{Michael Bernstein},
  {and} \bibinfo{person}{Diyi Yang}.} \bibinfo{year}{2023}\natexlab{}.
\newblock \showarticletitle{On Second Thought, Let{'}s Not Think Step by Step!
  Bias and Toxicity in Zero-Shot Reasoning}. In
  \bibinfo{booktitle}{\emph{Annual Meeting of the Association for Computational
  Linguistics}}.
\newblock


\bibitem[Sharma et~al\mbox{.}(2023)]%
        {sharma2023towards}
\bibfield{author}{\bibinfo{person}{Mrinank Sharma}, \bibinfo{person}{Meg Tong},
  \bibinfo{person}{Tomasz Korbak}, \bibinfo{person}{David Duvenaud},
  \bibinfo{person}{Amanda Askell}, \bibinfo{person}{Samuel~R Bowman},
  \bibinfo{person}{Newton Cheng}, \bibinfo{person}{Esin Durmus},
  \bibinfo{person}{Zac Hatfield-Dodds}, \bibinfo{person}{Scott~R Johnston},
  {et~al\mbox{.}}} \bibinfo{year}{2023}\natexlab{}.
\newblock \showarticletitle{Towards understanding sycophancy in language
  models}.
\newblock \bibinfo{journal}{\emph{arXiv}} (\bibinfo{year}{2023}).
\newblock


\bibitem[Shayegani et~al\mbox{.}(2023)]%
        {shayegani2023survey}
\bibfield{author}{\bibinfo{person}{Erfan Shayegani},
  \bibinfo{person}{Md~Abdullah~Al Mamun}, \bibinfo{person}{Yu Fu},
  \bibinfo{person}{Pedram Zaree}, \bibinfo{person}{Yue Dong}, {and}
  \bibinfo{person}{Nael Abu-Ghazaleh}.} \bibinfo{year}{2023}\natexlab{}.
\newblock \showarticletitle{Survey of vulnerabilities in large language models
  revealed by adversarial attacks}.
\newblock \bibinfo{journal}{\emph{arXiv}} (\bibinfo{year}{2023}).
\newblock


\bibitem[Shen et~al\mbox{.}(2024)]%
        {shen2024hugginggpt}
\bibfield{author}{\bibinfo{person}{Yongliang Shen}, \bibinfo{person}{Kaitao
  Song}, \bibinfo{person}{Xu Tan}, \bibinfo{person}{Dongsheng Li},
  \bibinfo{person}{Weiming Lu}, {and} \bibinfo{person}{Yueting Zhuang}.}
  \bibinfo{year}{2024}\natexlab{}.
\newblock \showarticletitle{Hugginggpt: Solving ai tasks with chatgpt and its
  friends in hugging face}.
\newblock \bibinfo{journal}{\emph{NeurIPS}}.
\newblock


\bibitem[Shuster et~al\mbox{.}(2021)]%
        {shuster2021retrieval}
\bibfield{author}{\bibinfo{person}{Kurt Shuster}, \bibinfo{person}{Spencer
  Poff}, \bibinfo{person}{Moya Chen}, \bibinfo{person}{Douwe Kiela}, {and}
  \bibinfo{person}{Jason Weston}.} \bibinfo{year}{2021}\natexlab{}.
\newblock \showarticletitle{Retrieval Augmentation Reduces Hallucination in
  Conversation}. In \bibinfo{booktitle}{\emph{Findings of EMNLP}}.
\newblock


\bibitem[Soice et~al\mbox{.}(2023)]%
        {soice2023can}
\bibfield{author}{\bibinfo{person}{Emily~H Soice}, \bibinfo{person}{Rafael
  Rocha}, \bibinfo{person}{Kimberlee Cordova}, \bibinfo{person}{Michael
  Specter}, {and} \bibinfo{person}{Kevin~M Esvelt}.}
  \bibinfo{year}{2023}\natexlab{}.
\newblock \showarticletitle{Can large language models democratize access to
  dual-use biotechnology?}
\newblock \bibinfo{journal}{\emph{arXiv}} (\bibinfo{year}{2023}).
\newblock


\bibitem[Song and Raghunathan(2020)]%
        {song2020information}
\bibfield{author}{\bibinfo{person}{Congzheng Song} {and}
  \bibinfo{person}{Ananth Raghunathan}.} \bibinfo{year}{2020}\natexlab{}.
\newblock \showarticletitle{Information leakage in embedding models}. In
  \bibinfo{booktitle}{\emph{ACM SIGSAC conference on computer and
  communications security}}. \bibinfo{pages}{377--390}.
\newblock


\bibitem[Song et~al\mbox{.}(2023)]%
        {song2023llm}
\bibfield{author}{\bibinfo{person}{Chan~Hee Song}, \bibinfo{person}{Jiaman Wu},
  \bibinfo{person}{Clayton Washington}, \bibinfo{person}{Brian~M Sadler},
  \bibinfo{person}{Wei-Lun Chao}, {and} \bibinfo{person}{Yu Su}.}
  \bibinfo{year}{2023}\natexlab{}.
\newblock \showarticletitle{Llm-planner: Few-shot grounded planning for
  embodied agents with large language models}. In
  \bibinfo{booktitle}{\emph{CVPR}}. \bibinfo{pages}{2998--3009}.
\newblock


\bibitem[Sudhakaran et~al\mbox{.}(2024)]%
        {sudhakaran2024mariogpt}
\bibfield{author}{\bibinfo{person}{Shyam Sudhakaran}, \bibinfo{person}{Miguel
  Gonz{\'a}lez-Duque}, \bibinfo{person}{Matthias Freiberger},
  \bibinfo{person}{Claire Glanois}, \bibinfo{person}{Elias Najarro}, {and}
  \bibinfo{person}{Sebastian Risi}.} \bibinfo{year}{2024}\natexlab{}.
\newblock \showarticletitle{Mariogpt: Open-ended text2level generation through
  large language models}.
\newblock \bibinfo{journal}{\emph{NeurIPS}}.
\newblock


\bibitem[Sun et~al\mbox{.}(2023a)]%
        {sun2023contrastive}
\bibfield{author}{\bibinfo{person}{Weiwei Sun}, \bibinfo{person}{Zhengliang
  Shi}, \bibinfo{person}{Shen Gao}, \bibinfo{person}{Pengjie Ren},
  \bibinfo{person}{Maarten de Rijke}, {and} \bibinfo{person}{Zhaochun Ren}.}
  \bibinfo{year}{2023}\natexlab{a}.
\newblock \showarticletitle{Contrastive learning reduces hallucination in
  conversations}. In \bibinfo{booktitle}{\emph{AAAI}}.
\newblock


\bibitem[Sun et~al\mbox{.}(2023b)]%
        {sun2023self}
\bibfield{author}{\bibinfo{person}{Yuxiang Sun}, \bibinfo{person}{Checheng Yu},
  \bibinfo{person}{Junjie Zhao}, \bibinfo{person}{Wei Wang}, {and}
  \bibinfo{person}{Xianzhong Zhou}.} \bibinfo{year}{2023}\natexlab{b}.
\newblock \showarticletitle{Self Generated Wargame AI: Double Layer Agent Task
  Planning Based on Large Language Model}.
\newblock \bibinfo{journal}{\emph{arXiv}} (\bibinfo{year}{2023}).
\newblock


\bibitem[Suresh and Guttag(2019)]%
        {suresh2019framework}
\bibfield{author}{\bibinfo{person}{Harini Suresh} {and} \bibinfo{person}{John~V
  Guttag}.} \bibinfo{year}{2019}\natexlab{}.
\newblock \showarticletitle{A framework for understanding unintended
  consequences of machine learning}.
\newblock \bibinfo{journal}{\emph{arXiv}} (\bibinfo{year}{2019}).
\newblock


\bibitem[Suri et~al\mbox{.}(2024)]%
        {suri2024large}
\bibfield{author}{\bibinfo{person}{Gaurav Suri}, \bibinfo{person}{Lily~R
  Slater}, \bibinfo{person}{Ali Ziaee}, {and} \bibinfo{person}{Morgan Nguyen}.}
  \bibinfo{year}{2024}\natexlab{}.
\newblock \showarticletitle{Do large language models show decision heuristics
  similar to humans? A case study using GPT-3.5.}
\newblock \bibinfo{journal}{\emph{Journal of Experimental Psychology: General}}
  (\bibinfo{year}{2024}).
\newblock


\bibitem[Tan et~al\mbox{.}(2024)]%
        {tan2024true}
\bibfield{author}{\bibinfo{person}{Weihao Tan}, \bibinfo{person}{Wentao Zhang},
  \bibinfo{person}{Shanqi Liu}, \bibinfo{person}{Longtao Zheng},
  \bibinfo{person}{Xinrun Wang}, {and} \bibinfo{person}{Bo An}.}
  \bibinfo{year}{2024}\natexlab{}.
\newblock \showarticletitle{True Knowledge Comes from Practice: Aligning Large
  Language Models with Embodied Environments via Reinforcement Learning}. In
  \bibinfo{booktitle}{\emph{ICLR}}.
\newblock


\bibitem[Tang et~al\mbox{.}(2024)]%
        {tang2024prioritizing}
\bibfield{author}{\bibinfo{person}{Xiangru Tang}, \bibinfo{person}{Qiao Jin},
  \bibinfo{person}{Kunlun Zhu}, \bibinfo{person}{Tongxin Yuan},
  \bibinfo{person}{Yichi Zhang}, \bibinfo{person}{Wangchunshu Zhou},
  \bibinfo{person}{Meng Qu}, \bibinfo{person}{Yilun Zhao},
  \bibinfo{person}{Jian Tang}, \bibinfo{person}{Zhuosheng Zhang},
  {et~al\mbox{.}}} \bibinfo{year}{2024}\natexlab{}.
\newblock \showarticletitle{Prioritizing Safeguarding Over Autonomy: Risks of
  LLM Agents for Science}.
\newblock \bibinfo{journal}{\emph{arXiv}} (\bibinfo{year}{2024}).
\newblock


\bibitem[Tian et~al\mbox{.}(2023)]%
        {tian2023evil}
\bibfield{author}{\bibinfo{person}{Yu Tian}, \bibinfo{person}{Xiao Yang},
  \bibinfo{person}{Jingyuan Zhang}, \bibinfo{person}{Yinpeng Dong}, {and}
  \bibinfo{person}{Hang Su}.} \bibinfo{year}{2023}\natexlab{}.
\newblock \showarticletitle{Evil geniuses: Delving into the safety of llm-based
  agents}.
\newblock \bibinfo{journal}{\emph{arXiv}} (\bibinfo{year}{2023}).
\newblock


\bibitem[Touvron et~al\mbox{.}(2023)]%
        {touvron2023llama}
\bibfield{author}{\bibinfo{person}{Hugo Touvron}, \bibinfo{person}{Thibaut
  Lavril}, \bibinfo{person}{Gautier Izacard}, \bibinfo{person}{Xavier
  Martinet}, \bibinfo{person}{Marie-Anne Lachaux},
  \bibinfo{person}{Timoth{\'e}e Lacroix}, \bibinfo{person}{Baptiste
  Rozi{\`e}re}, \bibinfo{person}{Naman Goyal}, \bibinfo{person}{Eric Hambro},
  \bibinfo{person}{Faisal Azhar}, {et~al\mbox{.}}}
  \bibinfo{year}{2023}\natexlab{}.
\newblock \showarticletitle{Llama: Open and efficient foundation language
  models}.
\newblock \bibinfo{journal}{\emph{arXiv}} (\bibinfo{year}{2023}).
\newblock


\bibitem[Toyer et~al\mbox{.}(2024)]%
        {toyer2023tensor}
\bibfield{author}{\bibinfo{person}{Sam Toyer}, \bibinfo{person}{Olivia
  Watkins}, \bibinfo{person}{Ethan~Adrian Mendes}, \bibinfo{person}{Justin
  Svegliato}, \bibinfo{person}{Luke Bailey}, \bibinfo{person}{Tiffany Wang},
  \bibinfo{person}{Isaac Ong}, \bibinfo{person}{Karim Elmaaroufi},
  \bibinfo{person}{Pieter Abbeel}, \bibinfo{person}{Trevor Darrell},
  \bibinfo{person}{Alan Ritter}, {and} \bibinfo{person}{Stuart Russell}.}
  \bibinfo{year}{2024}\natexlab{}.
\newblock \showarticletitle{Tensor Trust: Interpretable Prompt Injection
  Attacks from an Online Game}. In \bibinfo{booktitle}{\emph{ICLR}}.
\newblock


\bibitem[Ulmer et~al\mbox{.}(2024)]%
        {ulmer2024bootstrapping}
\bibfield{author}{\bibinfo{person}{Dennis Ulmer}, \bibinfo{person}{Elman
  Mansimov}, \bibinfo{person}{Kaixiang Lin}, \bibinfo{person}{Justin Sun},
  \bibinfo{person}{Xibin Gao}, {and} \bibinfo{person}{Yi Zhang}.}
  \bibinfo{year}{2024}\natexlab{}.
\newblock \showarticletitle{Bootstrapping llm-based task-oriented dialogue
  agents via self-talk}.
\newblock \bibinfo{journal}{\emph{arXiv}} (\bibinfo{year}{2024}).
\newblock


\bibitem[Vaswani et~al\mbox{.}(2017)]%
        {vaswani2017attention}
\bibfield{author}{\bibinfo{person}{Ashish Vaswani}, \bibinfo{person}{Noam
  Shazeer}, \bibinfo{person}{Niki Parmar}, \bibinfo{person}{Jakob Uszkoreit},
  \bibinfo{person}{Llion Jones}, \bibinfo{person}{Aidan~N Gomez},
  \bibinfo{person}{{\L}ukasz Kaiser}, {and} \bibinfo{person}{Illia
  Polosukhin}.} \bibinfo{year}{2017}\natexlab{}.
\newblock \showarticletitle{Attention is all you need}.
\newblock \bibinfo{journal}{\emph{NeurIPS}}.
\newblock


\bibitem[Vidgen et~al\mbox{.}(2024)]%
        {vidgen2024introducing}
\bibfield{author}{\bibinfo{person}{Bertie Vidgen}, \bibinfo{person}{Adarsh
  Agrawal}, \bibinfo{person}{Ahmed~M Ahmed}, \bibinfo{person}{Victor
  Akinwande}, \bibinfo{person}{Namir Al-Nuaimi}, \bibinfo{person}{Najla
  Alfaraj}, \bibinfo{person}{Elie Alhajjar}, \bibinfo{person}{Lora Aroyo},
  \bibinfo{person}{Trupti Bavalatti}, \bibinfo{person}{Borhane Blili-Hamelin},
  {et~al\mbox{.}}} \bibinfo{year}{2024}\natexlab{}.
\newblock \showarticletitle{Introducing v0. 5 of the AI Safety Benchmark from
  MLCommons}.
\newblock \bibinfo{journal}{\emph{arXiv}} (\bibinfo{year}{2024}).
\newblock


\bibitem[Wald and Pfahler(2023)]%
        {wald2023exposing}
\bibfield{author}{\bibinfo{person}{Celine Wald} {and} \bibinfo{person}{Lukas
  Pfahler}.} \bibinfo{year}{2023}\natexlab{}.
\newblock \showarticletitle{Exposing bias in online communities through
  large-scale language models}.
\newblock \bibinfo{journal}{\emph{arXiv}} (\bibinfo{year}{2023}).
\newblock


\bibitem[Wallace et~al\mbox{.}(2024)]%
        {wallace2024instruction}
\bibfield{author}{\bibinfo{person}{Eric Wallace}, \bibinfo{person}{Kai Xiao},
  \bibinfo{person}{Reimar Leike}, \bibinfo{person}{Lilian Weng},
  \bibinfo{person}{Johannes Heidecke}, {and} \bibinfo{person}{Alex Beutel}.}
  \bibinfo{year}{2024}\natexlab{}.
\newblock \showarticletitle{The Instruction Hierarchy: Training LLMs to
  Prioritize Privileged Instructions}.
\newblock \bibinfo{journal}{\emph{arXiv}} (\bibinfo{year}{2024}).
\newblock


\bibitem[Wan et~al\mbox{.}(2023)]%
        {wan2023poisoning}
\bibfield{author}{\bibinfo{person}{Alexander Wan}, \bibinfo{person}{Eric
  Wallace}, \bibinfo{person}{Sheng Shen}, {and} \bibinfo{person}{Dan Klein}.}
  \bibinfo{year}{2023}\natexlab{}.
\newblock \showarticletitle{Poisoning language models during instruction
  tuning}. In \bibinfo{booktitle}{\emph{ICML}}.
\newblock


\bibitem[Wang et~al\mbox{.}(2023a)]%
        {wang2023decodingtrust}
\bibfield{author}{\bibinfo{person}{Boxin Wang}, \bibinfo{person}{Weixin Chen},
  \bibinfo{person}{Hengzhi Pei}, \bibinfo{person}{Chulin Xie},
  \bibinfo{person}{Mintong Kang}, \bibinfo{person}{Chenhui Zhang},
  \bibinfo{person}{Chejian Xu}, \bibinfo{person}{Zidi Xiong},
  \bibinfo{person}{Ritik Dutta}, \bibinfo{person}{Rylan Schaeffer},
  \bibinfo{person}{Sang~T. Truong}, \bibinfo{person}{Simran Arora},
  \bibinfo{person}{Mantas Mazeika}, \bibinfo{person}{Dan Hendrycks},
  \bibinfo{person}{Zinan Lin}, \bibinfo{person}{Yu Cheng},
  \bibinfo{person}{Sanmi Koyejo}, \bibinfo{person}{Dawn Song}, {and}
  \bibinfo{person}{Bo Li}.} \bibinfo{year}{2023}\natexlab{a}.
\newblock \showarticletitle{DecodingTrust: A Comprehensive Assessment of
  Trustworthiness in {GPT} Models}. In \bibinfo{booktitle}{\emph{NeurIPS}}.
\newblock


\bibitem[Wang et~al\mbox{.}(2023b)]%
        {wang2023survey}
\bibfield{author}{\bibinfo{person}{Yuntao Wang}, \bibinfo{person}{Yanghe Pan},
  \bibinfo{person}{Miao Yan}, \bibinfo{person}{Zhou Su}, {and}
  \bibinfo{person}{Tom~H Luan}.} \bibinfo{year}{2023}\natexlab{b}.
\newblock \showarticletitle{A survey on ChatGPT: AI-generated contents,
  challenges, and solutions}.
\newblock \bibinfo{journal}{\emph{IEEE Open Journal of the Computer Society}}
  (\bibinfo{year}{2023}).
\newblock


\bibitem[Wang and Chang(2022)]%
        {wang2022toxicity}
\bibfield{author}{\bibinfo{person}{Yau-Shian Wang} {and}
  \bibinfo{person}{Yingshan Chang}.} \bibinfo{year}{2022}\natexlab{}.
\newblock \showarticletitle{Toxicity detection with generative prompt-based
  inference}.
\newblock \bibinfo{journal}{\emph{arXiv}} (\bibinfo{year}{2022}).
\newblock


\bibitem[Wang et~al\mbox{.}(2024)]%
        {wang2024describe}
\bibfield{author}{\bibinfo{person}{Zihao Wang}, \bibinfo{person}{Shaofei Cai},
  \bibinfo{person}{Guanzhou Chen}, \bibinfo{person}{Anji Liu},
  \bibinfo{person}{Xiaojian~Shawn Ma}, {and} \bibinfo{person}{Yitao Liang}.}
  \bibinfo{year}{2024}\natexlab{}.
\newblock \showarticletitle{Describe, explain, plan and select: interactive
  planning with LLMs enables open-world multi-task agents}.
\newblock \bibinfo{journal}{\emph{NeurIPS}}  \bibinfo{volume}{36}.
\newblock


\bibitem[Weeks et~al\mbox{.}(2023)]%
        {weeks2023first}
\bibfield{author}{\bibinfo{person}{Connor Weeks}, \bibinfo{person}{Aravind
  Cheruvu}, \bibinfo{person}{Sifat~Muhammad Abdullah}, \bibinfo{person}{Shravya
  Kanchi}, \bibinfo{person}{Daphne Yao}, {and} \bibinfo{person}{Bimal
  Viswanath}.} \bibinfo{year}{2023}\natexlab{}.
\newblock \showarticletitle{A first look at toxicity injection attacks on
  open-domain chatbots}. In \bibinfo{booktitle}{\emph{Annual Computer Security
  Applications Conference}}.
\newblock


\bibitem[Wei et~al\mbox{.}(2023a)]%
        {wei2023simple}
\bibfield{author}{\bibinfo{person}{Jerry Wei}, \bibinfo{person}{Da Huang},
  \bibinfo{person}{Yifeng Lu}, \bibinfo{person}{Denny Zhou}, {and}
  \bibinfo{person}{Quoc~V Le}.} \bibinfo{year}{2023}\natexlab{a}.
\newblock \showarticletitle{Simple synthetic data reduces sycophancy in large
  language models}.
\newblock \bibinfo{journal}{\emph{arXiv}} (\bibinfo{year}{2023}).
\newblock


\bibitem[Wei et~al\mbox{.}(2022)]%
        {wei2022chain}
\bibfield{author}{\bibinfo{person}{Jason Wei}, \bibinfo{person}{Xuezhi Wang},
  \bibinfo{person}{Dale Schuurmans}, \bibinfo{person}{Maarten Bosma},
  \bibinfo{person}{Fei Xia}, \bibinfo{person}{Ed Chi}, \bibinfo{person}{Quoc~V
  Le}, \bibinfo{person}{Denny Zhou}, {et~al\mbox{.}}}
  \bibinfo{year}{2022}\natexlab{}.
\newblock \showarticletitle{Chain-of-thought prompting elicits reasoning in
  large language models}.
\newblock \bibinfo{journal}{\emph{NeurIPS}}  \bibinfo{volume}{35},
  \bibinfo{pages}{24824--24837}.
\newblock


\bibitem[Wei et~al\mbox{.}(2024)]%
        {wei2024long}
\bibfield{author}{\bibinfo{person}{Jerry Wei}, \bibinfo{person}{Chengrun Yang},
  \bibinfo{person}{Xinying Song}, \bibinfo{person}{Yifeng Lu},
  \bibinfo{person}{Nathan Hu}, \bibinfo{person}{Dustin Tran},
  \bibinfo{person}{Daiyi Peng}, \bibinfo{person}{Ruibo Liu},
  \bibinfo{person}{Da Huang}, \bibinfo{person}{Cosmo Du}, {et~al\mbox{.}}}
  \bibinfo{year}{2024}\natexlab{}.
\newblock \showarticletitle{Long-form factuality in large language models}.
\newblock \bibinfo{journal}{\emph{arXiv}} (\bibinfo{year}{2024}).
\newblock


\bibitem[Wei et~al\mbox{.}(2023b)]%
        {wei2023jailbreak}
\bibfield{author}{\bibinfo{person}{Zeming Wei}, \bibinfo{person}{Yifei Wang},
  {and} \bibinfo{person}{Yisen Wang}.} \bibinfo{year}{2023}\natexlab{b}.
\newblock \showarticletitle{Jailbreak and guard aligned language models with
  only few in-context demonstrations}.
\newblock \bibinfo{journal}{\emph{arXiv}} (\bibinfo{year}{2023}).
\newblock


\bibitem[Weiss et~al\mbox{.}(2024)]%
        {weiss2024your}
\bibfield{author}{\bibinfo{person}{Roy Weiss}, \bibinfo{person}{Daniel
  Ayzenshteyn}, \bibinfo{person}{Guy Amit}, {and} \bibinfo{person}{Yisroel
  Mirsky}.} \bibinfo{year}{2024}\natexlab{}.
\newblock \showarticletitle{What Was Your Prompt? A Remote Keylogging Attack on
  AI Assistants}.
\newblock \bibinfo{journal}{\emph{arXiv}} (\bibinfo{year}{2024}).
\newblock


\bibitem[Welbl et~al\mbox{.}(2021)]%
        {welbl2021challenges}
\bibfield{author}{\bibinfo{person}{Johannes Welbl}, \bibinfo{person}{Amelia
  Glaese}, \bibinfo{person}{Jonathan Uesato}, \bibinfo{person}{Sumanth
  Dathathri}, \bibinfo{person}{John Mellor}, \bibinfo{person}{Lisa~Anne
  Hendricks}, \bibinfo{person}{Kirsty Anderson}, \bibinfo{person}{Pushmeet
  Kohli}, \bibinfo{person}{Ben Coppin}, {and} \bibinfo{person}{Po-Sen Huang}.}
  \bibinfo{year}{2021}\natexlab{}.
\newblock \showarticletitle{Challenges in Detoxifying Language Models}. In
  \bibinfo{booktitle}{\emph{Findings of EMNLP}}.
\newblock


\bibitem[Windridge et~al\mbox{.}(2021)]%
        {windridge2021utility}
\bibfield{author}{\bibinfo{person}{David Windridge}, \bibinfo{person}{Henrik
  Svensson}, {and} \bibinfo{person}{Serge Thill}.}
  \bibinfo{year}{2021}\natexlab{}.
\newblock \showarticletitle{On the utility of dreaming: A general model for how
  learning in artificial agents can benefit from data hallucination}.
\newblock \bibinfo{journal}{\emph{Adaptive Behavior}} (\bibinfo{year}{2021}).
\newblock


\bibitem[Wolf et~al\mbox{.}(2023)]%
        {wolf2023fundamental}
\bibfield{author}{\bibinfo{person}{Yotam Wolf}, \bibinfo{person}{Noam Wies},
  \bibinfo{person}{Yoav Levine}, {and} \bibinfo{person}{Amnon Shashua}.}
  \bibinfo{year}{2023}\natexlab{}.
\newblock \showarticletitle{Fundamental limitations of alignment in large
  language models}.
\newblock \bibinfo{journal}{\emph{arXiv}} (\bibinfo{year}{2023}).
\newblock


\bibitem[Wooldridge and Jennings(1995)]%
        {wooldridge1995intelligent}
\bibfield{author}{\bibinfo{person}{Michael Wooldridge} {and}
  \bibinfo{person}{Nicholas~R Jennings}.} \bibinfo{year}{1995}\natexlab{}.
\newblock \showarticletitle{Intelligent agents: Theory and practice}.
\newblock \bibinfo{journal}{\emph{The knowledge engineering review}}
  (\bibinfo{year}{1995}).
\newblock


\bibitem[Wu et~al\mbox{.}(2024a)]%
        {wu2024wipi}
\bibfield{author}{\bibinfo{person}{Fangzhou Wu}, \bibinfo{person}{Shutong Wu},
  \bibinfo{person}{Yulong Cao}, {and} \bibinfo{person}{Chaowei Xiao}.}
  \bibinfo{year}{2024}\natexlab{a}.
\newblock \showarticletitle{WIPI: A New Web Threat for LLM-Driven Web Agents}.
\newblock \bibinfo{journal}{\emph{arXiv}} (\bibinfo{year}{2024}).
\newblock


\bibitem[Wu et~al\mbox{.}(2024b)]%
        {wu2024new}
\bibfield{author}{\bibinfo{person}{Fangzhou Wu}, \bibinfo{person}{Ning Zhang},
  \bibinfo{person}{Somesh Jha}, \bibinfo{person}{Patrick McDaniel}, {and}
  \bibinfo{person}{Chaowei Xiao}.} \bibinfo{year}{2024}\natexlab{b}.
\newblock \showarticletitle{A New Era in LLM Security: Exploring Security
  Concerns in Real-World LLM-based Systems}.
\newblock \bibinfo{journal}{\emph{arXiv}} (\bibinfo{year}{2024}).
\newblock


\bibitem[Xi et~al\mbox{.}(2023)]%
        {xi2023rise}
\bibfield{author}{\bibinfo{person}{Zhiheng Xi}, \bibinfo{person}{Wenxiang
  Chen}, \bibinfo{person}{Xin Guo}, \bibinfo{person}{Wei He},
  \bibinfo{person}{Yiwen Ding}, \bibinfo{person}{Boyang Hong},
  \bibinfo{person}{Ming Zhang}, \bibinfo{person}{Junzhe Wang},
  \bibinfo{person}{Senjie Jin}, \bibinfo{person}{Enyu Zhou}, {et~al\mbox{.}}}
  \bibinfo{year}{2023}\natexlab{}.
\newblock \showarticletitle{The rise and potential of large language model
  based agents: A survey}.
\newblock \bibinfo{journal}{\emph{arXiv}} (\bibinfo{year}{2023}).
\newblock


\bibitem[Xie et~al\mbox{.}(2023)]%
        {xie2023brief}
\bibfield{author}{\bibinfo{person}{Xingrui Xie}, \bibinfo{person}{Han Liu},
  \bibinfo{person}{Wenzhe Hou}, {and} \bibinfo{person}{Hongbin Huang}.}
  \bibinfo{year}{2023}\natexlab{}.
\newblock \showarticletitle{A Brief Survey of Vector Databases}. In
  \bibinfo{booktitle}{\emph{International Conference on Big Data and
  Information Analytics (BigDIA)}}. IEEE.
\newblock


\bibitem[Xing(2024)]%
        {xing2024designing}
\bibfield{author}{\bibinfo{person}{Frank Xing}.}
  \bibinfo{year}{2024}\natexlab{}.
\newblock \showarticletitle{Designing Heterogeneous LLM Agents for Financial
  Sentiment Analysis}.
\newblock \bibinfo{journal}{\emph{arXiv}} (\bibinfo{year}{2024}).
\newblock


\bibitem[Xu et~al\mbox{.}(2023a)]%
        {xu2023exploring}
\bibfield{author}{\bibinfo{person}{Yuzhuang Xu}, \bibinfo{person}{Shuo Wang},
  \bibinfo{person}{Peng Li}, \bibinfo{person}{Fuwen Luo},
  \bibinfo{person}{Xiaolong Wang}, \bibinfo{person}{Weidong Liu}, {and}
  \bibinfo{person}{Yang Liu}.} \bibinfo{year}{2023}\natexlab{a}.
\newblock \showarticletitle{Exploring large language models for communication
  games: An empirical study on werewolf}.
\newblock \bibinfo{journal}{\emph{arXiv}} (\bibinfo{year}{2023}).
\newblock


\bibitem[Xu et~al\mbox{.}(2023b)]%
        {xu2023language}
\bibfield{author}{\bibinfo{person}{Zelai Xu}, \bibinfo{person}{Chao Yu},
  \bibinfo{person}{Fei Fang}, \bibinfo{person}{Yu Wang}, {and}
  \bibinfo{person}{Yi Wu}.} \bibinfo{year}{2023}\natexlab{b}.
\newblock \showarticletitle{Language agents with reinforcement learning for
  strategic play in the werewolf game}.
\newblock \bibinfo{journal}{\emph{arXiv}} (\bibinfo{year}{2023}).
\newblock


\bibitem[Yan et~al\mbox{.}(2023)]%
        {yan2023ask}
\bibfield{author}{\bibinfo{person}{Xue Yan}, \bibinfo{person}{Yan Song},
  \bibinfo{person}{Xinyu Cui}, \bibinfo{person}{Filippos Christianos},
  \bibinfo{person}{Haifeng Zhang}, \bibinfo{person}{David~Henry Mguni}, {and}
  \bibinfo{person}{Jun Wang}.} \bibinfo{year}{2023}\natexlab{}.
\newblock \showarticletitle{Ask more, know better: Reinforce-Learned Prompt
  Questions for Decision Making with Large Language Models}.
\newblock \bibinfo{journal}{\emph{arXiv}} (\bibinfo{year}{2023}).
\newblock


\bibitem[Yang et~al\mbox{.}(2024a)]%
        {yang2024watch}
\bibfield{author}{\bibinfo{person}{Wenkai Yang}, \bibinfo{person}{Xiaohan Bi},
  \bibinfo{person}{Yankai Lin}, \bibinfo{person}{Sishuo Chen},
  \bibinfo{person}{Jie Zhou}, {and} \bibinfo{person}{Xu Sun}.}
  \bibinfo{year}{2024}\natexlab{a}.
\newblock \showarticletitle{Watch Out for Your Agents! Investigating Backdoor
  Threats to LLM-Based Agents}.
\newblock \bibinfo{journal}{\emph{arXiv}} (\bibinfo{year}{2024}).
\newblock


\bibitem[Yang et~al\mbox{.}(2024b)]%
        {yang2024prsa}
\bibfield{author}{\bibinfo{person}{Yong Yang}, \bibinfo{person}{Xuhong Zhang},
  \bibinfo{person}{Yi Jiang}, \bibinfo{person}{Xi Chen}, \bibinfo{person}{Haoyu
  Wang}, \bibinfo{person}{Shouling Ji}, {and} \bibinfo{person}{Zonghui Wang}.}
  \bibinfo{year}{2024}\natexlab{b}.
\newblock \showarticletitle{PRSA: Prompt Reverse Stealing Attacks against Large
  Language Models}.
\newblock \bibinfo{journal}{\emph{arXiv}} (\bibinfo{year}{2024}).
\newblock


\bibitem[Yao et~al\mbox{.}(2024)]%
        {yao2024fuzzllm}
\bibfield{author}{\bibinfo{person}{Dongyu Yao}, \bibinfo{person}{Jianshu
  Zhang}, \bibinfo{person}{Ian~G Harris}, {and} \bibinfo{person}{Marcel
  Carlsson}.} \bibinfo{year}{2024}\natexlab{}.
\newblock \showarticletitle{Fuzzllm: A novel and universal fuzzing framework
  for proactively discovering jailbreak vulnerabilities in large language
  models}. In \bibinfo{booktitle}{\emph{ICASSP}}.
\newblock


\bibitem[Yao et~al\mbox{.}(2023)]%
        {yao2023react}
\bibfield{author}{\bibinfo{person}{Shunyu Yao}, \bibinfo{person}{Jeffrey Zhao},
  \bibinfo{person}{Dian Yu}, \bibinfo{person}{Nan Du}, \bibinfo{person}{Izhak
  Shafran}, \bibinfo{person}{Karthik Narasimhan}, {and} \bibinfo{person}{Yuan
  Cao}.} \bibinfo{year}{2023}\natexlab{}.
\newblock \showarticletitle{ReAct: Synergizing Reasoning and Acting in Language
  Models}. In \bibinfo{booktitle}{\emph{ICLR}}.
\newblock


\bibitem[Yi et~al\mbox{.}(2023)]%
        {yi2024benchmarking}
\bibfield{author}{\bibinfo{person}{Jingwei Yi}, \bibinfo{person}{Yueqi Xie},
  \bibinfo{person}{Bin Zhu}, \bibinfo{person}{Keegan Hines},
  \bibinfo{person}{Emre Kiciman}, \bibinfo{person}{Guangzhong Sun},
  \bibinfo{person}{Xing Xie}, {and} \bibinfo{person}{Fangzhao Wu}.}
  \bibinfo{year}{2023}\natexlab{}.
\newblock \showarticletitle{Benchmarking and defending against indirect prompt
  injection attacks on large language models}.
\newblock \bibinfo{journal}{\emph{arXiv}} (\bibinfo{year}{2023}).
\newblock


\bibitem[Yu et~al\mbox{.}(2023)]%
        {yu2023gptfuzzer}
\bibfield{author}{\bibinfo{person}{Jiahao Yu}, \bibinfo{person}{Xingwei Lin},
  {and} \bibinfo{person}{Xinyu Xing}.} \bibinfo{year}{2023}\natexlab{}.
\newblock \showarticletitle{Gptfuzzer: Red teaming large language models with
  auto-generated jailbreak prompts}.
\newblock \bibinfo{journal}{\emph{arXiv}} (\bibinfo{year}{2023}).
\newblock


\bibitem[Yuan et~al\mbox{.}(2024a)]%
        {yuan2024r}
\bibfield{author}{\bibinfo{person}{Tongxin Yuan}, \bibinfo{person}{Zhiwei He},
  \bibinfo{person}{Lingzhong Dong}, \bibinfo{person}{Yiming Wang},
  \bibinfo{person}{Ruijie Zhao}, \bibinfo{person}{Tian Xia},
  \bibinfo{person}{Lizhen Xu}, \bibinfo{person}{Binglin Zhou},
  \bibinfo{person}{Fangqi Li}, \bibinfo{person}{Zhuosheng Zhang},
  {et~al\mbox{.}}} \bibinfo{year}{2024}\natexlab{a}.
\newblock \showarticletitle{R-Judge: Benchmarking Safety Risk Awareness for LLM
  Agents}. In \bibinfo{booktitle}{\emph{ICLR}}.
\newblock


\bibitem[Yuan et~al\mbox{.}(2024b)]%
        {yuan2023gpt}
\bibfield{author}{\bibinfo{person}{Youliang Yuan}, \bibinfo{person}{Wenxiang
  Jiao}, \bibinfo{person}{Wenxuan Wang}, \bibinfo{person}{Jen tse Huang},
  \bibinfo{person}{Pinjia He}, \bibinfo{person}{Shuming Shi}, {and}
  \bibinfo{person}{Zhaopeng Tu}.} \bibinfo{year}{2024}\natexlab{b}.
\newblock \showarticletitle{{GPT}-4 Is Too Smart To Be Safe: Stealthy Chat with
  {LLM}s via Cipher}. In \bibinfo{booktitle}{\emph{ICLR}}.
\newblock
\urldef\tempurl%
\url{https://openreview.net/forum?id=MbfAK4s61A}
\showURL{%
\tempurl}


\bibitem[Yue et~al\mbox{.}(2023)]%
        {yue2023automatic}
\bibfield{author}{\bibinfo{person}{Xiang Yue}, \bibinfo{person}{Boshi Wang},
  \bibinfo{person}{Ziru Chen}, \bibinfo{person}{Kai Zhang}, \bibinfo{person}{Yu
  Su}, {and} \bibinfo{person}{Huan Sun}.} \bibinfo{year}{2023}\natexlab{}.
\newblock \showarticletitle{Automatic Evaluation of Attribution by Large
  Language Models}. In \bibinfo{booktitle}{\emph{EMNLP}}.
\newblock


\bibitem[Zaharia et~al\mbox{.}(2018)]%
        {zaharia2018accelerating}
\bibfield{author}{\bibinfo{person}{Matei Zaharia}, \bibinfo{person}{Andrew
  Chen}, \bibinfo{person}{Aaron Davidson}, \bibinfo{person}{Ali Ghodsi},
  \bibinfo{person}{Sue~Ann Hong}, \bibinfo{person}{Andy Konwinski},
  \bibinfo{person}{Siddharth Murching}, \bibinfo{person}{Tomas Nykodym},
  \bibinfo{person}{Paul Ogilvie}, \bibinfo{person}{Mani Parkhe},
  {et~al\mbox{.}}} \bibinfo{year}{2018}\natexlab{}.
\newblock \showarticletitle{Accelerating the machine learning lifecycle with
  MLflow.}
\newblock \bibinfo{journal}{\emph{IEEE Data Eng. Bull.}}
  (\bibinfo{year}{2018}).
\newblock


\bibitem[Zeng et~al\mbox{.}(2024b)]%
        {zeng2024good}
\bibfield{author}{\bibinfo{person}{Shenglai Zeng}, \bibinfo{person}{Jiankun
  Zhang}, \bibinfo{person}{Pengfei He}, \bibinfo{person}{Yue Xing},
  \bibinfo{person}{Yiding Liu}, \bibinfo{person}{Han Xu}, \bibinfo{person}{Jie
  Ren}, \bibinfo{person}{Shuaiqiang Wang}, \bibinfo{person}{Dawei Yin},
  \bibinfo{person}{Yi Chang}, {et~al\mbox{.}}}
  \bibinfo{year}{2024}\natexlab{b}.
\newblock \showarticletitle{The Good and The Bad: Exploring Privacy Issues in
  Retrieval-Augmented Generation (RAG)}.
\newblock \bibinfo{journal}{\emph{arXiv}} (\bibinfo{year}{2024}).
\newblock


\bibitem[Zeng et~al\mbox{.}(2024a)]%
        {zeng2024autodefense}
\bibfield{author}{\bibinfo{person}{Yifan Zeng}, \bibinfo{person}{Yiran Wu},
  \bibinfo{person}{Xiao Zhang}, \bibinfo{person}{Huazheng Wang}, {and}
  \bibinfo{person}{Qingyun Wu}.} \bibinfo{year}{2024}\natexlab{a}.
\newblock \showarticletitle{AutoDefense: Multi-Agent LLM Defense against
  Jailbreak Attacks}.
\newblock \bibinfo{journal}{\emph{arXiv}} (\bibinfo{year}{2024}).
\newblock


\bibitem[Zhan et~al\mbox{.}(2024)]%
        {zhan2024injecagent}
\bibfield{author}{\bibinfo{person}{Qiusi Zhan}, \bibinfo{person}{Zhixiang
  Liang}, \bibinfo{person}{Zifan Ying}, {and} \bibinfo{person}{Daniel Kang}.}
  \bibinfo{year}{2024}\natexlab{}.
\newblock \showarticletitle{Injecagent: Benchmarking indirect prompt injections
  in tool-integrated large language model agents}.
\newblock \bibinfo{journal}{\emph{arXiv}} (\bibinfo{year}{2024}).
\newblock


\bibitem[Zhang et~al\mbox{.}(2023a)]%
        {zhang2023proagent}
\bibfield{author}{\bibinfo{person}{Hongxin Zhang}, \bibinfo{person}{Weihua Du},
  \bibinfo{person}{Jiaming Shan}, \bibinfo{person}{Qinhong Zhou},
  \bibinfo{person}{Yilun Du}, \bibinfo{person}{Joshua Tenenbaum},
  \bibinfo{person}{Tianmin Shu}, {and} \bibinfo{person}{Chuang Gan}.}
  \bibinfo{year}{2023}\natexlab{a}.
\newblock \showarticletitle{Building Cooperative Embodied Agents Modularly with
  Large Language Models}. In \bibinfo{booktitle}{\emph{NeurIPS Workshop}}.
\newblock


\bibitem[Zhang and Lu(2023)]%
        {zhang2023rladapter}
\bibfield{author}{\bibinfo{person}{Wanpeng Zhang} {and}
  \bibinfo{person}{Zongqing Lu}.} \bibinfo{year}{2023}\natexlab{}.
\newblock \showarticletitle{Rladapter: Bridging large language models to
  reinforcement learning in open worlds}.
\newblock \bibinfo{journal}{\emph{arXiv}} (\bibinfo{year}{2023}).
\newblock


\bibitem[Zhang et~al\mbox{.}(2024d)]%
        {zhang2024privacyasst}
\bibfield{author}{\bibinfo{person}{Xinyu Zhang}, \bibinfo{person}{Huiyu Xu},
  \bibinfo{person}{Zhongjie Ba}, \bibinfo{person}{Zhibo Wang},
  \bibinfo{person}{Yuan Hong}, \bibinfo{person}{Jian Liu},
  \bibinfo{person}{Zhan Qin}, {and} \bibinfo{person}{Kui Ren}.}
  \bibinfo{year}{2024}\natexlab{d}.
\newblock \showarticletitle{Privacyasst: Safeguarding user privacy in
  tool-using large language model agents}.
\newblock \bibinfo{journal}{\emph{IEEE Transactions on Dependable and Secure
  Computing}} (\bibinfo{year}{2024}).
\newblock


\bibitem[Zhang et~al\mbox{.}(2024b)]%
        {zhang2024effective}
\bibfield{author}{\bibinfo{person}{Yiming Zhang}, \bibinfo{person}{Nicholas
  Carlini}, {and} \bibinfo{person}{Daphne Ippolito}.}
  \bibinfo{year}{2024}\natexlab{b}.
\newblock \showarticletitle{Effective Prompt Extraction from Language Models}.
\newblock \bibinfo{journal}{\emph{arXiv}} (\bibinfo{year}{2024}).
\newblock


\bibitem[Zhang et~al\mbox{.}(2024c)]%
        {zhang2024intention}
\bibfield{author}{\bibinfo{person}{Yuqi Zhang}, \bibinfo{person}{Liang Ding},
  \bibinfo{person}{Lefei Zhang}, {and} \bibinfo{person}{Dacheng Tao}.}
  \bibinfo{year}{2024}\natexlab{c}.
\newblock \showarticletitle{Intention analysis prompting makes large language
  models a good jailbreak defender}.
\newblock \bibinfo{journal}{\emph{arXiv}} (\bibinfo{year}{2024}).
\newblock


\bibitem[Zhang et~al\mbox{.}(2023b)]%
        {zhang2023siren}
\bibfield{author}{\bibinfo{person}{Yue Zhang}, \bibinfo{person}{Yafu Li},
  \bibinfo{person}{Leyang Cui}, \bibinfo{person}{Deng Cai},
  \bibinfo{person}{Lemao Liu}, \bibinfo{person}{Tingchen Fu},
  \bibinfo{person}{Xinting Huang}, \bibinfo{person}{Enbo Zhao},
  \bibinfo{person}{Yu Zhang}, \bibinfo{person}{Yulong Chen}, {et~al\mbox{.}}}
  \bibinfo{year}{2023}\natexlab{b}.
\newblock \showarticletitle{Siren’s song in the AI ocean: A survey on
  hallucination in large language models}.
\newblock \bibinfo{journal}{\emph{arXiv}} (\bibinfo{year}{2023}).
\newblock


\bibitem[Zhang et~al\mbox{.}(2024a)]%
        {zhang2024survey}
\bibfield{author}{\bibinfo{person}{Zeyu Zhang}, \bibinfo{person}{Xiaohe Bo},
  \bibinfo{person}{Chen Ma}, \bibinfo{person}{Rui Li}, \bibinfo{person}{Xu
  Chen}, \bibinfo{person}{Quanyu Dai}, \bibinfo{person}{Jieming Zhu},
  \bibinfo{person}{Zhenhua Dong}, {and} \bibinfo{person}{Ji-Rong Wen}.}
  \bibinfo{year}{2024}\natexlab{a}.
\newblock \showarticletitle{A Survey on the Memory Mechanism of Large Language
  Model based Agents}.
\newblock \bibinfo{journal}{\emph{arXiv}} (\bibinfo{year}{2024}).
\newblock


\bibitem[Zhang et~al\mbox{.}(2023c)]%
        {zhang2023defending}
\bibfield{author}{\bibinfo{person}{Zhexin Zhang}, \bibinfo{person}{Junxiao
  Yang}, \bibinfo{person}{Pei Ke}, {and} \bibinfo{person}{Minlie Huang}.}
  \bibinfo{year}{2023}\natexlab{c}.
\newblock \showarticletitle{Defending large language models against
  jailbreaking attacks through goal prioritization}.
\newblock \bibinfo{journal}{\emph{arXiv}} (\bibinfo{year}{2023}).
\newblock


\bibitem[Zhao et~al\mbox{.}(2023)]%
        {zhao2023competeai}
\bibfield{author}{\bibinfo{person}{Qinlin Zhao}, \bibinfo{person}{Jindong
  Wang}, \bibinfo{person}{Yixuan Zhang}, \bibinfo{person}{Yiqiao Jin},
  \bibinfo{person}{Kaijie Zhu}, \bibinfo{person}{Hao Chen}, {and}
  \bibinfo{person}{Xing Xie}.} \bibinfo{year}{2023}\natexlab{}.
\newblock \showarticletitle{Competeai: Understanding the competition behaviors
  in large language model-based agents}.
\newblock \bibinfo{journal}{\emph{arXiv}} (\bibinfo{year}{2023}).
\newblock


\bibitem[Zhou et~al\mbox{.}(2024)]%
        {zhou2023analyzing}
\bibfield{author}{\bibinfo{person}{Yiyang Zhou}, \bibinfo{person}{Chenhang
  Cui}, \bibinfo{person}{Jaehong Yoon}, \bibinfo{person}{Linjun Zhang},
  \bibinfo{person}{Zhun Deng}, \bibinfo{person}{Chelsea Finn},
  \bibinfo{person}{Mohit Bansal}, {and} \bibinfo{person}{Huaxiu Yao}.}
  \bibinfo{year}{2024}\natexlab{}.
\newblock \showarticletitle{Analyzing and Mitigating Object Hallucination in
  Large Vision-Language Models}. In \bibinfo{booktitle}{\emph{ICLR}}.
\newblock


\bibitem[Zou et~al\mbox{.}(2024)]%
        {zou2024poisonedrag}
\bibfield{author}{\bibinfo{person}{Wei Zou}, \bibinfo{person}{Runpeng Geng},
  \bibinfo{person}{Binghui Wang}, {and} \bibinfo{person}{Jinyuan Jia}.}
  \bibinfo{year}{2024}\natexlab{}.
\newblock \showarticletitle{PoisonedRAG: Knowledge Poisoning Attacks to
  Retrieval-Augmented Generation of Large Language Models}.
\newblock \bibinfo{journal}{\emph{arXiv}} (\bibinfo{year}{2024}).
\newblock


\end{thebibliography}

\appendix

\end{document}